\documentclass[aps,prd,10pt,amsmath,amssymb,tightenlines,preprint,nofootinbib,groupedaddress,superscriptaddress,floatfix]{revtex4-2}

\usepackage{graphicx} 
\usepackage{amsmath,amssymb}
\usepackage{enumerate}
\usepackage{dcolumn}
\usepackage{bm}
\usepackage{microtype}
\usepackage{xcolor}
\usepackage{bbold}
\usepackage{soul}

\definecolor{navyblue}{rgb}{0.0, 0.0, 0.5}

\usepackage{hyperref}
\usepackage{cleveref}
\crefname{equation}{Eq.}{Eqs.}
\Crefname{equation}{Equation}{Equations}
\crefname{figure}{Fig.}{Figs.}
\Crefname{figure}{Figure}{Figures}
\crefname{appendix}{appendix}{appendices}
\Crefname{appendix}{Appendix}{Appendices}

\makeatletter
\AddToHook{cmd/appendix/before}{\def\cref@section@alias{appendix}}
\makeatother

\hypersetup{
    colorlinks=true,
    linktoc=all,
    allcolors=navyblue
}

\definecolor{mulberry}{rgb}{0.5,0,0.5}
\definecolor{elderberry}{rgb}{0.3,0,0.7}
\definecolor{raspberry}{RGB}{255, 87, 51}
\definecolor{gooseberry}{RGB}{0, 171, 79}
\definecolor{blueberry}{RGB}{75, 134, 247}
\definecolor{winterberry}{RGB}{255, 135, 0}

\definecolor{DarkSlateNavy}{rgb}{0.106, 0.106, 0.227}
\definecolor{DeepPurple}{rgb}{0.294, 0.180, 0.514}
\definecolor{HypotheticBlue}{rgb}{0.216, 0.494, 0.722}
\definecolor{HypotheticGreen}{rgb}{0.302, 0.686, 0.290}
\definecolor{HypotheticOrange}{rgb}{1.000, 0.498, 0.000}

\makeatletter
\def\l@subsubsection#1#2{}%
\def\l@paragraph#1#2{}%
\def\l@subparagraph#1#2{}%
\makeatother
\begin{document}

\title{
Axiverse Lampposts}
\author{Masha Baryakhtar}
\email{mbaryakh@uw.edu}
\affiliation{Department of Physics, University of Washington, Seattle, WA 98195, U.S.A.}
\author{David Cyncynates}
\email{davidcyn@ictp.it}
\affiliation{Department of Physics, University of Washington, Seattle, WA 98195, U.S.A.}
\affiliation{International Centre for Theoretical Physics (ICTP), Strada Costiera 11, 34151 Trieste, Italy}
\affiliation{Istituto Nazionale di Fisica Nucleare (INFN), Sezione di Trieste, I-34127 Trieste, Italy}
\author{Ella Henry}
\email{ech43@uw.edu}
\affiliation{Department of Physics, University of Washington, Seattle, WA 98195, U.S.A.}
\date{\today}

\begin{abstract}
The string axiverse predicts a unique connection between the high scales approachable only through theory and the low energies within reach of experimental verification: a multitude of light, feebly interacting axions. In order to capture the collective effects of such an axion ensemble, we model the string axiverse by $N$ coupled axions with a simple assumption: hierarchical axion masses that arise from hierarchical instantons with statistically distributed axion couplings. In this limit, we find that axion field ranges, which determine late-time cosmological abundances, shrink as $1/\sqrt{N}$ as the number of axions grows. Moreover, the heaviest modes tend to align with the smallest kinetic eigenvalues, further reducing their field ranges. Interactions with the Standard Model (SM) are largely set by the kinetic structure and do not grow with $N$, thus suppressing detection prospects relative to the individual-axion expectation. The main exception is the QCD axion, whose coupling is tied to its potential and is therefore unsuppressed. The heaviest and lightest axions can also avoid the typical suppression in certain limits. We further find that coupled axiverse dark matter 
has parametrically relaxed tuning on initial conditions when produced via long, low-scale inflation relative to independent axions and high-scale inflation. 
Taken together, these results sharpen the observational outlook: the most accessible signals typically come from the QCD axion and from heavy axions that make up small dark matter subcomponents. An anthropic plateau of comparable energy density states produces subdominant signals; meanwhile, if light axions have SM interactions independent of QCD, they can also be within reach of future direct-detection experiments.
\end{abstract}
\maketitle
\newpage
\tableofcontents

\section{Introduction}
\label{sec:intro}

One of the most surprising predictions of string theory is that we may live in a universe with more than just the four dimensions of space and time. In practice, nature must hide these extra dimensions, rendering many of their consequences out of experimental reach. However, some remnants can survive into the four-dimensional effective description of our world~\cite{Taylor:1988nw,Damour:1994ya,Arkani-Hamed:1998jmv, Randall:1999ee, Svrcek:2006yi,Arvanitaki:2009fg, Arvanitaki:2009hb,Goodsell:2009xc}. A leading example is provided by axions---light, feebly interacting, shift-symmetric pseudoscalars---which descend from higher-dimensional gauge fields in the compactified space of perturbative string theory~\cite{Svrcek:2006yi}. In the effective theory, the underlying gauge redundancy appears as a (typically discrete) shift symmetry, protecting axion masses from perturbative corrections. This makes axions compelling targets for experimental tests of physics beyond the Standard Model across many scales~\cite{Arvanitaki:2009fg}.

Axions were first introduced in Refs.~\cite{Peccei:1977hh,Weinberg:1977ma,Wilczek:1977pj} to resolve the strong CP problem of Quantum Chromodynamics (QCD). In the simplest effective description, the QCD axion carries an approximate shift symmetry that is broken predominantly by nonperturbative strong dynamics, generating an effective potential with a minimum at the CP-conserving point,
\begin{align}\label{eqn:QCD_single_axion}
    {\cal L} \approx \frac12 (\partial a)^2 - \chi_c\left(1 - \cos \frac{a}{f_a}\right)\,,
\end{align}
where $\chi_c$ is the QCD topological susceptibility ($\simeq (75\,{\rm MeV})^4$ at zero temperature), $f_a$ is the axion decay constant, and the axion mass is $m = \chi_c^{1/2}/f_a$~\cite{Witten:1979vv}.
Soon after its introduction, it was realized that this same particle can play the role of the dark matter: after inflation, the axion generically begins displaced from its minimum and later undergoes coherent oscillations whose energy density redshifts as nonrelativistic matter~\cite{Abbott:1982af,Dine:1982ah,Preskill:1982cy}. Because the QCD axion couples to Standard Model fields, it has a rich experimental and astrophysical phenomenology and is a prime target for direct and indirect searches.

More broadly, string compactifications generically contain many axionlike fields. In addition to QCD instantons, nonperturbative effects, such as Euclidean brane instantons in the extra dimensions and strong hidden-sector dynamics, can generate axion potentials. The number of axions is controlled by the topology of the compactification, which can be very complex, motivating the expectation of tens to hundreds of axions with a wide range of masses and couplings: the string axiverse~\cite{Arvanitaki:2009fg}.

As a baseline, it is useful to begin with an ``independent-axion'' picture in which the axiverse consists of $N$ decoupled fields,
\begin{align}\label{eqn:decoupled_Lagrangian}
    {\cal L}
    = \frac12(\partial a_i)^2
    - \Lambda_i^4\left(1 - \cos\frac{a_i}{f_{i}}\right)\,,
\end{align}
where $i=1,\dots,N$, $f_{i}$ is the decay constant, and $\Lambda_i$ is the instanton scale (with $\Lambda_{i_c}^4=\chi_c$ for the QCD axion).\footnote{Summation over repeated indices is implied throughout this manuscript unless otherwise stated; repeated non-summed indices will appear in parentheses, e.g.\ $(i)$.}
The scales $\Lambda_i$ are typically exponentially sensitive to instanton actions,
\begin{align}
    \Lambda_i^4 \sim M_{\rm UV}^4 e^{-S_i}\,,
\end{align}
so axion masses can naturally populate many decades even when decay constants vary only polynomially~\cite{Arvanitaki:2009fg}. As a consequence, axions can play a role across a wide range of scales, from strong CP and dark matter~\cite{Peccei:1977hh,Wilczek:1977pj,Weinberg:1977ma,Abbott:1982af,Sikivie:1983ip,Dine:1982ah,Preskill:1982cy}, to inflation~\cite{Freese:1990rb,Dimopoulos:2005ac}, baryogenesis~\cite{Co:2019wyp,Asadi:2025cvm}, and anthropic scanning~\cite{Bachlechner:2017zpb,Bachlechner:2017hsj,Bachlechner:2018gew,Bachlechner:2019vcb}.

In the decoupled picture, each axion can contribute a relic abundance set primarily by its initial misalignment angle and microscopic parameters
\begin{align}
    \Omega_i \sim \left(\frac{m_{i}}{H_{\rm eq}}\right)^{1/2}
    \left(\frac{f_{i}}{M_{\rm pl}}\right)^2\theta_{0,i}^2\,,
\end{align}
where $H_{\rm eq}$ is the Hubble rate at matter--radiation equality, $M_{\rm pl}$ is the reduced Planck mass, and $\theta_{0,i}\equiv a_i(0)/f_{i}$ is set by inflationary dynamics.
With many axions and a wide range of possibilities for their field ranges and potentials, $(f_{i},\Lambda_i)$, this immediately raises a basic question: what experimental consequences should we generically expect from an axiverse, and which regions of parameter space are actually populated once inflationary initial conditions and observed dark matter abundance are taken into account?

The same freedom appears in couplings. A ``generic axionlike particle'' has couplings to photons~\cite{Kim:1979if,Zhitnitsky:1980tq,Dine:1981rt} or fermions~\cite{Srednicki:1985xd,Georgi:1986df} which scale as $1/f_{i}$ up to ${\cal O}(1)$ coefficients, e.g.\
\begin{align}
\label{eqn:axion_photon_coupling}
    {\cal L}_{\rm int}\supset
    C_{a_i\gamma\gamma}\frac{\alpha_{\rm EM}}{2\pi}\frac{a_i}{f_{i}}\frac{1}{4}\tilde F_{\mu\nu}F^{\mu\nu}
    \quad{\rm or}\quad
    C_{a_if}\frac{1}{2 f_{i}}(\partial_\mu a_i)\bar\psi_f\gamma^\mu\gamma^5\psi_f\,,
\end{align}
where $F_{\mu\nu}$ is the photon field strength tensor, $\alpha_{\rm EM}$ is the electromagnetic fine structure constant, $\psi_f$ is the field of a Standard Model fermion, and $C_{a_i\gamma\gamma},C_{a_if}$ are model-dependent, dimensionless coefficients.
However, the QCD coupling is qualitatively different,
\begin{align}
    {\cal L}_{\rm int}\supset
    \frac{\alpha_c}{2\pi}\frac{a_{i_c}}{f_{i_c}}\frac{1}{2}{\rm Tr}\,\tilde G_{\mu\nu} G^{\mu\nu}\,,
\end{align}
since it both controls axion-gluon interactions and generates the QCD instanton potential, where here $G$ is the gluon field strength, and $\alpha_c$ is the QCD fine structure constant. 
In an unmixed picture, exactly one axion couples to QCD. But in realistic multi-axion effective theories, kinetic mixing and overlapping instanton charge vectors are generic, and the identification of ``the'' QCD axion is basis-dependent~\cite{Arvanitaki:2009fg}. As a result, the decoupled baseline can miss important physics: mixing can reshape both relic abundances and the strength with which axions couple to the Standard Model.

Several lines of work have investigated these effects from complementary perspectives.
Top-down studies in explicit Calabi--Yau compactifications compute kinetic mixing and leading nonperturbative contributions to the potential in controlled corners of moduli space, yielding concrete lessons about typical masses and couplings~\cite{Cicoli:2012sz,Demirtas:2018akl,Halverson:2019kna,Halverson:2019cmy,Demirtas:2021gsq,Mehta:2021pwf,Demirtas:2022hqf,Gendler:2023kjt,Sheridan:2024vtt,Gendler:2024adn,Fallon:2025lvn,Cheng:2025ggf}. 
Bottom-up approaches instead model ensembles of multi-axion EFTs to identify robust large-$N$ phenomena under minimal assumptions~\cite{Kim:2007bc,Dimopoulos:2005ac,Bachlechner:2014hsa,Stott:2017hvl,Bachlechner:2017hsj,Bachlechner:2017zpb,Bachlechner:2018gew,Bachlechner:2019vcb,Reig:2021ipa}. These studies reveal that many-axion dynamics can yield qualitatively new effects, including modified experimental signatures in laboratories \cite{deGiorgi:2025ldc}, cosmological stasis~\cite{Dienes:2011ja,Dienes:2011sa,Dienes:2012jb,Halverson:2024oir,Dienes:2021woi,Dienes:2023ziv,Dienes:2024wnu,Dienes:2025tox}, non-thermal energy transfer~\cite{Cyncynates:2021xzw,Cyncynates:2022wlq,Kitajima:2014xla,Daido:2015cba,Murai:2023xjn,Cyncynates:2023esj,Murai:2024nsp,Li:2024okl,Li:2024kdy,Li:2025cep,Nakagawa:2020eeg}, topological defects~\cite{Daido:2015bva,Agrawal:2019lkr,Agrawal:2020euj,Petrossian-Byrne:2025mto,Cyncynates:2021xzw}, quintessence~\cite{Kamionkowski:2014zda,Karwal:2016vyq,Emami:2016mrt,Reig:2021ipa,Katewongveerachart:2026ovj}, and dark radiation~\cite{Acharya:2015zfk,Dessert:2025yvk}, and that priors on kinetic structure and charges can strongly control field ranges and couplings.

At the same time, explicit compactification studies indicate that axion potentials in controlled regimes can be substantially less rich than suggested by the most general bottom-up ensembles. In bottom-up effective theories, the charge matrix can be made essentially arbitrarily rich: one may introduce instantons through generalized KSVZ-type matter sectors that couple to different axion combinations, thereby populating the charge matrix with few restrictions (see e.g.\ the construction in Ref.~\cite{Cyncynates:2023esj}). On the other hand, Ref.~\cite{Gendler:2023hwg} finds that in a large sample of type IIB orientifolds in a perturbatively controlled regime, the number of distinct axion minima per geometry is typically small, reflecting the sparsity of the leading charge matrix and the exponential suppression of subleading terms. This does not mean that axion landscapes are generically sparse, but rather that richness is not automatic in the currently best-controlled corners of string moduli space. Moreover, these conclusions are tied to specific corners of moduli space and simplified treatments of moduli stabilization~\cite{Halverson:2018cio,Halverson:2019cmy,Gendler:2023kjt}. Whether more complete stabilization mechanisms populate ensembles with different degrees of hierarchy, sparsity, and mixing remains an open question, motivating intermediate approaches that incorporate robust UV features while retaining flexibility in the low-energy statistical structure.

In this paper we take such an intermediate, ensemble-based view. We consider the most general multi-axion Lagrangian consistent with the discrete axion shift symmetries,
\begin{align}\label{eqn:generic_axion_potential}
    {\cal L}
    =  \frac12(\partial_\mu\bm \theta)^T \bm K (\partial^\mu \bm\theta)
    - \Lambda_i^4 \!\left[1-\cos(\bm r_i^T\bm\theta + \delta_i)\right]\,,
\end{align}
neglecting field dependence of the positive-definite kinetic matrix $\bm K$. Here $\bm r_i$ are integer charge vectors, $\delta_i$ are constant phases, and $\bm \theta$ denotes the $N$ dimensionless axions in the fundamental basis where $2\pi$ periodicities are manifest.

Our goal is to develop a tractable, theory-motivated baseline for axiverse phenomenology. We adopt a minimal UV prior: the instanton actions are sufficiently spread such that $\Lambda_i\gg\Lambda_{i+1}$ and the potential is hierarchical. In this regime the spectrum can be organized by sequentially integrating out heavy modes, and the resulting mass eigenstates track a simple orthogonalization of the charge directions (the Gram-Schmidt basis)~\cite{Demirtas:2021gsq,Gendler:2023kjt}. We then ask what this hierarchy implies for relic abundances, couplings, and discovery prospects.

Given the hierarchy, we treat the remaining ingredients statistically. Concretely, we assume the instanton charge vectors to have fixed, arbitrary length with orientations in field space that are independent and identically distributed (i.i.d.), while allowing for general kinetic structure and phases.\footnote{In the numerical examples, we also allow the lengths to fluctuate according to the specified component distributions, which mainly introduces radial scatter around the fixed-length estimates.} Within this setup we derive analytic estimates for the effective field ranges of the hierarchical eigenstates, the resulting relic abundance scalings across the spectrum (including when and why they depart from the independent axion expectation $\Omega_i\propto m_i^{1/2}$), and the couplings of the mass eigenstates to Standard Model operators. We further study initial condition priors in long inflation models and anthropic weighting of the resulting energy densities to quantify the probability of achieving the observed dark matter abundance and its distribution across the axion ensemble. We then combine these ingredients to identify which regions of parameter space are populated and which are most visible in experiments. In this minimal hierarchical axiverse, the best discovery prospects tend to come from the QCD axion in direct searches and from heavy axion subcomponents in indirect searches such as decays.

This paper is organized as follows. In \cref{sec:review}, we review standard results which will be useful throughout the remainder of the text: single axion dark matter and detection, the enhancement of axion effective field ranges through collective motion, and the Gram-Schmidt orthogonalization procedure. In \cref{sec:Statistics}, we present statistical results about the axion field ranges and couplings of both generic axionlike particles and the QCD axion in a hierarchical axiverse. In \cref{sec:anthropics}, we review our scheme for determining the axion initial conditions and the assignment of an anthropic probability to a given axion ensemble. We then show how this probability can be easily estimated by counting the number of ``relevant axions''. In \cref{sec:results}, we present our results as a unified picture by performing a Monte Carlo sampling of example axion ensembles. We discuss the resulting distribution of the dark matter abundance among the axions as well as the detection prospects for these relics. We conclude in \cref{sec:conclusion}. In this manuscript, we use ``axion'' and ``axionlike particle'' (ALP) interchangeably, and distinguish both from the QCD axion, i.e.\ the axion that solves the strong CP problem.

\section{Review}\label{sec:review}

This section collects a few standard results that we will use repeatedly. We begin with the single-axion framework, emphasizing the parametric scalings that control misalignment production and the resulting direct and indirect detection signals. We then summarize several results on axion inflation that will serve as useful reference points when we discuss multi-axion effective field ranges. Finally, we review the Gram-Schmidt procedure and its connection to the approximate mass eigenbasis in the hierarchical limit.

\subsection{Single-axion production and detection}
\label{subsec:single_axion}

We start by reviewing the single-axion picture and fixing conventions. Our goal is twofold: (i) to recall how misalignment sets the relic abundance in terms of the axion mass and its initial displacement, and (ii) to highlight the corresponding parametric dependence of direct and indirect signatures on the axion decay constant. These scalings will provide the baseline which the multi-axion theory will alter.

Including the ${\cal O}(1)$ factors that we suppressed in \cref{sec:intro}, the misalignment abundance of a single axion is given by
\begin{align}
\label{eqn:axion-relic-abundance}
    \Omega_a \approx \frac{2^{5/4}\Gamma^2(5/4)}{3\pi}\left(\frac{g_{\star s}(T_{\rm eq})}{g_{\star s}(T_{\rm osc})}\right)\left(\frac{g_\star(T_{\rm osc})}{g_\star(T_{\rm eq})}\right)^{3/4}\left(\frac{m}{H_{\rm eq}}\right)^{1/2} \left(\frac{f_a}{M_{\rm pl}}\right)^2F(\theta_0)\,.
\end{align}
Here, $M_{\rm pl}\approx 2.44\times 10^{18}\,{\rm GeV}$ is the reduced Planck mass, $g_\star$ is the number of relativistic degrees of freedom, $g_{\star s}$ the effective number of degrees of freedom in entropy, and $T_{\rm osc}$ is the temperature of the Universe when axion oscillations begin (around $3 H\approx m$). This equation presumes that $m>H_{\rm eq}\approx 2.25\times 10^{-28}\,{\rm eV}$ is the Hubble rate at matter-radiation equality, and the corresponding photon temperature is $T_{\rm eq}\approx 0.796\,{\rm eV}$~\cite{Planck:2018vyg}. The axion initial condition enters as
\begin{align}
\label{eqn:Ftheta}
    F(\theta_{0}) \approx \theta_{0}^2\log\left[\frac{e}{1 - \theta_{0}^2/\pi^2}\right]\,,
\end{align}
where the logarithm accounts for the delay in axion oscillations due to the anharmonicity of the axion potential near the top of the cosine \cite{Visinelli:2009zm}.

Assuming the Peccei-Quinn (PQ) symmetry is broken during inflation and not subsequently restored, the distribution of the initial misalignment angle $\theta_0$ depends on the duration of inflation and its scale relative to the size of the axion potential $V(\theta)$. For a sufficiently long inflationary period, the axion undergoes stochastic fluctuations with a stationary probability distribution $p(\theta_0)$ given by~\cite{Starobinsky:1986fx,Dimopoulos:1988pw,Starobinsky:1994bd}
\begin{align}
    p(\theta_0)\ \propto\ \exp\left[-\frac{8\pi^2}{3H_I^4}V(\theta_0)\right]\,.
\end{align}
Absent additional dynamics during inflation (see e.g.\ Refs.~\cite{Co:2018mho,Huang:2020etx}), one typically expects $\theta_0 \sim \mathcal{O}(1)$. In particular, in the limiting case where the axion potential is everywhere subdominant, $V(\theta)\ll H_I^4$, the initial misalignment angle is approximately uniformly distributed on $\theta_0\in[-\pi,\pi]$. In comparison, for a shorter inflationary epoch (e.g.\ the last $\sim 50 - 60$ $e$-folds) a sufficient condition for the field not to relax appreciably is that it be effectively overdamped, $\sqrt{V'(\theta)}/f_a \ll 3H_I$, so that classical drift is negligible over the duration of inflation. In either case, in a single-axion theory under minimal assumptions, it is typical for $\theta_0\sim 1$ and the maximum available excursion is set by the periodicity, $\Delta a\sim 2\pi f_a$.

The decay constant also sets the scale of axion interactions, for instance with photons in \cref{eqn:axion_photon_coupling},
\begin{align}
   {\cal L} \supset  \frac{g_{a\gamma\gamma} a}{4}\tilde F_{\mu\nu}F^{\mu\nu}\,,
\end{align}
where the axion-photon coupling $g_{a\gamma\gamma}$ is related to the decay constant via
\begin{align}
\label{eqn:gagg}
    g_{a\gamma\gamma}= C_{a\gamma\gamma} \frac{\alpha_{\rm EM}}{2\pi f_a}\,.
\end{align}
The power deposited by the axion in a laboratory experiment such as a haloscope cavity will scale with the axion parameters as~\cite{Sikivie:1983ip}
\begin{align}
\label{eq:P_det}
    P_{\rm det} \propto g_{a\gamma\gamma}^2 \rho_a\,,
\end{align}
where $\rho_a\propto\Omega_a$ is the dark matter energy density. Using \cref{eqn:axion-relic-abundance,eqn:gagg}, we find
\begin{align}
    P_{\rm det}  \propto \frac{1}{f_a^2}\times m^{1/2}f_a^2\theta_0^2\propto m^{1/2}\theta_0^2\,,
\end{align}
so that the leading dependence on $f_a$ cancels up to the $\mathcal{O}(1)$ model-dependent coefficients. This illustrates a useful point: in the simplest single-axion picture, the direct-detection prospects at fixed mass are controlled mainly by the initial misalignment angle (or equivalently the axion energy density), rather than by the microscopic scale $f_a$ itself~\cite{Cyncynates:2021xzw,Cyncynates:2022wlq}.

The same cancellation also appears in decay-based signatures. For the two-photon interaction above, the vacuum decay rate scales as $\Gamma_{a\to\gamma\gamma}\propto g_{a\gamma\gamma}^2 m^3 \propto m^3/f_a^2$, so that the decay power per unit volume, $\rho_a \Gamma_{a\to\gamma\gamma}$, and thus the photon energy flux incident on a detector $\Phi_{\rm det} $, is proportional to
\begin{align}
    \Phi_{\rm det} \propto \rho_a \Gamma_{a\to\gamma\gamma}\ \propto\ (m^{1/2} f_a^2 \theta_0^2)\times (m^3/f_a^2)
    \ \propto\ m^{7/2}\,\theta_0^2\,,
\end{align}
and is likewise independent of $f_a$ at fixed $m$ in the naive single-axion misalignment picture. 
It is worth noting that this cancellation extends to other non-thermal production mechanisms, e.g.\ postinflationary production.

In the axiverse, however, the notion of a single ``decay constant'' becomes ambiguous: both the maximum excursion of a particular axionlike particle and the strength with which it couples to matter depend on the geometry of field space, kinetic mixing, and the structure of the couplings. The issue of separating the axion field range and its coupling constant is at the heart of \cref{sec:Statistics}.

\subsection{$N$-flation and axion alignment}
\label{subsec:N-flation}

Multi-field dynamics have long played a central role in inflationary model building. In large-field scenarios, such as natural inflation, slow roll over $\sim 50$ -- $60$ $e$-folds typically favors a super-Planckian canonical excursion in order to maintain sufficient flatness. Early examples of many fields working collectively to sustain slow roll go back to Ref.~\cite{Liddle:1998jc}. In Ref.~\cite{Dimopoulos:2005ac}, axions were particularly well-motivated because their approximate shift symmetries make their potentials technically natural, and because string compactifications often furnish many such fields.

For a single axion with periodicity $2\pi f_a$, the field range per period is $\Delta a \sim 2\pi f_a$, while simple natural-inflation potentials typically prefer an effective range $\Delta a \gtrsim M_{\rm pl}$. Realizing $f_a \gtrsim M_{\rm pl}$ is widely viewed as challenging in quantum gravity~\cite{Arkani-Hamed:2006emk,Rudelius:2014wla,Brown:2015lia,Heidenreich:2015wga}, motivating mechanisms in which multiple axions combine to produce an effective super-Planckian excursion even when each individual periodicity is sub-Planckian, although such enhancements may be constrained by the weak gravity conjecture~\cite{Heidenreich:2015wga,Bachlechner:2015qja,Hebecker:2015rya,Brown:2015lia}.

A useful way to see the origin of the collective enhancement is to expand a multi-axion theory about a minimum,
\begin{align}
    {\cal L}
    = \frac12\,\partial_\mu a_i\,\partial^\mu a_i
    - \frac12 m_i^2 a_i^2
    + \frac{1}{24}\lambda_i a_i^4 + \dots\,,
\end{align}
where $m_i$ and $\lambda_i$ encode the local curvature and self-interactions of the potential. If the masses and quartics are approximately degenerate, $m_i=m$ and $\lambda_i=\lambda$, then along the radial trajectory $\rho^2=\sum_i a_i^2$ one finds
\begin{align}
    {\cal L}
    = \frac12(\partial \rho)^2
    - \frac12 m^2\rho^2
    + \frac{\lambda}{24N}\rho^4 + \dots\,,
\end{align}
so that the self-interactions along the radial direction are suppressed by $1/N$. For axions with comparable periodicities and approximately isotropic kinetic terms, the same line of argument implies that the radial direction spans an enhanced distance: if $ a_i = f_a \theta_i$ with $\theta_i \sim \theta$ along the radial direction, then $\rho = \sqrt{N}\,f_a\,\theta$, and one period corresponds to
\begin{align}
    \Delta \rho \sim 2\pi \sqrt{N}\,f_a\,.
\end{align}
Equivalently, the radial direction behaves as an axion with an effective decay constant enhanced by $\sqrt{N}$~\cite{Dimopoulos:2005ac}. This is the origin of the $\sqrt{N}$ scaling expectation in multi-axion models with approximately isotropic kinetic terms and comparable potential terms.

This reasoning does not require exact degeneracy: as argued in Ref.~\cite{Dimopoulos:2005ac}, a broad set of light fields can still yield an enhancement scaling roughly as $\sqrt{N_{\rm eff}}$, where $N_{\rm eff}$ is the number of fields that remain dynamical over the relevant epoch. At its core, however, the argument relies on an (approximately) diagonal, isotropic kinetic term, ${\bm K}\propto f_a^2 {\bm I}$, together with potential terms of comparable size so that an effective radial mode can be identified. As we will show in \cref{sec:Statistics}, in the regime of interest for the axiverse where the axion potentials are hierarchical, this intuition can fail dramatically, and the $\sqrt{N}$ scaling is turned on its head. 

The broader lesson---that many light fields can behave collectively---along with related ideas such as alignment~\cite{Kim:2004rp}, motivated a complementary line of work that treats the axiverse from an ensemble perspective.
Early work along these lines employed random-matrix methods to study multi-axion effective theories of the form \cref{eqn:generic_axion_potential}. In particular, Ref.~\cite{Bachlechner:2017hsj} (see also Refs.~\cite{Bachlechner:2017zpb,Bachlechner:2018gew,Bachlechner:2019vcb}) argued that axion field ranges, quantified by the ``diameter'' of a smooth patch of the potential, can be enhanced up to $N^{3/2}$ in certain classes of random ensembles. Ref.~\cite{Bachlechner:2017hsj} interprets this scaling as the product of three distinct $\sqrt{N}$ factors with different origins: (i) a ``kinetic alignment'' factor associated with the typical orientation of the long direction in the kinetic metric relative to the instanton constraints, (ii) a ``lattice alignment'' factor arising from the small singular values of the (near-square) random charge matrix whose rows are $\bm r_i$, which can yield an anomalously long direction in the fundamental domain, and (iii) an additional $\sqrt{N}$ enhancement when the charge matrix is sparse, so that along delocalized directions the argument of each cosine varies only as $1/\sqrt{N}$.

A general lesson from these random-ensemble studies is that even modest assumptions about ``generic'' ultraviolet data tend to produce substantial structure in the infrared. Much of that literature (e.g.\ Ref.~\cite{Bachlechner:2017hsj} and related work) considers ensembles in which many nonperturbative terms compete, and asks when kinetic and random matrix effects can yield large field-range enhancements. Our perspective is complementary: rather than optimizing for the largest achievable range, we seek a tractable baseline for the full spectrum, including the typical field ranges and Standard-Model couplings of all axions.

To obtain concrete statements we adopt two assumptions about the ensemble. The first, weaker assumption is that the axion potentials are hierarchical, $\Lambda_i\gg\Lambda_{i + 1}$, so the spectrum organizes sequentially and the approximate eigenbasis assumes a particularly simple form. The second, stronger assumption is that the instanton charge vector orientations are i.i.d. By leaving the magnitudes of the instanton charge vectors as fixed, free parameters, we allow for possible correlations between the instanton actions $S_i$ and the charge vector magnitudes. Such correlations are anticipated in string axiverse constructions: instanton charges correspond to the $p$-cycles wrapped by $p$-form gauge fields, where more complicated instanton charges correspond to larger cycle volumes and thus longer instanton charge vectors.

\subsection{The Gram-Schmidt basis}
\label{subsec:gs-review}

Here we review the Gram-Schmidt (GS) orthogonalization procedure and its connection to axion effective potentials in the hierarchical-instanton limit as introduced in Refs.~\cite{Demirtas:2021gsq,Gendler:2023kjt}. The utility of the GS basis is that it enables an approximate independent axion description in the hierarchical instanton limit: heavy modes are stabilized primarily by the largest instanton terms, and can be integrated out sequentially.

Without loss of generality we order the instanton scales as $\Lambda_i\gg\Lambda_{i+1}$. If the first $N$ instanton charge vectors $\bm r_i$ are linearly independent,\footnote{If the $\hat {\bm r}_i$ are i.i.d.\ vectors, linear independence becomes exponentially likely at large $N$.} then the first $N$ phases may be absorbed by a shift of $\bm\theta$, so the potential of \cref{eqn:generic_axion_potential} may be written
\begin{align}
    V(\bm\theta)
    =
    \Lambda_{i\leq N}^4\!\left[1-\cos(\bm r_{i\leq N}^T\bm\theta)\right]
    + \Lambda_{j> N}^4\!\left[1-\cos(\bm r_{j>N}^T\bm\theta + \delta_{j>N})\right]\,.
\end{align}
In the hierarchical limit $\Lambda_{i+1}/\Lambda_i\to 0$, the  terms with instanton scales $\Lambda_{j>N}^4$ are negligible at leading order.

It is convenient to work with canonically normalized fields. We diagonalize the kinetic matrix as
\begin{align}
    \bm K \equiv \bm R_K \bm D_K^2 \bm R_K^T\,,
\end{align}
where $\bm D_K^2 = {\rm diag}(f_1^2,\dots,f_N^2)$ collects the kinetic eigenvalues and $\bm R_K$ is an orthogonal matrix satisfying $\bm R_K^T\bm R_K=\bm I$. The canonically normalized fields $ \bm \phi$ and corresponding (dimensionful) charge vectors $\bm\rho_i$ are
\begin{align}
\label{eqn:canon_normalization}
    \bm \phi =  \bm D_K\bm R_K^T\bm\theta\,,\qquad
    \bm\rho_i = \bm D_K^{-1}\bm R_K^T\bm r_i\,,
\end{align}
so that the leading potential takes the form
\begin{align}
    {\cal L}
    = \frac12(\partial\bm \phi)^2
    - \Lambda_i^4\!\left(1-\cos(\bm \rho_i^T\bm\phi)\right)\,.
\end{align}
The heaviest axion is stabilized primarily by the largest instanton, which we take to be $\Lambda_1^4$, and corresponds to the direction in field space parallel to $\bm\rho_1$. Its fundamental period in the canonically normalized coordinate is set by the inverse length $2\pi/|\bm\rho_1|$. After integrating out this heaviest mode, the next heaviest axion is stabilized primarily by $\Lambda_2^4$ and corresponds to the component of $\bm\rho_2$ orthogonal to $\bm\rho_1$, denoted $\bm\rho_{2\perp}$. Proceeding iteratively, $\bm\rho_{i\perp}$ is defined as the component of $\bm\rho_i$ orthogonal to the span of the previous vectors, i.e.\ the GS orthogonalization procedure:
\begin{align}
    \bm \rho_{i\perp}\equiv \left[\bm I - \sum_{j = 1}^{i - 1}\hat{\bm\rho}_{j\perp}\hat{\bm\rho}_{j\perp}^T\right]\bm\rho_i\,.
\end{align}
To leading order in the hierarchy, the resulting effective dynamics are well-approximated by a set of decoupled axions with altered decay constants,
\begin{align}\label{eqn:GS_Lagrangian}
    {\cal L}_{\rm GS}
    = \frac12(\partial a_i)^2
    - \Lambda_i^4\left(1 - \cos\frac{a_i}{f_{{\rm GS},i}}\right)\,,
\end{align}
where $a_i \equiv \hat{\bm\rho}_{i\perp}\cdot \bm\phi$ and we define the Gram--Schmidt decay constant,
\begin{align}
    f_{{\rm GS},i}\equiv\frac{1}{|\bm \rho_{i\perp}|}\,.
\end{align}
We emphasize that because each $\bm\rho_{i\perp}$ is orthogonal to the subspace spanned by the previous $i - 1$ charge vectors, the tendency of the GS procedure is to shorten each successive $|\bm\rho_{i\perp}|$, and consequently lighter axions will tend to have relatively larger GS decay constants as a pure consequence of geometry. The $\bm \rho$ vector magnitudes will further affect the effective field ranges in a way which will depend on their specific distribution.

From \cref{eqn:GS_Lagrangian} it is clear that in the hierarchical limit the effective degrees of freedom $a_i$ decouple and have field ranges $2\pi f_{{\rm GS},i}$. Consequently, it is the GS decay constants that enter the misalignment relic abundance when generalizing \cref{eqn:axion-relic-abundance}. Understanding the statistics of the GS procedure in physically motivated limits is therefore one of the focuses of this work (\cref{subsec:field-ranges}).

This construction also highlights that the relation between the GS decay constants and the axion--SM couplings is not straightforward: the same mixing and kinetic structure that determine $f_{{\rm GS},i}$ also control how the light eigenstates inherit couplings in the canonically normalized basis. We return to this point in \cref{subsec:couplings}.

\section{Field ranges and couplings}
\label{sec:Statistics}

In this section, we focus on a simplified set of examples that illustrate the salient features of the field ranges and couplings in a hierarchical axiverse. The purpose of the main text is not to present the most general statistical treatment, but rather to make the physical mechanism transparent: the GS procedure typically assigns the smallest field range to the heaviest axion, the next smallest to the next heaviest, and so on down the spectrum. We leave the more general treatment, in which only the orientations of the instanton charge vectors are assumed to be i.i.d.\ while their lengths may be fixed independently, to \cref{app:General_Statistics}.

The main results do not require the full instanton charge vectors to be i.i.d.; they only require that the charge-vector orientations are drawn independently from a common angular distribution, though as we argue in appendix \ref{subsec:iidnt} even this assumption can be relaxed. The lengths of the instanton charge vectors may instead be fixed independently, and may therefore be correlated with the instanton actions. This naturally accommodates the expectation that instantons with different charge-vector lengths can appear at different places in the instanton hierarchy, rather than being randomly ordered with respect to their actions. In the simplified examples below, we often use rounded Gaussian charge ensembles because they give a simple concrete illustration of the effect; the generalization beyond these examples is given in \cref{app:General_Statistics}.

In \cref{subsec:field-ranges}, we quantify how strongly the heaviest-mode field range is suppressed and show that the suppression is controlled not only by the number of lighter axions in the spectrum but also, crucially, by the eigenvalue spectrum of the kinetic matrix. We also show that in some regimes the resulting scaling of the GS decay constants can outpace the comparatively mild $m^{1/2}$ dependence of the misalignment relic abundance. In such regimes, the field-range suppression competes directly with (and can dominate over) the mass dependence in setting cosmological abundances. We illustrate these effects in representative ensembles, while emphasizing that the full space of possibilities is considerably broader.

We then turn to the statistics of axion couplings to matter in \cref{subsec:couplings}. We find an analogous ordering: the smallest effective ``coupling decay constant'' (i.e.\ the largest coupling) tends to align with the heaviest mode, and so on. An important difference, however, is that the couplings do not generically inherit the same additional GS projection suppression that controls field ranges in high dimension. This leads to the expectation of a $\sqrt{N}$ suppression of the typical product of abundances times couplings, relevant for detection [see \cref{eq:P_det}], relative to naive single-axion intuition. The possible exceptions to this are the lightest and heaviest axions. On the other hand, in the case of the QCD axion, this generic $\sqrt{N}$ suppression is undone: the coupling to matter is tied to the QCD instanton charge vector $\bm r_{i_c}$ so the inherited couplings recover the same dimensional scaling that controls the GS ranges. This further motivates a focused look at the QCD axion parameter space: its direct-detection signatures are comparatively robust even in large axion ensembles, whereas generic axions tend to exhibit suppressed couplings to matter. Even so, axions coupled to matter in directions orthogonal to the QCD coupling may also exhibit promising direct detection signatures.

\subsection{Field ranges}
\label{subsec:field-ranges}

The GS construction as reviewed in \cref{subsec:gs-review} gives a simple dynamical description of a hierarchical axiverse: as successively smaller instanton terms become relevant, one integrates out heavy modes in descending order of $\Lambda_i$. The resulting GS decay constants, however, are not free parameters. They are determined by the underlying fundamental data, namely the integer charge vectors in the $2\pi$-periodic basis $\bm\theta$ together with the kinetic matrix, and therefore exhibit strong statistical structure.

In this section we illustrate how the GS decay constants obey highly constrained patterns when the instanton charge-vector orientations are statistically similar across the instanton hierarchy. First, they are typically ordered with the mass hierarchy: the heaviest mode (rank $i=1$ in the mass ordering) tends to have the smallest GS decay constant, the next heaviest the next smallest, and so on, up to variations associated with the charge-vector lengths. Second, the typical size of the $i$th GS decay constant is controlled by the dimension of the orthogonal complement after integrating out the previous $i-1$ modes, leading in the isotropic limit to the scaling $f_{{\rm GS},i}\propto (N-i+1)^{-1/2}$. We then show that anisotropy further suppresses the field ranges of the heaviest axions. Intuitively, anisotropy biases the GS projections so that the most probable directions in canonically normalized field space are preferentially removed early in the procedure, causing the heavy end of the spectrum to inherit the smallest effective field ranges. We derive analytic expressions for the mean GS decay constants as a function of mass rank in a simple Gaussian instanton charge ensemble, quantify the typical scatter about these means, and illustrate how the resulting  correlations between masses and decay constants can qualitatively modify the relic-abundance versus mass scaling reviewed in \cref{subsec:single_axion}. The more general derivation, with fixed charge-vector lengths and i.i.d.\ orientations, is given in \cref{app:General_Statistics}.

In order to derive analytic expressions for the GS decay constants, it is useful to express the length of the $i$th orthogonalized vector geometrically. At the $i$th stage of the GS process, the previous $i-1$ vectors span a subspace defined by an $(i-1)$-dimensional parallelepiped. The projection of $\bm\rho_i$ onto the orthogonal subspace is given by a ratio of parallelepiped volumes,
\begin{align}
    |\bm\rho_{i\perp}| = \frac{{\rm vol}_{i}(\bm\rho_1,\dots,\bm\rho_{i})}{{\rm vol}_{i-1}(\bm\rho_1,\dots,\bm\rho_{i-1})}\,.
\end{align}
Low-dimensional versions are familiar:
\begin{align}
    |\bm\rho_{2\perp}| = \frac{|\bm\rho_1\times\bm\rho_2|}{|\bm\rho_1|}\,,
    \qquad
    |\bm\rho_{3\perp}| = \frac{|(\bm\rho_{1}\times\bm\rho_2)\cdot\bm\rho_3|}{|\bm\rho_1\times\bm\rho_2|}\,,
\end{align}
and in general one may write $|\bm\rho_{i\perp}| = |\bm\rho_1\wedge\cdots\wedge\bm\rho_i|\,/\,|\bm\rho_1\wedge\cdots\wedge\bm\rho_{i-1}|$, where the higher-dimensional expressions involve the exterior product and an appropriate norm on $i$-vectors. Fortunately, the parallelepiped volumes admit a standard representation in terms of Gram determinants,
\begin{align}\label{eqn:gramanian}
    f_{\text{GS},i}^{-2} = |\bm \rho_{i\perp}|^2 = \frac{\det \bm G_{i}}{\det \bm G_{i - 1}}\,.
\end{align}
Here, $\bm G_i$ is the $i\times i$ Gram matrix
\begin{align}
    \bm G_i\equiv \bm Q_i \bm Q_i^T\,,
    \qquad
    \bm Q_i^T \equiv \left[\begin{array}{cccc}\bm\rho_1&\bm\rho_2&\cdots&\bm\rho_i\end{array}\right]\,,
\end{align}
i.e.\ $\bm Q_i$ is the matrix whose rows are the first $i$ vectors $\bm\rho_1,\dots,\bm\rho_i$.
While many equivalent descriptions exist, the ratio-of-volumes form is particularly convenient for analytic calculations.

\begin{figure}
    \centering
    \includegraphics[width=\columnwidth]{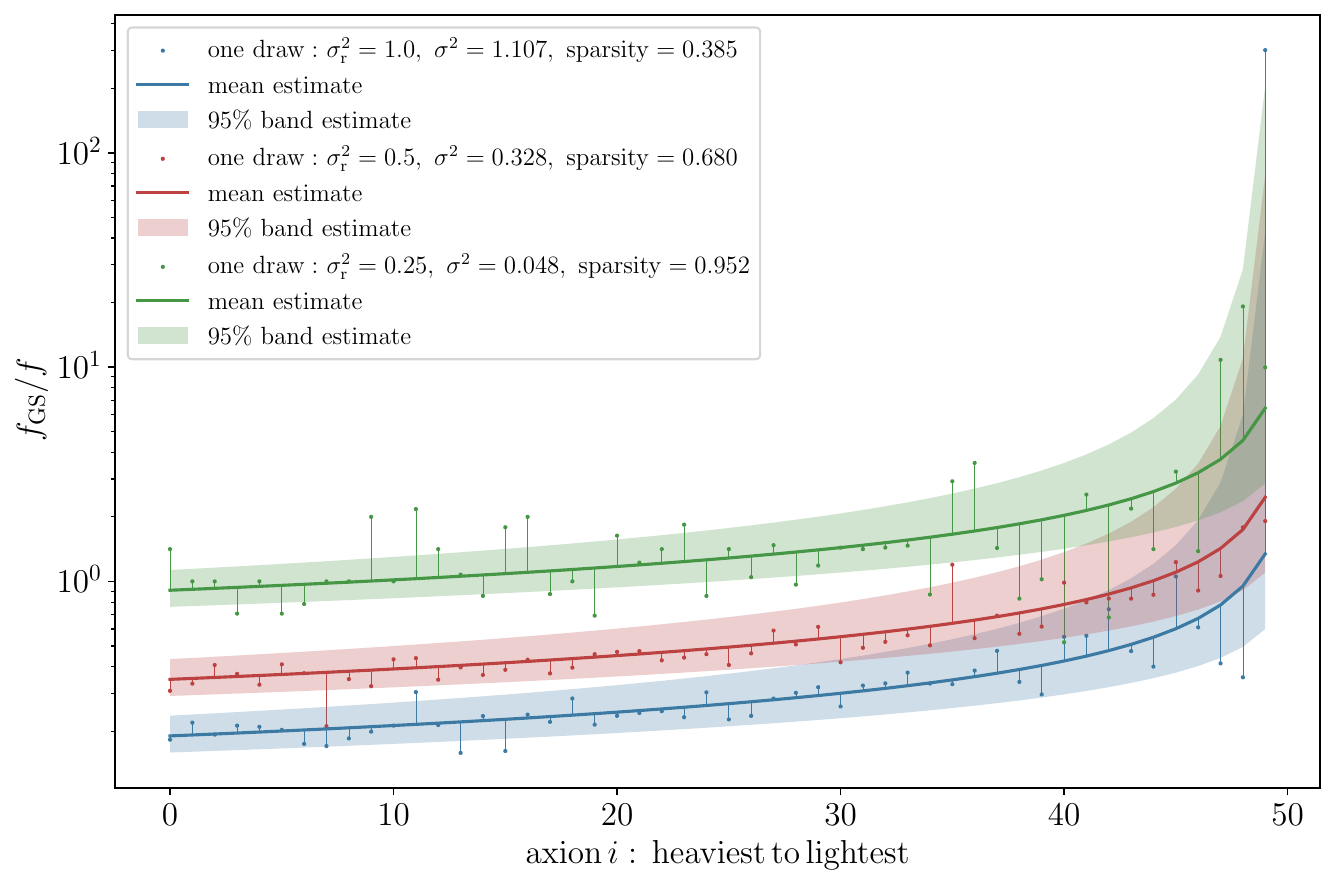}
    \caption{Effective field ranges (GS decay constants $f_{\rm GS}$) relative to the nominal UV scale $f$ for an ensemble of $N=50$ axions with an isotropic kinetic matrix, $\bm K=f^2\bm I$, and isotropic instanton charges drawn as $[\bm Q]_{ij}=[\bm r_i]_j\sim {\rm round}[{\cal N}(0,\sigma_r^2)]$, where ${\rm round}[\cdot]$ denotes rounding to the nearest integer. Points show a single draw of this ensemble. Solid curves show the mean prediction from [\cref{eqn:meanfield-ni2}], and shaded regions the central $95\%$ interval from [\cref{eqn:fGS_variance_estimate}]. All curves exhibit the trend expected from the isotropic scaling \cref{eqn:fGS_isotropic}: heavier axions have field ranges suppressed by $\sqrt{N-i+1}$, while the lightest axions have unsuppressed [and, for sufficiently small ${\rm var}([\bm r_i]_j)$, enhanced] field ranges relative to $f$. The spread also grows toward lighter axions, so typical realizations contain light modes with $f_{\rm GS}$ well above the mean. As $\sigma_r$ decreases from 1.0 to 0.25 (see legend), more entries round to zero and the charge matrix becomes sparse. In this sparse regime the mean prediction continues to capture the overall trend, while the spread estimate becomes less accurate as fluctuations become increasingly dominated by shot noise. In the extreme sparse limit (green curve), the isotropic scaling with ${\rm var}([\bm r_i]_j)\sim 1/N$ yields the enhancement $f_{{\rm GS},N}\propto \sqrt{N}\,f$ for the lightest mode described in Ref.~\cite{Bachlechner:2017hsj} (see text around \cref{eqn:sparse}]. Because rounding distorts the variance when $\sigma_r$ is small, analytic curves use the post-rounding variance $\sigma_{\rm eff}^2 \equiv {\rm var}([\bm r_i]_j)$ (measured from the rounded charge matrix) in place of the input $\sigma_r^2$.} 
    \label{fig:GS_isotropic}
\end{figure}

Under simplifying assumptions about the underlying distribution of parameters, we can now derive analytic expressions for the statistical properties of the axion field ranges $f_{{\rm GS},i}=1/|\bm\rho_{i\perp}|$. For the main-text illustration, we specialize to a Gaussian ensemble with mean zero and standard deviation $\sigma_r$ for the instanton charges in the fundamental basis,
\begin{align}
\label{eqn:fundChargeMatrix-dist}
    [\bm r_i]_j \sim {\cal N}(0,\sigma_r^2)\,,
\end{align}
with independent draws across $i$. This should be understood as a simple example rather than a necessary assumption: in the more general treatment of \cref{app:General_Statistics}, the charge-vector lengths may be fixed independently, while only the orientations are taken to be i.i.d. Although $[\bm r_i]_j$ is formally continuous in this approximation, we find that the resulting predictions remain accurate when one rounds $[\bm r_i]_j$ to the nearest integer, even for $\sigma^2<1$ (corresponding to sparse $\mathcal{O}(1)$ charges) as illustrated in \cref{fig:GS_isotropic}.

We work in the kinetic eigenbasis, $\bm \rho_i = \bm D_K^{-1}\bm R_K^T \bm r_i$, so the components of $\bm\rho_i$ are Gaussian with variances set by the kinetic eigenvalues,
\begin{align}
\label{eqn:rho-distribution}
    [\bm\rho_i]_j\sim {\cal N}(0,\sigma_j^2)\,,\qquad \sigma_j\equiv \sigma_r/f_j\,.
\end{align}
We then make the following mean-field approximation, which we find accurately reproduces the mean behavior of the GS decay constants:
\begin{align}\label{eqn:meanfield-ni2}
    \langle f_{{\rm GS},i}^{-2}\rangle_{r}
    = \left\langle\frac{\det \bm G_{i}}{\det \bm G_{i - 1}}\right\rangle_{r}
    \approx \frac{\langle\det \bm G_{i}\rangle_{r}}{\langle\det \bm G_{i - 1}\rangle_{r}}\,,
\end{align}
where the subscript $r$ indicates that the average is taken over the row vectors.

Note that the mean-field approximation is not rigorous, but it becomes exact in the isotropic limit and remains accurate in the regimes of interest. 
To see why, consider \cref{eqn:gramanian}, but take the expectation value only over the $i$th vector with the remaining vectors held fixed. Using the Gram-determinant identity, the equation can be rewritten as
\begin{align}
    \frac{\langle\det\bm G_i\rangle_{\bm \rho_i}}{\det \bm G_{i - 1}}
    = \langle|\bm\rho_{i\perp}|^2\rangle_{\bm \rho_i}
    = {\rm Tr}\!\left[\bm P_{i - 1}^\perp\,\bm\Sigma_i\right]\,,
\end{align}
where $\bm\Sigma_i \equiv \langle \bm\rho_i\bm\rho_i^T\rangle$ is the covariance of $\bm\rho_i$ and $\bm P_{i - 1}^\perp$ is the projector onto the subspace orthogonal to the span of the first $i-1$ vectors. In the isotropic limit $\bm\Sigma_i=\sigma_r^2\bm I$, rotational invariance implies that ${\rm Tr}[\bm P_{i-1}^\perp\bm\Sigma_i]=\sigma_r^2\,{\rm Tr}[\bm P_{i-1}^\perp]$ is independent of the orientation of the projector, so the conditional expectation of the ratio becomes independent of the earlier vectors and the mean-of-a-ratio equals the ratio-of-means. In the strongly anisotropic limit, the GS procedure preferentially removes the highest-variance directions first, so the orientation of $\bm P_{i-1}^\perp$ becomes effectively deterministic (set by the variance ordering), and again the averages at each GS stage decouple.

To evaluate $\langle \det \bm G_i\rangle$, we use the Cauchy--Binet formula,
\begin{equation}
\det(\bm Q_i \bm Q_i^{T})
=\sum_{\text{all $i\times i$ column-submatrices } \bm M \text{ of } \bm Q_i} (\det \bm M)^2 \,,
\end{equation}
together with the independence and Gaussianity of the columns. For an $i\times i$ submatrix $\bm M$ whose columns have variances $\sigma_{a_1}^2,\dots,\sigma_{a_i}^2$, one has $\langle(\det \bm M)^2\rangle = i!\,\sigma_{a_1}^2\cdots\sigma_{a_i}^2$. Summing over all choices yields
\begin{equation}
\label{eqn:detGi-expectation}
    \langle \det \bm G_i\rangle_{r}
    = i!\sum_{1\le a_1<\cdots<a_i\le N}\sigma_{a_1}^2\cdots\sigma_{a_i}^2
    \equiv i!\,e_i(\sigma_1^2,\dots,\sigma_N^2)\,,
\end{equation}
where $e_i$ is the $i$-th elementary symmetric polynomial in $N$ symbols. Combining \eqref{eqn:detGi-expectation}
and \eqref{eqn:meanfield-ni2} gives the compact estimate
\begin{equation}
\label{eqn:fGS-ESP}
    \langle f_{{\rm GS},i}^{-2}\rangle_{r}\approx
    i\,\frac{e_i(\sigma_1^2,\dots,\sigma_N^2)}{e_{i-1}(\sigma_1^2,\dots,\sigma_N^2)}\,.
\end{equation}

Let us now consider limiting cases of \cref{eqn:fGS-ESP}. If we assume that the kinetic term is proportional to the identity $\bm K = f^2 \bm I$, so that $\sigma_i = \sigma_r/f$, we can isolate the effect of dimensionality. Using $e_i(\sigma_r^2,\dots,\sigma_r^2) = \sigma_r^{2i}\binom{N}{i}$, the ratio of elementary symmetric polynomials simplifies to
\begin{align}\label{eqn:fGS_isotropic}
    \langle f_{{\rm GS},i}^{-2}\rangle_{r} = \frac{\sigma_r^2}{f^2}(N - i + 1)\,,\qquad ({\rm isotropic})\,.
\end{align}
Thus, the field range of the $i$th axion, $f_{{\rm GS},i}$, is typically suppressed by the square root of the number of lighter fields relative to the single-field expectation. This is purely a consequence of dimensionality: on average, each of the remaining $N - i + 1$ light axions couples with strength $\sigma_r$ to the $i$th instanton potential. We illustrate the isotropic limit in \cref{fig:GS_isotropic} for several choices of $\sigma_r$ using a Gaussian ensemble rounded to the nearest integer, $[\bm r_i]_j\sim {\rm round}[{\cal N}(0,\sigma_r^2)]$. This captures the case of sparse $\mathcal O(1)$ integer charges when $\sigma_r^2\lesssim 1$. 
Because rounding distorts the variance at small $\sigma_r^2$, it is convenient to define an effective variance
\begin{align}
    \sigma_{\rm eff}^2 \equiv {\rm var}([\bm r_i]_j)\,,
\end{align}
measured after rounding in each sample, and to compare to \cref{eqn:fGS_isotropic} with $\sigma_r^2\to\sigma_{\rm eff}^2$. 
Even for $\sigma_r^2=0.25$, where $\approx95\%$ of entries are zero, this variance-corrected estimate accurately reproduces the mean behavior of the GS decay constants, as illustrated by the green line in \cref{fig:GS_isotropic}.

\begin{figure}
    \centering
    \includegraphics[width=\columnwidth]{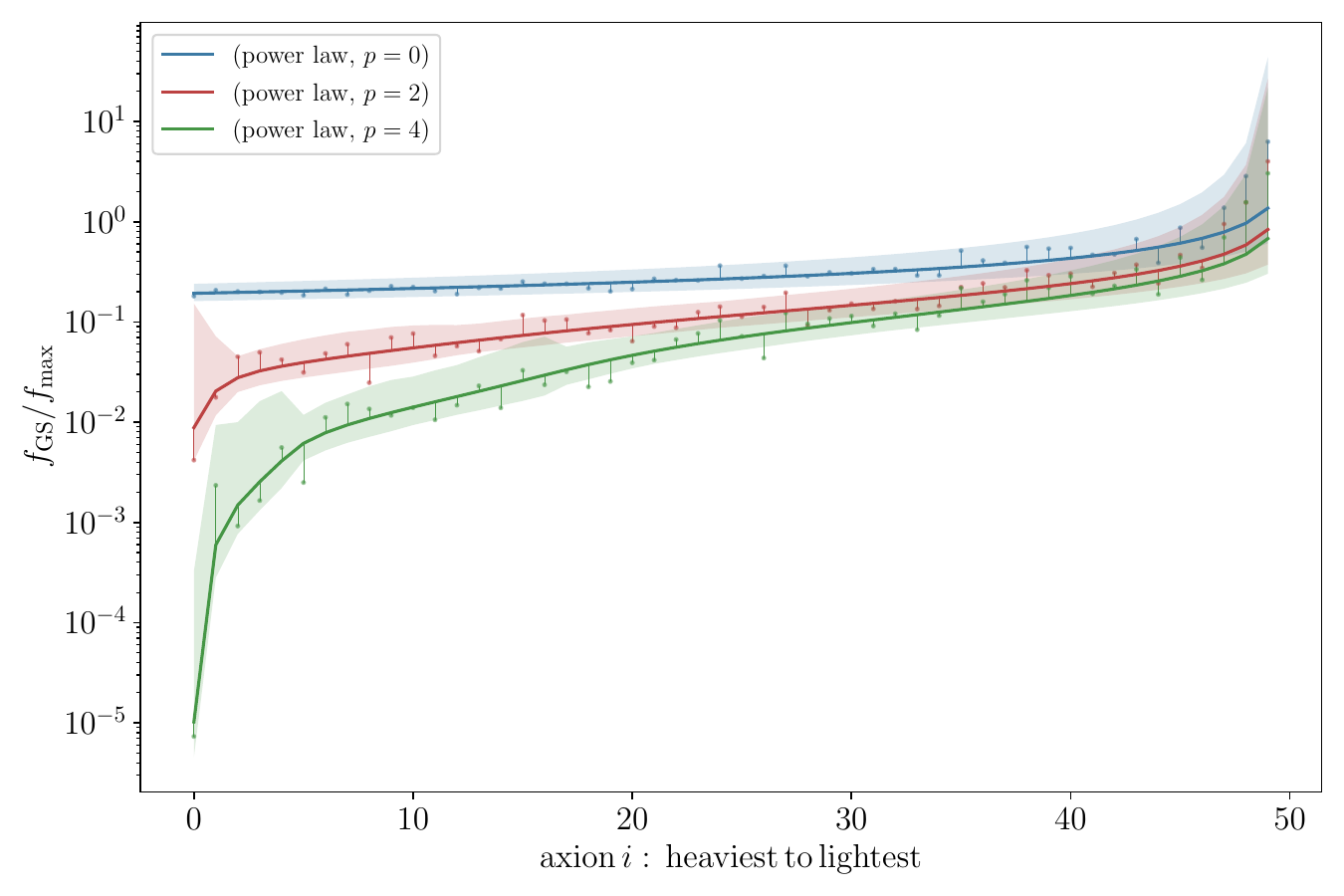}
    \caption{Gram-Schmidt decay constants $f_{\rm GS}$ for ensembles of $N=50$ axions with instanton charges drawn in the fundamental basis as $[\bm r_i]_j \sim {\rm round}[{\cal N}(0,1)]$, where ${\rm round}[\cdot]$ denotes rounding to the nearest integer. The kinetic matrix eigenvalues are sampled from a power law,
    $f_i^2 = f_{\rm max}^2\, x^p$ with $x\sim{\cal U}(0,1)$, where $f_{\rm max}$ sets the UV scale in the fundamental basis. The case $p=0$ (blue) corresponds to an isotropic kinetic matrix, consistent with the behavior shown in \cref{fig:GS_isotropic}. Increasing $p$ introduces anisotropy: for $p=2$ (red) the distribution shifts downward and, as the heaviest axions are integrated out first, they preferentially absorb the smallest kinetic eigenvalues, further reducing their field ranges, as expected from the strongly anisotropic limit [\cref{eqn:fGS_anisotropic}]. For $p=4$ (green) this effect is more pronounced, as a few especially small kinetic eigenvalues lead to a stronger suppression of the heaviest-mode field ranges. As the highly anisotropic directions are pruned, the kinetic term becomes more isotropic for lighter axions, returning to the scaling expected from \cref{eqn:fGS_isotropic}. The solid lines indicate the mean estimate [\cref{eqn:meanfield-ni2}] while the shaded regions indicate the central $95\%$ interval [\cref{eqn:fGS_variance_estimate}].}
    \label{fig:GS_anisotropic}
\end{figure}

It is also useful to note how \cref{eqn:fGS_isotropic} connects to the ``delocalization'' intuition emphasized in Ref.~\cite{Bachlechner:2017hsj}: for isotropic random directions, the mass eigenstate typically has support spread over many fundamental fields, so individual components in each cosine are parametrically small, leading to a $\sqrt{N}$ enhancement of the effective field range. Concretely, if the charge vectors $\bm r_i$ have support on ${\cal O}(1)$ fields (i.e.\ ${\cal O}(1)$ non-zero entries), then $\sigma_{\rm eff}^2\sim 1/N$. Using this variance in \cref{eqn:fGS_isotropic}, one immediately finds for the lightest mode ($i=N$) that
\begin{align}\label{eqn:sparse}
    \langle f_{{\rm GS},N}^{-2}\rangle_r \sim \frac{1}{N f^2}\,,\qquad ({\rm sparse})\,,
\end{align}
so that $f_{{\rm GS},N}\sim \sqrt{N}\,f$, reproducing the $\sqrt{N}$ enhancement discussed in Ref.~\cite{Bachlechner:2017hsj}. In \cref{fig:GS_isotropic}, this trend is visible in the sparsest (green) case.

On the other hand, we can isolate the effect of the kinetic eigenvalues by taking the hierarchical limit where $f_1\ll f_2\dots$ corresponding to $\sigma_1\gg\sigma_2\dots$. In this case, the expression for the GS decay constants simplifies
\begin{align}\label{eqn:fGS_anisotropic}
    \langle f_{{\rm GS},i}^{-2}\rangle_{r}\approx i\sigma_i^2 = i\frac{\sigma_r^2}{f_i^2}\,,\qquad({\rm strongly\,\,anisotropic})\,.
\end{align}
Recall that $\sigma_i$ is inversely proportional to the kinetic eigenvalue: the heaviest axions absorb the smallest kinetic eigenvalues, and have further-suppressed field ranges. This effect is clearly visible in \cref{fig:GS_anisotropic}, where we have taken the kinetic eigenvalues themselves to come from the distribution $f_i^2= f_{\rm max}^2 x^p$ where $x\sim {\cal U}(0,1)$ is uniformly distributed. The isotropic case corresponds to $p = 0$, and is plotted in blue. As we introduce anisotropy by taking $p > 0$, the polynomial distribution tends to produce more, smaller kinetic eigenvalues. As expected from \cref{eqn:fGS_anisotropic}, the tendency is for the heaviest axions to favor the directions with the largest variance, i.e. the smallest kinetic eigenvalues, suppressing their decay constants. This effect is most pronounced in \cref{fig:GS_anisotropic} for $p = 4$, where the heaviest axion has a decay constant suppressed by ${\cal O}(10^{-2})$ relative to its next-lightest neighbor.

To summarize, the mean behavior of axion field space, characterized by the GS decay constants, is to self-organize into an increasing sequence, so that the lightest axions also have the largest possible field excursions.

We now characterize the typical spread about the mean. In the strongly anisotropic regime, $f_{{\rm GS},i}^{-2}=|\bm\rho_{i\perp}|^2$ is dominated by the largest variance of the remaining directions, so its fractional fluctuations are order unity. In the isotropic regime, the $N-i+1$ remaining directions contribute comparably, and the spread is suppressed by self-averaging. Since $|\bm\rho_{i\perp}|^2$ is a quadratic form in Gaussian variables, we model its distribution as a scaled $\chi^2$ with an effective number of degrees of freedom $\nu_{{\rm eff},i}$. A simple estimate that interpolates between the isotropic and hierarchical regimes is
\begin{align}
    \nu_{{\rm eff},i}\approx \frac{\left(\sum_{k = i}^N\sigma_k^{2}\right)^2}{\sum_{k = i}^N\sigma_k^{4}}\,.
\end{align}
Consequently, we have the estimated fluctuations
\begin{align}\label{eqn:fGS_variance_estimate}
    \frac{{\rm var}(f_{{\rm GS},i}^{-2})}{(\langle f_{{\rm GS},i}^{-2}\rangle)^2} \sim \frac{2}{\nu_{{\rm eff},i}}\,.
\end{align}
This estimated spread corresponds to the shaded regions in \cref{fig:GS_isotropic,fig:GS_anisotropic}, where we plot the corresponding 95-percentile intervals for a $\chi^2$-distribution with $\nu_{{\rm eff},i}$ degrees of freedom about the mean estimate \cref{eqn:fGS-ESP}. This estimate becomes unreliable when the instanton charges become sparse, at which point the fluctuations become dominated by shot noise of individual charges, as illustrated by the $\sigma_r^2 = 0.25$ example in \cref{fig:GS_isotropic}.

\subsubsection*{Relic abundance scaling with mass}
\label{subsubsec:GS_relic_abundance_scaling}
Our discussion in this section has illustrated how the field space available to each axion is organized by the GS procedure in the hierarchical-instanton regime: as one integrates out heavy modes in descending order of $\Lambda_i$, the remaining light modes inherit progressively larger effective field ranges. This has direct implications for misalignment production, which in the hierarchical limit takes the form of \cref{eqn:axion-relic-abundance} with the substitutions $f_a\to f_{{\rm GS},i}$, $m\to m_i$, and $\theta_0\to\theta_{0,i}$ where $\theta_{0,i}$ denotes the initial misalignment of $a_i$ in units of $f_{{\rm GS},i}$. Since $f_{{\rm GS},i}$ is typically a decreasing function of mass rank, it can compete with or overwhelm the slow $m_i^{1/2}$ dependence of the relic abundance.

To illustrate what kinds of $\Omega_i$ versus $m_i$ scalings are possible, we consider a simple toy model in which the kinetic eigenvalues and instanton scales are exponentially distributed:
\begin{align}
    f_i = e^{-y i}\, f_{\rm max}\,, \qquad \Lambda_i^4 = e^{-2z i}\,\Lambda_{\rm max}^4\,,
\end{align}
with $y,z>0$ and large enough that the hierarchical limits apply (note that $i$ is again an index and not the complex unit). In the strongly anisotropic regime of \cref{eqn:fGS_anisotropic}, the projected charge variance at GS stage $i$ is dominated by the largest remaining variance direction, giving
\begin{align}
\label{eqn:fGS-exponential}
    f_{{\rm GS},i}^{-2} \sim i\sigma_r^2e^{-2yi}f_{\rm max}^{-2}\,,
\end{align}
where we have used the fact that the GS procedure tends to sort the GS decay constants so that the heaviest axion has the smallest decay constant.
Using $m_i^2\simeq \Lambda_i^4/f_{{\rm GS},i}^2$, we obtain
\begin{align}
\label{eqn:mGS-exponential}
    m_i^2 \sim i\sigma_r^2\,e^{-2(y+z)i}\Lambda_{\rm max}^4f_{\rm max}^{-2}\,.
\end{align}
It is worth observing here that the GS ordering of the decay constants imposes an even steeper axion mass hierarchy than imposed by the instanton hierarchy alone. 
Combining \cref{eqn:mGS-exponential} and \cref{eqn:fGS-exponential} with \cref{eqn:axion-relic-abundance} (with the appropriate substitutions $f_a\to f_{{\rm GS},i}$, $m\to m_{i}$, $\theta_{0}\to \theta_{0,i}$ and taking $\theta_{0,i}$ to be rank-independent for this estimate), we find
\begin{align}
\label{eqn:Omega-toy}
    \Omega \propto m^{\frac12(\frac{z - 3 y}{z + y})}\,,
\end{align}
where we have ignored logarithmic scaling with mass.
The case where the decay constants are all equal corresponds to $y = 0$. Typically, one may expect $y$ should not be too large, so $\Omega(m)$ is often an increasing function of mass. Nonetheless, even modest hierarchies in the kinetic eigenvalues are able to overcome the independent axion mass scaling, e.g.\ $y = z/3$. In summary, any $\Omega(m)\propto m^p$ power law is possible with $p < 1/2$.

\subsection{Couplings}
\label{subsec:couplings}

In this section, we describe the coupling of the GS basis states (mass eigenstates in the hierarchical limit $\Lambda_i\gg\Lambda_{i + 1}$) to the Standard Model, assuming a generic interaction of the form
\begin{align}
\label{eqn:generic_coupling}
    {\cal L}_{\rm int} = \bm q^T\bm\theta{\cal O}_{\rm SM}\,,
\end{align}
specified in the fundamental basis, where $\bm q$ is a dimensionless coupling vector which can be thought of as (up to overall normalization) a vector of anomaly coefficients. Like the GS decay constants, which tend to sort themselves so that the heaviest axion has the smallest field space available, the coupling decay constants, defined by writing the GS-basis interaction as
\begin{align}
    {\cal L}_{{\rm GS,int}} = g_ia_i{\cal O}_{\rm SM}
    \equiv \frac{a_i}{f_{{\cal O},i}}\,{\cal O}_{\rm SM}\,,
\end{align}
(with $f_{{\cal O},i}^{-1}\equiv g_i$), arrange themselves so that the heaviest axion typically has the smallest coupling decay constant (i.e.\ the largest coupling). However, unlike the GS decay constants, whose sizes are further suppressed by dimensionality, the typical $f_{{\cal O},i}$ do not inherit an additional $(N-i+1)^{-1/2}$ factor. Therefore, for generic axions the combination relevant to direct-detection and decay searches, $f_{{\rm GS},i}/f_{{\cal O},i}$, is typically suppressed as $N^{-1/2}$ for mid-spectrum modes, weakening detection prospects in large axiverses for generic axionlike particles, except for possibly the lightest and heaviest ALPs. We then study the specific case of the QCD axion, which has the special property that its coupling vector is aligned with one of the instanton charge vectors. As a result, we find that the QCD axion itself does not incur the $\sqrt{N}$ suppression making it alone especially visible in direct detection experiments.

\subsubsection{Axionlike couplings}
\label{subsubsec:axionlike_couplings}

We now consider the coupling of the axions to the Standard Model in the fundamental basis, as specified in \cref{eqn:generic_coupling}, where ${\cal O}_{\rm SM}$ is any Standard Model operator except for $\tilde G G$, which we treat separately in \cref{subsubsec:QCD_axion_coupling}, and $\bm q$ is a coupling vector which is not related to any particular instanton charge vector $\bm r_i$. We move from the fundamental basis to the GS basis via the following transformation
\begin{align}
    \bm a = \bm A^{-1}\bm\theta\,,
\end{align}
which combines both the step of canonically normalizing the fields and the GS procedure:
\begin{align}
    \bm A = \bm R_K \bm D_K^{-1}\bm R_Q\,,
\end{align}
where $\bm R_K \bm D_K^{-1}$ encodes canonical normalization as in \cref{eqn:canon_normalization} and $\bm R_Q$ encodes the GS process. In the GS basis, the axion couplings to the SM then take the form
\begin{align}
    {\cal L}_{\rm GS,int} = \bm q^T\bm A\bm a{\cal O}_{\rm SM}\,,
\end{align}
hence the couplings are
\begin{align}
\label{eqn:couplings-in-GS-basis}
    \bm g^T \equiv \bm q^T \bm A\,.
\end{align}
At this stage, this expression does not admit an obvious scaling in terms of the fundamental parameters. Phrased in terms of an expectation value over coupling vectors, however, its interpretation becomes clear.

Suppose that $\bm q$ is a Gaussian random vector
\begin{align}
\label{eqn:q-distribution}
    \bm q\sim {\cal N}(0,\sigma_q^2)\,,
\end{align}
so that the mean square expectation value of the couplings can be written
\begin{align}
\label{eqn:coupling_decay_constant_average}
    \langle f_{{\cal O},i}^{-2}\rangle_q
    = \langle g_i^2\rangle_q
    = \sigma_q^2\, [\bm A^T\bm A]_{ii}
    = \sigma_q^2\, [\bm R_Q^T\bm D_K^{-2}\bm R_Q]_{ii}\,.
\end{align}
This expression is far simpler to interpret, and we do so in two limits.

First, when the kinetic term is isotropic, we can write $\bm D_K^{-2} = f^{-2}\bm I$, and so the expression for the couplings simplifies
\begin{align}
    \langle f_{{\cal O},i}^{-2}\rangle_q = \sigma_q^2 f^{-2}\,,\qquad{\rm (isotropic)}\,.
\end{align}
Comparing to the corresponding isotropic limit of the GS decay constants, we find that the ratio of their expectation values over the coupling vector and the instanton charges is 
\begin{align}
\label{eqn:signal-strength-isotropic}
    \frac{\langle f_{{\cal O},i}^{-2}\rangle_q}{\langle f_{{\rm GS},i}^{-2}\rangle_r} = \frac{\sigma_q^2}{\sigma_r^2}\frac{1}{N - i + 1}\,,\qquad{\rm (isotropic)}.
\end{align}

\begin{figure}
    \centering
    \includegraphics[width=\columnwidth]{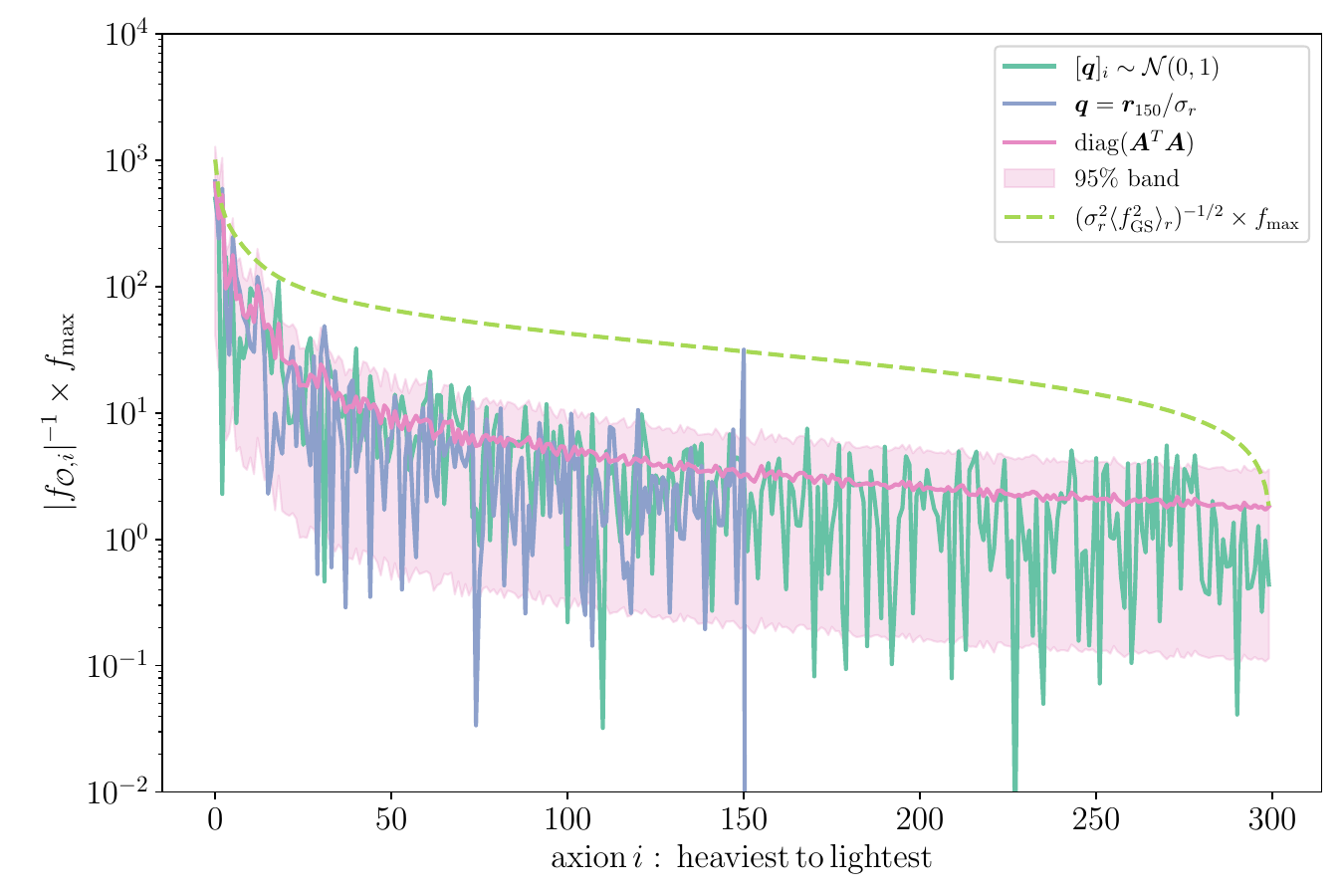}
    \caption{Couplings $|f_{{\cal O},i}|$ of $N=300$ axions to a Standard Model operator written in the fundamental basis as $\bm q^{T}\bm\theta\,{\cal O}_{\rm SM}$, shown for two choices of the coupling vector $\bm q$. 
    For the solid green curve, the components of $\bm q$ are drawn i.i.d.\  as $q_j\sim{\cal N}(0,1)$, which we take as a simple model for couplings (e.g.\ to photons) that are not directly tied to an instanton term in the axion potential. The solid pink curve shows the mean coupling after marginalizing over draws of $\bm q$, with the shaded pink band indicating the central $95\%$ interval. For the blue curve, $\bm q$ is identified with the QCD charge vector, so that the same combination $\bm q^{T}\bm\theta$ appears in the potential via $\bm q^{T}\bm\theta\,\tilde G G$. We choose $\Lambda_{\rm QCD}$ to coincide with the 150th instanton scale in the hierarchy and set $\bm q=\bm r_{150}/\sigma_r$, where the division by $\sigma_r$ normalizes the component variance of $\bm q$ to unity for a direct comparison with the Gaussian case. The blue curve exhibits three regimes: for large masses, the QCD-induced term is negligible and the couplings are statistically indistinguishable from those of a random $\bm q$ (hence agreement with the green curve). At the QCD scale ($i=150$), the axion aligned with the $150$th GS direction couples in proportion to the corresponding GS decay constant (up to trivial rescalings), leading to an enhancement relative to heavier modes by a factor $\sqrt{N-i+1}\simeq 12$. The dashed lime-green curve shows the analytic expectation for the inverse GS decay constants [\cref{eqn:meanfield-ni2}], illustrating the correspondence between the QCD axion's GS decay constant and coupling depending on where it falls in the instanton potential hierarchy. Finally, for axions lighter than the $150$th axion, their couplings are exactly zero in the limit of infinite hierarchy. We discuss finite hierarchy corrections in \cref{subsubsec:QCD_axion_coupling}.
    In this example, the kinetic matrix is chosen with eigenvalues $f_i/ f_{\rm max} \sim{\cal U}(0,1)$, and the instanton charges are drawn as $[\bm r_i]_j\sim {\rm round}[{\cal N}(0,\sigma_r^2)]$.}
    \label{fig:couplings_isotropic}
\end{figure}

Away from the isotropic limit, no such clean simplification occurs. However, using our earlier experience with the GS vectors, we may gain some intuition for the $\bm R_Q$ transformation, which encodes the GS process. In particular, in the strongly anisotropic limit, where $\sigma_1\gg\sigma_2\dots$, we found that the effect of GS was to sort the kinetic eigenvectors from smallest to largest, i.e.\ the $i$th GS direction is approximately aligned with the $i$th largest entry of $\bm D_K^{-2}$. Therefore, we can infer that in the strongly anisotropic limit, the effect of $\bm R_Q$ on $\bm D_K^{-1}$ is to sort it from largest to smallest, so that
\begin{align}
    \langle f_{{\cal O},i}^{-2}\rangle_q \sim  \frac{\sigma_q^2}{f_i^2}\qquad{(\rm strongly\,\,anisotropic)}\,,
\end{align}
where the kinetic eigenvalues $f_i$ have been sorted from smallest to largest. Comparing to the GS decay constants, we find
\begin{align}
\label{eqn:signal-strength-anisotropic}
    \frac{\langle f_{{\cal O},i}^{-2}\rangle_q}{\langle f_{{\rm GS},i}^{-2}\rangle_r} \sim \frac{\sigma_q^2}{\sigma_r^2}\frac{1}{i}\,,\qquad{\rm (strongly\,\,anisotropic)}.
\end{align}
As in the isotropic case, direct and decay detection prospects are suppressed by dimensionality, though the precise relationship between rank and the dimensionality suppression is reversed. Nonetheless, ``typical'' axions (meaning $i$ is not close to 1 or $N$) will be suppressed by a factor of order $\sqrt{N}$. While these results were derived here for a specific example of charge vector and coupling vector distributions, the dimensionality factors appearing in \cref{eqn:signal-strength-isotropic} and \cref{eqn:signal-strength-anisotropic} are general, as we show in \cref{eqn:app_signal_isotropic} and \cref{eqn:app_signal_anisotropic}.

We provide an explicit example of the coupling decay constants in \cref{fig:couplings_isotropic}. The green line illustrates the coupling decay constants for a coupling vector $\bm q$ selected from the distribution \cref{eqn:q-distribution}, and the kinetic eigenvalues selected from a uniform distribution $f_i/f_{\rm max}\sim{\cal U}(0,1)$. The pink line corresponds to the average over $\bm q$, namely \cref{eqn:coupling_decay_constant_average}, while the pink shaded region corresponds to the 95-percentile intervals. Since each $f_{{\cal O},i}^{-1}$ is a linear transformation of a Gaussian random vector, each component is itself Gaussian
\begin{align}
    f_{{\cal O},i}^{-1}\sim {\cal N}(0,\sigma_q^2[\bm A^T\bm A]_{ii})\,,
\end{align}
and its spread is determined by a $\chi^2$ distribution of a single gaussian random variable. In particular
\begin{align}
    {\rm var}(f_{{\cal O},i}^{-2}) = 2\sigma_q^4[\bm A^T\bm A]_{ii}^2\,.
\end{align}

\begin{figure}
    \centering
    \includegraphics[width=\columnwidth]{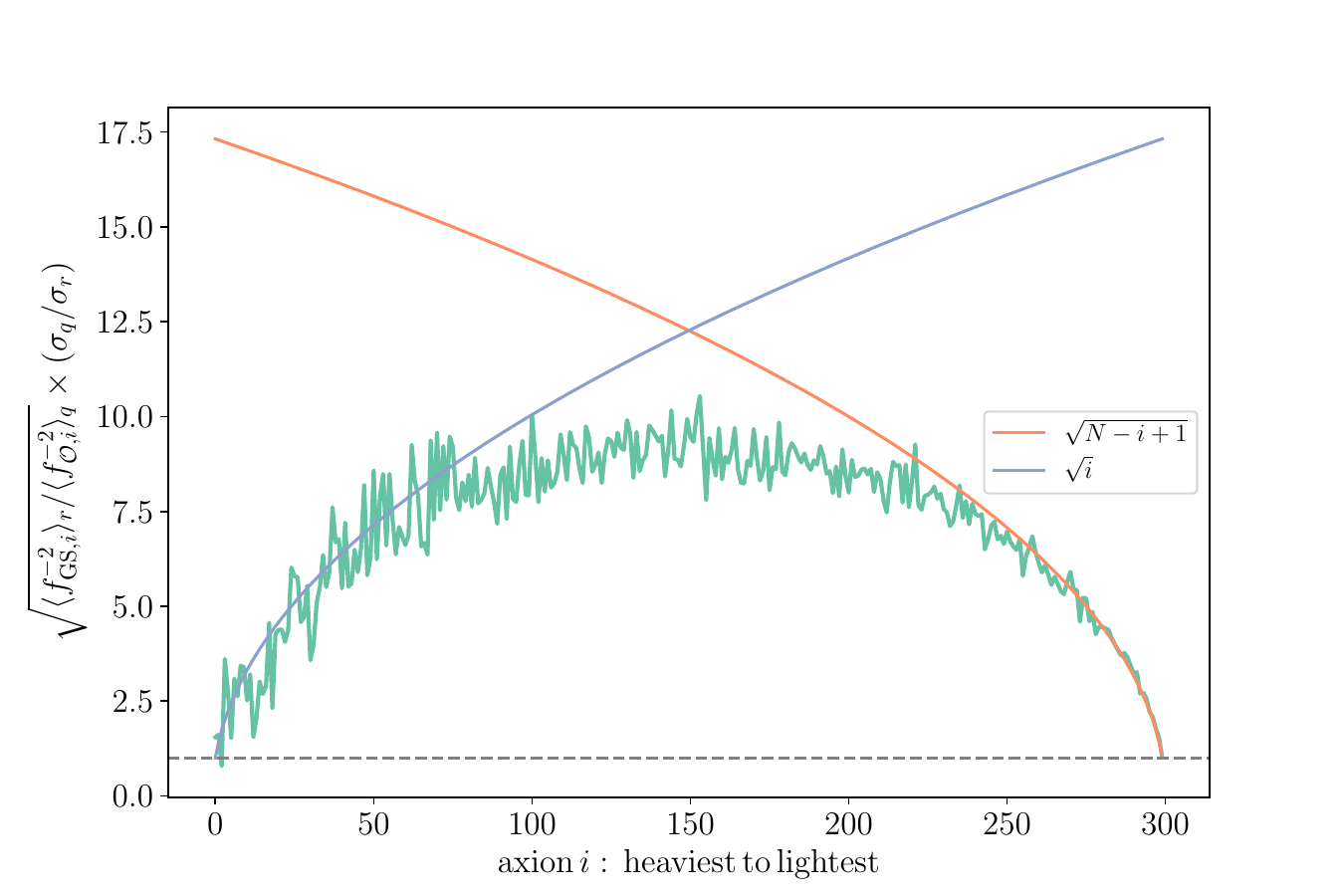}
    \caption{The ratio of the GS decay constants to the corresponding coupling decay constants normalized by the ratio of the variances of the instanton charges and the coupling vector to enable direct comparison to the analytic expressions \cref{eqn:signal-strength-isotropic,eqn:signal-strength-anisotropic} assuming kinetic matrices which are isotropic versus strongly anisotropic, respectively. The green line in this figure corresponds exactly to the ratio of the dashed lime green line to the pink line in \cref{fig:couplings_isotropic}, i.e.\ the ratio of the mean coupling decay constants \cref{eqn:coupling_decay_constant_average} to the analytic expression for the GS decay constants \cref{eqn:fGS-ESP}. The orange line corresponds to the scaling expected in the isotropic limit \cref{eqn:signal-strength-isotropic}, while the blue line corresponds to the expected scaling in the anisotropic limit \cref{eqn:signal-strength-anisotropic}. Values of the green curve larger than $1$ (gray dashed) represent a relative suppression of detection prospects.}
    \label{fig:couplings_ratio}
\end{figure}

As a point of comparison, the dashed line corresponds to the inverse GS decay constants (in the appropriate units), calculated using \cref{eqn:fGS-ESP}. Over the majority of mass ranks, the ratio of the GS decay constants to the coupling decay constants is roughly 10, as expected by the estimates \cref{eqn:signal-strength-anisotropic,eqn:signal-strength-isotropic}: taking the typical value of $i = N/2$, the suppression estimate is $\sqrt{N/2}\sim 12$. At small $i$, the GS process is dominated by a few very small kinetic eigenvalues, so the anisotropic estimate \cref{eqn:signal-strength-anisotropic} applies, and the expected ratio approaches 1, as expected. As the GS process proceeds, anisotropies are eliminated: eventually the isotropic estimate \cref{eqn:signal-strength-isotropic} applies, and as $i$ approaches $N$, the ratio again approaches 1, as expected from \cref{eqn:signal-strength-isotropic}. In \cref{fig:couplings_ratio}, we illustrate that the expected scalings are indeed realized by the example in \cref{fig:couplings_isotropic}.

\subsubsection{QCD axion}
\label{subsubsec:QCD_axion_coupling}
The QCD axion is special, in that its coupling to QCD
\begin{align}
    \frac{1}{32\pi^2}\bm q^T\bm\theta\tilde G G\,,   
\end{align}
also induces a term in the axion effective potential\footnote{We approximate the QCD axion potential by a cosine for simplicity and because it makes little difference for the analysis performed here. For the full potential calculated from chiral perturbation theory, see e.g.\ \cite{GrillidiCortona:2015jxo}.}
\begin{align}
    V\supset\chi_c(1 - \cos \bm q^T\bm\theta)\,,
\end{align}
where $\chi_c$ is the QCD topological susceptibility. 
We can therefore identify a row of the instanton charge matrix with the coupling vector, whose rank $i_{c}$ is determined by the size of the QCD topological susceptibility $\Lambda_{i_{c}}^4 = \chi_c$ relative to the other stringy instanton contributions:
\begin{align}
    \bm q = \bm r_{i_{c}}\,.
\end{align}
As a result, we will find that the QCD axion is special among axions in that it has an unsuppressed coupling to the Standard Model.

In the discussion of \cref{subsec:field-ranges}, we were concerned with the magnitude of the $i$th Gram vector $|\bm\rho_{i\perp}|$, which sets the effective field range of the $i$th axion. Phrased in terms of the combined transformation $\bm A$, we have
\begin{align}
\label{eqn:GS-length-A}
    f_{{\rm GS},i_{c}}^{-1} =|\bm\rho_{i\perp}| = [\bm r_{(i)}^T\bm A]_{(i)}\,.
\end{align}
Setting $\bm q = \bm r_{i_c}$ the expression for the axion-matter couplings in the GS basis \cref{eqn:couplings-in-GS-basis} and $\bm r_{i} = \bm r_{i_c}$ in \cref{eqn:GS-length-A}, we find
\begin{align}
    f_{{\cal O},i_{c}} = f_{{\rm GS},i_{c}}\,.
\end{align}
Recall from above that, except for perhaps the lightest and heaviest axions, the typical suppression of an axion coupling to the Standard Model scaled with $\sqrt{N}$. The QCD axion, on the other hand, satisfies the naive relationship between its field range and its decay constant, meaning that it incurs no suppression in its coupling to the Standard Model.

The couplings of the other axions to the Standard Model are determined by the remaining elements of the coupling vector:
\begin{align}
    [\bm \xi_{i}^T]_{(j)} f_{{\rm GS},(j)}^{-1} \equiv [\bm r_{i}^T \bm A]_{j}\,,
\end{align}
where we have chosen to normalize $\bm \xi_i$ such that $[\bm \xi_{(i)}]_{(i)} = 1$, and in particular\ $\bm \xi_{i_{c}}$ measures couplings relative to the QCD axion in units of the corresponding axion's GS decay constant.

Axions heavier than the QCD axion point in directions determined by the instanton charge vectors (and kinetic term) that are agnostic to the QCD coupling vector. Therefore, in the case of heavy axions, the situation is identical to that of coupling through a random vector $\bm q$. In other words, \cref{eqn:signal-strength-isotropic,eqn:signal-strength-anisotropic} apply, and we find
\begin{align}
\label{eqn:heavy-qcd-coupling}
    [\bm\xi_{i_{c}}]^2_{i<i_{c}}\sim\left\{\begin{array}{cc}(N - i + 1)^{-1}&{\rm(isotropic)}\\ i^{-1}&{\rm(strongly\,\,anisotropic).}\end{array}\right.
\end{align}

On the other hand, the GS process ensures that $[\bm r_i^T \bm A]_j = 0$ for $j>i$, since the $i$th GS basis vector is determined only by those vectors with lower rank. Therefore, in the hierarchical limit where the GS basis coincides exactly with the mass basis, the axions lighter than the QCD axion have exactly zero coupling~\cite{Agrawal:2022lsp,Gendler:2023kjt}:
\begin{align}
    [\bm\xi_{i_{c}}]_{i>i_{c}} = 0\,,\qquad{\rm (hierarchical\,\,limit)}\,.
\end{align}

We can account for finite hierarchy corrections to the light-ALP QCD coupling by computing the first order correction to the mass basis away from the GS basis. Because of the hierarchical nature of the GS procedure, the corrections to the $i$th axion direction will be principally controlled by the $i+1$th axion. Therefore, consider the following two-axion theory in the GS basis:
\begin{align}
    {\cal L}_2 = \frac12(\partial a_i)^2 + \frac12(\partial a_{i + 1})^2 - \Lambda_i^4\left(1 - \cos\left(\frac{a_i}{f_{{\rm GS},i}}\right)\right) - \Lambda_{i + 1}^4\left(1 - \cos\left([\bm\xi_{i+1}]_{i }\frac{a_{i}}{f_{{\rm GS},i}} + \frac{a_{i + 1}}{f_{{\rm GS},i+1}}\right)\right)\,.
\end{align}
Then, at leading order in the hierarchy $\Lambda_{i+1}/\Lambda_i$, the mass basis is:
\begin{align}
    \tilde a_i &= a_i - [\bm\xi_{i+1}]_{i}\frac{m_{i+1}^2}{m_i^2}\frac{f_{{\rm GS},i+1}}{f_{{\rm GS},i}}a_{i + 1}\,,\\
    \tilde a_{i+ 1}&=a_{i+1} + [\bm\xi_{i+1}]_{i}\frac{m_{i+1}^2}{m_i^2}\frac{f_{{\rm GS},i+1}}{f_{{\rm GS},i}}a_{i}\,,
\end{align}
where $m_i$ and $f_{{\rm GS},i}$ are the masses and GS decay constants of the axions in the unperturbed GS basis. Because of the pairwise nature of this perturbative correction, we may straightforwardly generalize to the case of $N$ axions:
\begin{align}
    \tilde{\bm a} = (\bm I + \bm \Xi)\bm a\,,
\end{align}
where at leading order in the mass hierarchy $\bm\Xi$ is antisymmetric:
\begin{align}
    [\bm\Xi]_{i,j} = -[\bm\xi_j]_{i}\frac{m_{j}^2}{m_i^2}\frac{f_{{\rm GS},j}}{f_{{\rm GS},i}}\,,\qquad j>i\,,
\end{align}
where we have also assumed $[\bm \xi_j]_i$ is small (i.e.\ there are corrections to this expression that are the same order in the $\Lambda_i$ hierarchy, but involve higher powers of $[\bm\xi_j]_i$), as indeed, its off-diagonal elements are either 0 or suppressed by the root of the number of axions.

Returning to our question of how strongly axions lighter than the QCD axion couple to the Standard Model, we see that their couplings are parametrically suppressed. The first correction to $\bm\xi_{i_{c}}$ is of the form~\cite{Agrawal:2022lsp,Gendler:2023kjt}:
\begin{align}
\label{eqn:light_axion_QCD_coupling}
    [\bm\xi^{(1)}_{i_{c}}]_{i > i_{c}}\sim [\bm\xi_{i_{c}}]_{i< i_{c}}\frac{m_i^2}{m_{i_{c}}^2}\frac{f_{{\rm GS},i>i_c}^2}{f_{{\rm GS},i_c}^2} =  [\bm\xi_{i_{c}}]_{i< i_{c}}\frac{\Lambda_{i>i_c}^4}{\Lambda_{i_c}^4}\,.
\end{align}
Thus, light-axion couplings to QCD are suppressed by both the instanton potential ratio, and by the dimensional factors that suppress the heavy axions' couplings to QCD \cref{eqn:heavy-qcd-coupling}. The suppressed interaction of the light axions is visible in \cref{fig:relic_signatures} in \cref{sec:results}.

\section{Initial conditions}
\label{sec:anthropics}

In addition to the statistical effects of field ranges and couplings discussed in \Cref{sec:Statistics}, it is natural to ask what a large axiverse implies for cosmology and direct detection if axions make up the dark matter~\cite{Arvanitaki:2010sy}. In particular, a theory that predicts a distribution of dark matter abundances raises an immediate question: how typical is the observed value of the cold dark matter energy density fraction $\Omega_{\rm cdm}$, given the large number of matter-like fields? We quantify this using the anthropic probability $\mathcal P$, defined schematically as the observer-weighted fraction of Universes whose dark matter-to-baryon ratio is no larger than the value we observe.

This line of reasoning is motivated in part by the success of anthropic arguments for the cosmological constant. Interestingly, the same large number of configurations that is responsible for a large number of string vacua and the possibility of finely scanning the cosmological constant~\cite{Bousso:2000xa,Polchinski:2006gy} also leads to the string axiverse~\cite{Arvanitaki:2009fg}. Weinberg provided an anthropic explanation for the selection of a small-but-nonzero cosmological constant by arguing that much larger values would dramatically alter structure formation \cite{Weinberg:1987dv}. Similar logic applies to axion dark matter. For generic axion parameters and priors on initial conditions, the total dark matter abundance can easily overshoot (or undershoot) the range compatible with a hospitable structure formation history \cite{Tegmark:2005dy}. Anthropic weighting therefore provides a concrete way to identify which parts of axion parameter space require tuned initial conditions and, correspondingly, which axions are most likely to carry an appreciable fraction of the dark matter and be experimentally visible.

Anthropic considerations are especially tractable for axions because the prior on the dark matter abundance is set by the prior on the initial misalignment angles, $p(\theta_0)$, which is in turn determined by the inflationary history. Moreover, once an anthropic weight is specified, the same framework both quantifies the degree of tuning needed to obtain the observed $\Omega_{\rm cdm}$ and predicts how the dark matter energy density is distributed across the axions in the ensemble, directly shaping axion relic detection prospects. 

The rest of this section is organized as follows. In \Cref{subsec:anthropic-probability}, we review the ingredients entering $\mathcal P$: the inflationary prior $p(\theta_0)$ (and its dependence on the scale of inflation $H_I$) and our choice of anthropic weight, taken to be the number of observers $N_{\rm obs}$. We then write $\mathcal P$ as an integral over misalignment angles and show in \Cref{subsec:relevant-axions} that, for hierarchical axiverses, it is dominated by the subset of axions with sufficiently large typical energy densities, motivating a definition of ``relevant'' axions. We review the result of Ref.~\cite{Arvanitaki:2010sy} in the limit where all axions are relevant, and then generalize to arbitrary (long) inflation and hierarchical spectra by counting only the relevant subset of axions. 
We find that lowering $H_I$  substantially weakens the exponential suppression of $\mathcal P$ with $N$, and that axion mixing  further enhances anthropic probabilities. Implications for relic abundances, including the emergence of an anthropic plateau and the parametric suppression of heavy-axion contributions, are discussed in \cref{sec:results}.

\subsection{Anthropic probability}
\label{subsec:anthropic-probability}

Quantitatively, the anthropic weight we use depends on the  dark matter abundance through its ratio to baryons,
\begin{align}
    \zeta \equiv \frac{\Omega_{\rm cdm}}{\Omega_b}\,.
\end{align}
 Our goal in this subsection is to express the anthropic probability $\mathcal P$ as an integral over inflationary initial conditions, which themselves are calculable via Fokker-Planck equilibrium~\cite{Graham:2018jyp,Takahashi:2018tdu,Reig:2021ipa}: schematically, $\mathcal P$ is obtained by weighting the prior distribution for $\zeta$ by an anthropic weight $W(\zeta)$.

For axion dark matter, the probability distribution of $\zeta$  is inherited from the probability distribution of the initial misalignment angles, $p(\theta_{0,i})$. To make this explicit, we rewrite the misalignment abundance \cref{eqn:axion-relic-abundance} by combining all factors other than the $\theta_{0,i}$-dependence into a single coefficient $c(m_i,f_{{\rm GS},i})$ as in Ref.~\cite{Arvanitaki:2010sy}. For dark matter composed of $N$ axions,
\begin{equation}
\label{eqn:axion-zeta}
    \zeta(\theta_{0,i},m_i,f_{{\rm GS},i}) \sim \sum_{i=1}^{N} c(m_i,f_{{\rm GS},i}) F(\theta_{0,i}),
\end{equation}
where we drop factors of $g_\star$ for simplicity, and
\begin{align}
    c(m_i,f_{{\rm GS},i}) \approx \left(\frac{m_i}{H_{\rm eq}}\right)^{1/2}\left(\frac{f_{{\rm GS},i}}{M_{\rm pl}}\right)^2\,.
\end{align}
We neglect the anharmonicity of the axion potential [i.e. the logarithm in $F(\theta_0)$; see \cref{eqn:Ftheta}] when performing analytical calculations as it has at most an order unity effect, but retain it in the numerical analysis.
For fixed choices of the axion masses $m_i$ and decay constants $f_{{\rm GS},i}$, the initial conditions $\theta_{0,i}$ are the only variables. Thus, the  probability distribution on $\zeta$ is given by that of $\theta_{0,i}$.

In terms of $\zeta$, the anthropic probability is the fraction of inflationary initial conditions that produce an axion dark matter abundance consistent with our observed value weighted by how many observers are expected to form in the resulting cosmology. We encode the observer bias through an anthropic weight $W(\zeta)$, which implements the idea that values of $\zeta$ far outside a window $[\zeta_{\rm min},\zeta_{\rm max}]$ compatible with galaxy and star formation should contribute negligibly to the inferred probability:
\begin{equation}
\label{eqn:anthrop1}
    \mathcal{P} =
    \mathcal{N}^{-1} \int_{\zeta_{\rm min}<\zeta(\theta_{0})<\zeta_{\rm obs}} d\theta_{0} \ p(\theta_{0}) \cdot W(\zeta),
\end{equation}
where $\zeta_{\rm obs}$ is the observed value of $\approx 5$~\cite{Planck:2018vyg} and the normalization is
\begin{equation}
    \mathcal{N} = \int_{\zeta_{\rm min}<\zeta(\theta_{_{0}})<\zeta_{\rm max}} d\theta_{0} \ p(\theta_{0}) \cdot W(\zeta).
\end{equation}
The normalization integral is evaluated over all values of $\zeta$ that would allow for observer formation, while the compatible cosmologies that appear in the integral \cref{eqn:anthrop1} 
are those for which axions do not produce more than the observed dark matter.
As there is much uncertainty regarding the threshold $\zeta$ values, we provide our results as a function of $\zeta_{\rm min}$ and $\zeta_{\rm max}$, though when evaluating our results numerically, we fix 
$\zeta_{\rm min} = 2.5$ and $\zeta_{\rm max} = 100$ consistent with prior literature \cite{Freivogel:2008qc,Arvanitaki:2010sy}. These threshold values are motivated by structure formation (used as a proxy for observer formation): roughly, $\zeta$ smaller than $\zeta_{\rm min}$ delays or suppresses the growth of galactic halos, while $\zeta$ larger than  $\zeta_{\rm max}$ reduces the baryon fraction enough to hinder efficient cooling and star formation~\cite{Tegmark:2005dy}.

With \cref{eqn:anthrop1} in hand, the calculation of $\mathcal P$ reduces to specifying the two independent ingredients: the prior on initial conditions $p(\theta_{0,i})$ and the anthropic weight $W(\zeta)$. We discuss these in turn.

\subsubsection{Inflationary prior: $p(\theta_{0,i})$}
\label{subsubsec:inf-priors-theta0}

\begin{figure}[t!]
    \centering
    \includegraphics[width=\columnwidth]{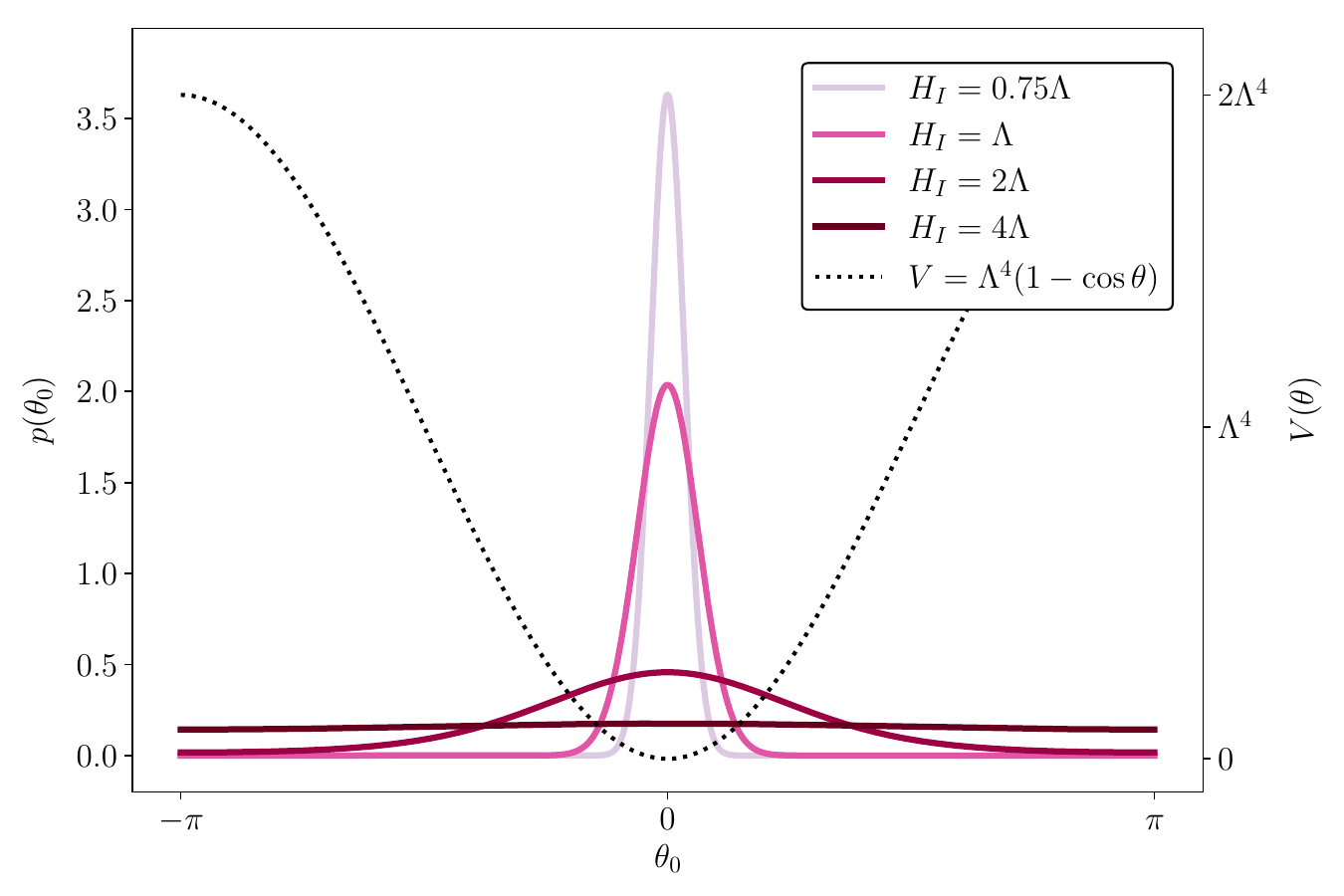}
    \caption{The probability density $p(\theta_0)$ for an axion's initial misalignment angle $\theta_0$, assuming a cosine potential $V(\theta) =\Lambda^4(1 - \cos\theta)$ and accounting for a long period of inflation at the scale $H_I$. Provided inflation lasts long enough [\cref{eqn:long-inflation}], the axion distribution $p(\theta_0)$ relaxes to a von Mises distribution, whose width is determined by the scale of inflation, as given in \cref{eqn:inflationtheta0-dist}. The lower the scale of inflation, the more Gaussian and peaked at $\theta_0=0$ the the distribution will be. The higher the scale of inflation, the more closely the distribution for the initial conditions will resemble a flat distribution.}
    \label{fig:theta0prior}
\end{figure}

The prior on the initial conditions is intimately connected to cosmological history. We focus on the pre-inflationary scenario, where the scale of inflation is low enough $H_I \ll f$ that the axions remain the only relevant degrees of freedom i.e.\ axion radial modes or moduli remain fixed at their vacuum expectation values. Under this assumption, the prior probability distribution $p(\theta_{0,i})$ is determined by the duration and scale of inflation, and by the axion potentials themselves.

Two competing effects contribute to the evolution of the axion field value during inflation: diffusion caused by inflationary fluctuations and slow-roll down the potential. These effects are captured mathematically by the Fokker-Planck equation, which describes the time evolution of the axion distribution under these two effects, and was studied in detail for a single axion in Ref.~\cite{Graham:2018jyp}. Under the assumption of hierarchical instantons in \cref{sec:Statistics}, each axion can be approximated by a single-axion effective theory \cref{eqn:GS_Lagrangian}, and therefore the results of Ref.~\cite{Graham:2018jyp} apply without modification to each axion individually. We recapitulate the relevant results here.

If inflation lasts long enough, then the misalignment angle distribution $p(\theta_{0,i})$ relaxes to an equilibrium state: the Fokker-Planck equilibrium. The relaxation time in e-folds can be estimated in terms of the axion parameters and the scale of inflation as \cite{Graham:2018jyp}
\begin{equation}
\label{eqn:long-inflation}
    N_e \sim \begin{cases}
    f_{{\rm GS},i}^2/H_I^2 &  \ \Lambda_i < H_I\,, \\
    H_I^2/m_i^2 &  \ \Lambda_i > H_I\,.
    \end{cases}
\end{equation}
We assume that inflation lasts at least as long as required by \cref{eqn:long-inflation}, so that the probability distribution for the axion field's value at the end of inflation is given by:
\begin{equation}
    \label{eqn:inflationtheta0-dist}
    p(\theta_{0,i}) \propto \exp{\left( -\frac{8\pi^2V(\theta_{0,i})}{3H_I^4} \right)}\,.
\end{equation}
Because the axion potentials are quadratic about their minima, $p(\theta_{0,i})$ exhibits two qualitatively different behaviors depending on the scale of inflation relative to the height of the potential $\Lambda_i^4$.
If $H_I^4\gg\Lambda_i^4$, the distribution flattens such that we recover the usual flat prior on the axion misalignment $\theta_{0,i}\sim {\cal U}(-\pi,\pi)$. On the other hand, as the scale of inflation becomes smaller than the axion's potential, the distribution becomes more sharply peaked at $\theta_{0,i}=0$. More precisely, the distribution approximates a normal distribution: $\theta_{0,i}\sim {\cal N}(0,3 H_I^4/8\pi^2\Lambda_i^4)$. We illustrate these limiting cases and the transition between them in \cref{fig:theta0prior}.

While we focus on sufficiently long inflation such that the Fokker-Planck distribution accurately describes the misalignment prior, it is worth noting that the minimum duration of inflation required by the CMB is $\lesssim 60$ $e$-folds, which for sufficiently light axions can be far shorter than the relaxation time [see \cref{eqn:long-inflation}] needed to erase pre-inflationary initial conditions. In this regime, the misalignment prior is no longer universal and depends on additional assumptions about the state of the axion fields at the onset of inflation.

\subsubsection{Anthropic weight: $W(\zeta)$}
\label{subsubsec:anthropic-weight}

While the inflationary prior $p(\theta_{0,i})$ is comparatively robust, much of the challenge in anthropic reasoning lies in specifying the weight $W(\zeta)$: relating a dark matter abundance to the formation of ``observers'' necessarily involves coarse-graining, astrophysical assumptions, and a choice of measure. Here we follow Ref.~\cite{Freivogel:2008qc} and take the anthropic weight to be proportional to the number of observers $N_{\rm obs}$ in a causal patch. Ref.~\cite{Freivogel:2008qc} used this choice to evaluate the anthropic probability for a single axion constituting the dark matter. To use $N_{\rm obs}$ as a weight on $\zeta$, one must determine its $\zeta$-dependence. While the conditions on $\zeta$ for observer formation are speculative, Ref.~\cite{Freivogel:2008qc} isolates the most uncertain astrophysical inputs by factorizing
\begin{align}
\label{eqn:n_obs_ratio}
    N_{\rm obs}(\zeta) = N_b(\zeta)\,\frac{N_{\rm obs}}{N_b}(\zeta)\,,
\end{align}
where $N_b$ is the number of baryons in the causal patch. The quantity $N_b(\zeta)$ can be estimated within a chosen measure, while the remaining factor $N_{\rm obs}/N_b$---the number of observers per baryon---is taken to be approximately constant within an anthropically allowed window in $\zeta$ and negligible outside it.

In the context of eternal inflation in which an exponentially large volume is continuously being produced, the number of baryons in the Universe is infinite. To render the anthropic weight calculable, one must choose how to regularize $N_b$. One such scheme is the causal diamond measure introduced by Bousso \cite{Freivogel:2008qc,Bousso:2006ev}. The causal diamond measure restricts the counting of $N_b$ to one causal patch, decoupling $N_b$ from the exponentially growing volume outside our horizon. Here a causal patch, or diamond, refers to the largest area of spacetime with which an observer is in causal contact. Geometrically, a causal patch is delimited by the diamond formed by the intersection of a past lightcone at a late-time event (taken to be the time of observation), and the future lightcone at an early-time event (taken to be reheating). The anthropic weight can therefore be calculated by counting the number of observers per baryon within a single causal patch. By regulating the infinite volume, the number of baryons that enters the anthropic weight is finite.

To isolate the effect of varying the dark matter abundance itself, we follow \cite{Freivogel:2008qc,Arvanitaki:2010sy} and hold all other cosmological parameters fixed while scanning $\zeta$; in particular, we fix the baryon-to-photon ratio (equivalently the baryon energy density at fixed radiation density) so that changes in $\zeta$ exactly correspond to changing the dark matter density rather than simultaneously re-tuning the baryon sector.
The key result is that
\begin{equation}
\label{eqn:Nb}
    N_b \propto \frac{1}{1+\zeta},
\end{equation}
at late times, when the radiation energy density is negligible relative to the matter density $\Omega_m\equiv \Omega_{b}+\Omega_{\rm cdm} $. This dependence can be understood by considering the amount of baryonic mass contained within a causal patch. 
At matter-dark energy equality, defined by $\Omega_m = \Omega_\Lambda$, we have
\begin{equation}
    \Omega_b = \frac{\Omega_m}{1+\zeta}=\frac{\Omega_\Lambda}{1+\zeta}\,,
\end{equation}
where we have used that $\Omega_\Lambda$ is held fixed as $\zeta$ is varied.\footnote{Note that while very light axions can also contribute to dark energy, the energy density in these axions is negligible for the parameters considered in this work and we neglect it here.} In the causal diamond measure, the physical size of the causal patch at equality is set by the Hubble rate at matter-dark energy equality, which depends only on $\Omega_\Lambda$ and is therefore independent of $\zeta$ by assumption. It follows that the total baryonic mass in the patch scales as $M_b\propto \Omega_b$, and hence the number of baryons obeys $N_b\propto M_b \propto (1+\zeta)^{-1}$, giving \cref{eqn:Nb}.

The above reasoning justifies the form of \cref{eqn:Nb} specifically at matter-dark energy equality, but this scaling persists throughout dark energy domination, during which the volume of the causal patch remains independent of $\zeta$. Furthermore, the total matter density is fixed by the condition at matter–dark energy equality and so the total mass in the patch does not change with $\zeta$. Thus, increasing the dark matter fraction necessarily reduces the baryonic mass by the same factor as measured in the dark energy dominated epoch. 

Note that at early times, for example near matter-radiation equality, the energy density in radiation is not negligible. Because we fix the baryon-to-photon ratio and the radiation density is observationally constrained, the baryon density at these times does not vary with $\zeta$. Instead, increasing $\zeta$ increases the total matter density, causing matter–radiation equality to occur earlier. Nevertheless, as the Universe expands and radiation becomes negligible, the late-time scaling of \cref{eqn:Nb} is recovered. Since $W(\zeta)$ is determined by late-time observers (and by the late-time baryon content of a causal patch), we use the late-time scaling \cref{eqn:Nb} and ignore early-time departures near matter-radiation equality.

Returning to \cref{eqn:n_obs_ratio}, we follow \cite{Tegmark:2005dy,Freivogel:2008qc} and take $N_{\rm obs}/N_b$ to be approximately constant within the range $\zeta_{\rm min}<\zeta<\zeta_{\rm max}$.
Outside of this range, $N_{\rm obs}/N_b \approx 0$, resulting in the simple proportionality
\begin{align}
\label{eqn:n_obs_zeta}
    N_{\rm obs} \propto \left\{\begin{array}{cc}(1 + \zeta)^{-1}\,,& \zeta\in[\zeta_{\rm min},\zeta_{\rm max}]\,,\\0&{\rm otherwise}\,.\end{array}\right.
\end{align} 
Therefore, under the simple, standard assumption that the efficiency of forming observers from baryons is constant in this range, the anthropic weight $N_{\rm obs}$ favors scenarios with a larger number of baryons.

This discussion has several caveats. In particular, the choice of anthropic weight is not unique, nor is there an optimal choice, as the requirements on the measure for observer formation cannot be determined. Indeed, many simplifying assumptions were required to obtain the simple relationship \cref{eqn:n_obs_zeta}. The hope is that in spite of these simplifications, the anthropic weight retains some level of realism at the qualitative level. Indeed, while the specific mathematical relationship \cref{eqn:n_obs_zeta} can be debated, the fact that $N_{\rm obs}\to 0$ as $\zeta\to\infty$ seems like a robust expectation since it corresponds to an absence of baryons in the patch. Nevertheless, there are many other possible choices for the anthropic weight \cite{Bousso:2007kq}, but as pointed out in \cite{Arvanitaki:2010sy}, the general trend of the results for the anthropic probability is largely independent of the choice of weight. Instead, as discussed in \Cref{subsec:relevant-axions}, it is a consequence of the geometry of the integrals and their integration regions.

Putting the ingredients of this discussion together, we can write down a compact form for the anthropic probability. Using the expression for $\zeta$ given in \cref{eqn:axion-zeta}, the anthropic weight $W(\zeta)\propto N_{\rm obs}(\zeta)$ from \cref{eqn:Nb}, and the inflationary priors on $\theta_{0,i}$ given in \cref{eqn:inflationtheta0-dist}, the probability in \cref{eqn:anthrop1} becomes:
\begin{equation}
\label{eqn:anthrop2}
 \mathcal{P} =
    \mathcal{N}^{-1} \int_{\zeta_{\rm min}<\sum_ic_i\theta_{0,i}^2<\zeta_{\rm obs}} \frac{\prod_{i}d\theta_{0,i} \ p_i(\theta_{0,i})}{1+\sum_ic_iF(\theta_{0,i})},    
\end{equation}
with $c_i \equiv c(m_i,f_{{\rm GS},i})$. While this is a challenging integral to compute in general, there is one clear feature that appears from the functional form of the integrand: if an axion's maximum typical energy density $c_iF(\theta_{0,i}) \ll 1$, it will be irrelevant to the integral. On the other hand, all axions for which the maximum typical $c_iF(\theta_{0,i}) \gtrsim 1$ will contribute and tend to suppress the anthropic probability. This feature will be leveraged in \Cref{subsec:relevant-axions} to simplify the evaluation of \cref{eqn:anthrop2} for generic choices of axion parameters and for an arbitrary scale of inflation.

\subsection{Relevant axions}
\label{subsec:relevant-axions}

The anthropic probability for $N$ independent axions was determined in~\cite{Arvanitaki:2010sy}, for the same choice of anthropic weight and measure as discussed in \Cref{subsubsec:anthropic-weight}. The authors consider axions at the GUT scale $f\sim 10^{16}\,{\rm GeV}$, with masses $\geq 10^{-19}$ eV, and a high scale of inflation so that $\theta_{0,i} \sim {\cal U}(-\pi,\pi)$. With such assumptions, every axion's typical relic density can overclose the Universe $c_i\pi^2 \gtrsim 5$ so every axion contributes to the integral \cref{eqn:anthrop2}. The integral then no longer depends on the masses and one can safely make a change of variables $\theta_{0,i} \rightarrow \vartheta_{0,i} c_i^{-1/2}$ so that $\zeta=\vartheta_{0,i}^2$ is a radial coordinate in $\vartheta_{0,i}$-space. The probability becomes straightforward to compute in spherical coordinates:
\begin{equation}
\label{eqn:bookkeeping-result}
    \mathcal P = \frac{\int_{\zeta_{\rm min}}^{\zeta_{\rm obs}} \frac{d\zeta \zeta^{(N-2)/2}}{1+\zeta}}{\int_{\zeta_{\rm min}}^{\zeta_{\rm max}}\frac{d\zeta \zeta^{(N-2)/2}}{1+\zeta}}\approx \text{0.3, 0.16, 0.06, 0.02, ...}
\end{equation}
where we have listed the result for the first few values of $N$ and for the $\zeta$ thresholds at their nominal values. The result for a single axion, $\mathcal P = 0.3$, matches the result of Ref.~\cite{Freivogel:2008qc}, which interprets this probability as ``30\% of observers form in regions with less dark matter than we observe, while 70\% of observers form in regions with more dark matter.'' \cref{eqn:bookkeeping-result} is shown as a function of $N$ in \cref{fig:prob-vs-Na}. Analytically, the anthropic probability for such a scenario is approximately given by
\begin{equation}
\label{eqn:anthrop-prob-heavy}
    \mathcal P 
     \lesssim  
    \left(\frac{\zeta_{\rm obs}}{\zeta_{\rm max}}\right)^{\frac{N-2}{2}}\,.
\end{equation}
The dependence of $\mathcal P$ on $\zeta_{\rm min}$ quickly becomes irrelevant for $N$ more than a few, so the anthropic probability depends largely on the ratio $\sqrt{\zeta_{\rm obs}/\zeta_{\rm max}} \sim 0.22$ and drops exponentially with $N$. It is important to point out that, as noted in \cite{Arvanitaki:2010sy}, while the choice of anthropic weight does appear as the factor of $(1+\zeta)^{-1}$ in both the numerator and denominator of \cref{eqn:bookkeeping-result}, the exponential scaling does not depend on this factor; it depends solely on the geometric factor of $\zeta^{(N-2)/2}$, and on the bounds of integration.

\begin{figure}[t!]
    \centering
    \includegraphics[width=\columnwidth]{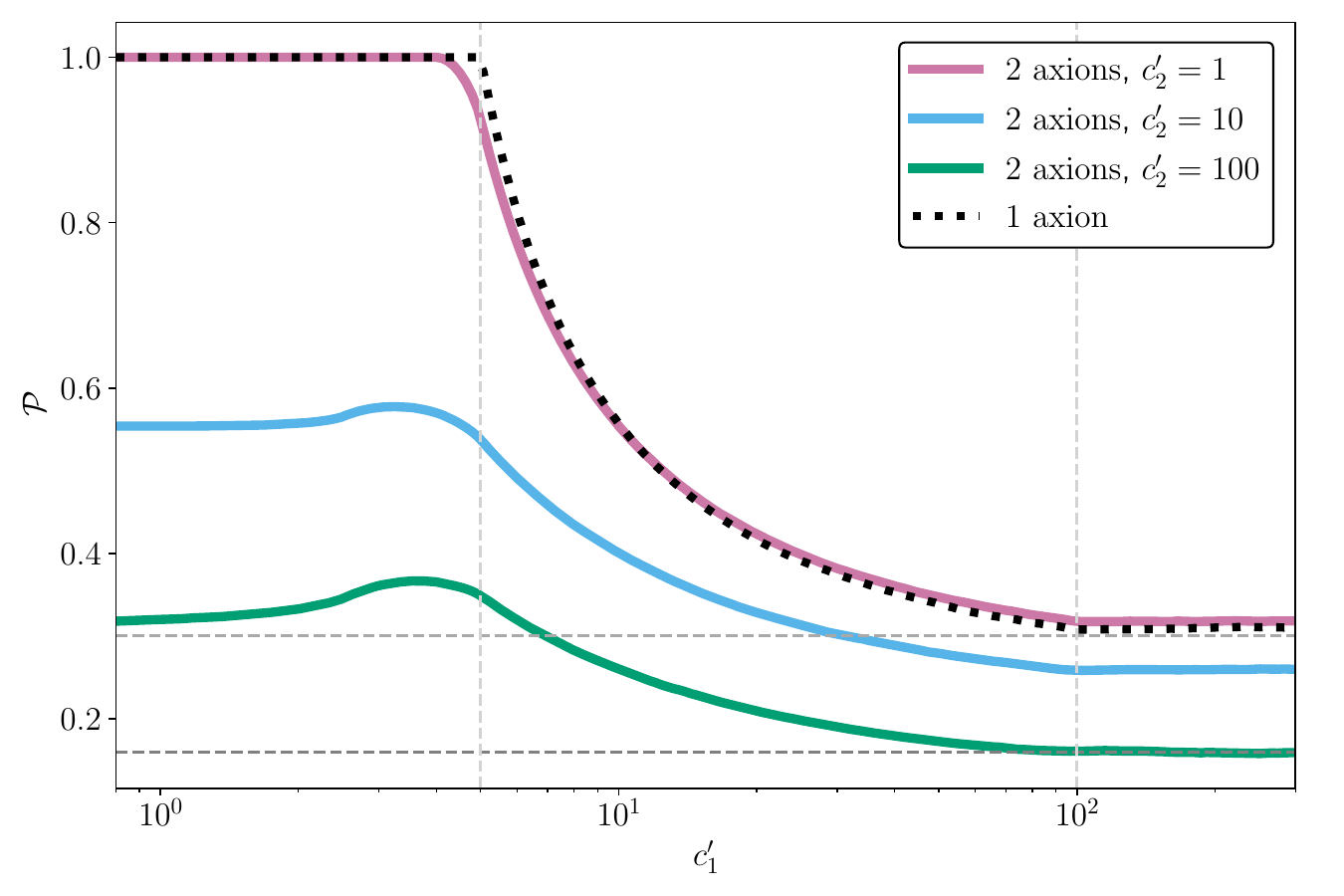}
    \caption{The anthropic probability $\mathcal P$ as a function of $c_1'$ in the one axion (black dotted) and two axion (color) scenarios as calculated with \cref{eqn:prob-theta-bounds}. For $c_1' \geq 100$, the anthropic probability reaches a constant value of $\mathcal P = 0.3$. For $c_1'<2.5$, $\mathcal P \rightarrow  1$. 
    Each curve fixes the value of $c_2'$, and scans over the value of $c_1'$. If $c_2'=1$ (pink curve), the second axion does not contribute significantly to the anthropic probability, and so the pink curve follows the single axion scenario (black, dotted curve). The effect of the second axion is in the rounded edge of the pink curve near $c_1' \sim 5$, which is otherwise sharp for the single-axion scenario, and to modestly alter the asymptotic ${\cal P}$ at large $c_1'$. For $c_2'=100$ (green curve), the second axion is relevant to the anthropic probability. As $c_1' \rightarrow0$, the first axion becomes irrelevant and thus the green curve plateaus at the single-axion value $\mathcal P = 0.3$. Once $c_1'>100$, $\mathcal P = 0.16$, corresponding to the two axion result in \cref{eqn:bookkeeping-result}. There is a nontrivial transition region for the value of $c_i'$, between when an axion is irrelevant and when it becomes relevant, that exhibits two main features: the first is the bumps right before $c_1'=5$, and the second is the broad transition region, illustrated both by the intermediate values taken by the blue curve (where $c_2'=10$) and the slow decline of the colored contours between $c_1' = 5$ and $100$. While in principle an accurate tally of the anthropic probability requires account for these transition regions, they are marginal in the full landscape of possible axion parameters and inflationary values, as illustrated in \cref{fig:anthropic-parameter-space}. 
    }
    \label{fig:1-2-axion-probs}
\end{figure}

We now generalize \cref{eqn:bookkeeping-result} to an axiverse with hierarchical masses and to long-lasting inflation at an arbitrary scale. The main new ingredient is that the coefficients $c_i$ need not all be $\gtrsim 1$, so the rescaling $\theta_{0,i}\to c_i^{-1/2}\vartheta_{0,i}$ that made $\zeta$ a simple radial coordinate is no longer correct. Instead, we proceed by making the inflationary prior explicit. When inflation lasts longer than the relaxation time in \cref{eqn:long-inflation}, the misalignment angles are drawn from the equilibrium distribution \cref{eqn:inflationtheta0-dist}. For analytic control, we approximate this equilibrium prior by a top-hat distribution supported on $\theta_{0,i}\in[-\Sigma_i,\Sigma_i]$, with width $\Sigma_i$ set by $H_I$, $m_i$, and $f_{{\rm GS},i}$:
\begin{equation}
\label{eqn:pdf-sigma}
    \Sigma_i^2 = \begin{cases}
        \frac{3H_I^4}{8\pi^2m_i^2f_{{\rm GS},i}^2} & \text{for } \ H_I\ll \Lambda_i\,, \\
        \hspace{0.7cm}\pi^2 & \text{for } \  H_I \gg \Lambda_i\,.
    \end{cases}
\end{equation}
Then, it is convenient to change variables $\theta_{0,i} \rightarrow \vartheta_{0,i}/\Sigma_i$ and rewrite \cref{eqn:anthrop2} [again, dropping the logarithm in \cref{eqn:Ftheta}] as,
\begin{equation}
\label{eqn:prob-theta-bounds}
    \mathcal{P} =
    \mathcal{N}^{-1} \int_{-1}^1 \frac{\prod_{i=1}^{N}d\vartheta_{0,i}}{1+c_j'\vartheta_{0,j}^2}\Theta(c_k'\vartheta_{0,k}^2-\zeta_{\rm min})\Theta(\zeta_{\rm obs}-c_\ell'\vartheta_{0,\ell}^2).
\end{equation}
Here, repeated indices are summed over, $\Theta$ is the Heaviside function, which provides the cutoffs on $\zeta$, and
\begin{align}
    c_i'\equiv\Sigma_i^2c_i\,,
\end{align}
is interpreted as the typical energy density of the $i$-th axion
and is a function of $H_I, m_i \ \text{and} \ f_i$.

While the bounds $\vartheta_{0,i} \in [-1,1]$ corresponds to a hypercube in $\vartheta_{0,i}$-space of side length 2, the Heaviside functions pick out a hyper-ellipsoidal shell over which to integrate. The shell is delimited by an inner and outer hyper-ellipsoid, whose $N$ principal axes have squared lengths given by $\zeta_{\rm min}/c_i'$ and $\zeta_{\rm obs}/c_i'$ respectively, for $i \in \{1,\dots,N\}$. This integral cannot be computed analytically for $N>1$. However, we can gain intuition for how the anthropic probability depends on $c_i'$ by studying the two simplest cases: $N=1$ and $N=2$. The numerical evaluation of \cref{eqn:prob-theta-bounds} for $N=2$ is shown in \cref{fig:1-2-axion-probs} as a function of $c_1'$ for fixed choices of $c_2'$. The black dotted line shows the anthropic probability evaluated analytically for $N=1$, as a function of $c_1'$. In \cref{fig:1-2-axion-probs}, we choose the values $\zeta_{\rm min} = 2.5$, $\zeta_{\rm obs} = 5$, and $\zeta_{\rm max} = 100$. The vertical dashed lines indicate $\zeta_{\rm obs}$ and $\zeta_{\rm max}$.

A main feature of \cref{fig:1-2-axion-probs} is the behavior for $c_1'>\zeta_{\rm max}$. In this regime, the anthropic probability is independent of the value of $c_1'$: once the typical energy density associated with the first axion exceeds the upper anthropic threshold, the allowed region is dominated by initial conditions in which $\vartheta_{0,1}$ is tuned small enough to satisfy the bound. This behavior is visible in both the one- and two-axion cases. More generally, any axion with $c_i'>\zeta_{\rm max}$ contributes to $\mathcal P$ in the same way, which is precisely what makes \cref{eqn:bookkeeping-result} tractable: in that limit every axion satisfies $c_i'>\zeta_{\rm max}$, so the result becomes independent of the individual values of $c_i'$ and we recover \cref{eqn:bookkeeping-result}.

In the opposite limit, when $c_1'\lesssim \zeta_{\rm min}$, axion $\vartheta_1$ is effectively irrelevant for satisfying the anthropic bounds in a multi-axion theory where other relevant axions exist. The single-axion case illustrates an important limiting behavior: as $c_1'\to 0$, one finds $\mathcal P\to 1$, reflecting the fact that essentially all initial conditions lie below $\zeta_{\rm obs}$. For this axion to be consistent with observer formation, initial conditions near the top of the potential are therefore required to raise $\zeta$ into the anthropically allowed window. Although such an initial condition is dynamically fine-tuned, an axion that accounts for the observed dark matter only by starting near the hilltop would, absent a mechanism that prepares it there~\cite{Co:2018mho,Huang:2020etx}, be a strong indication of anthropic selection. Thus, one way to interpret ${\cal P}$ is as a proxy for the degree to which anthropic selection is consistent with the observed dark matter abundance.

Finally, the nontrivial features in the two-axion case for $5\lesssim c_1'\lesssim 100$ arise from the geometry of the integration region in \cref{eqn:prob-theta-bounds}. As $c_1'$ increases, the ellipsoidal shell selected by the Heaviside functions stretches along the $\vartheta_{0,1}$ direction and begins to intersect (and then protrude beyond) the square set by the integration bounds, leading to the non-monotonicity seen in \cref{fig:1-2-axion-probs}.

This analysis extends straightforwardly to $N$ axions by generalizing the square and ellipses of the two-dimensional case to an $N$-dimensional hypercube and an ellipsoidal shell. While the transition regime
$\zeta_{\rm min}\lesssim c_i' \lesssim \zeta_{\rm max}$
generally requires numerical evaluation for $N\ge 2$, the limiting cases $c_i'\ll \zeta_{\rm min}$ and $c_i'\gg \zeta_{\rm max}$ extend cleanly to arbitrary $N$. Explicitly, if $c_i'\gg \zeta_{\rm max}$, then satisfying the anthropic upper bound forces $\theta_{0,i}$ to be tuned small, and the resulting contribution to $\mathcal P$ is parametrically the same as in the single-axion case appearing in \cref{eqn:bookkeeping-result}. We refer to such axions as relevant (for the anthropic integral), since they contribute the characteristic suppression associated with anthropic tuning. Conversely, axions with $c_i'\ll \zeta_{\rm min}$ cannot contribute appreciably to $\zeta$ even at maximal misalignment and are therefore irrelevant for the anthropic probability provided there is at least one relevant axion.

\begin{figure}[t!]
    \centering
    \includegraphics[width=\columnwidth]{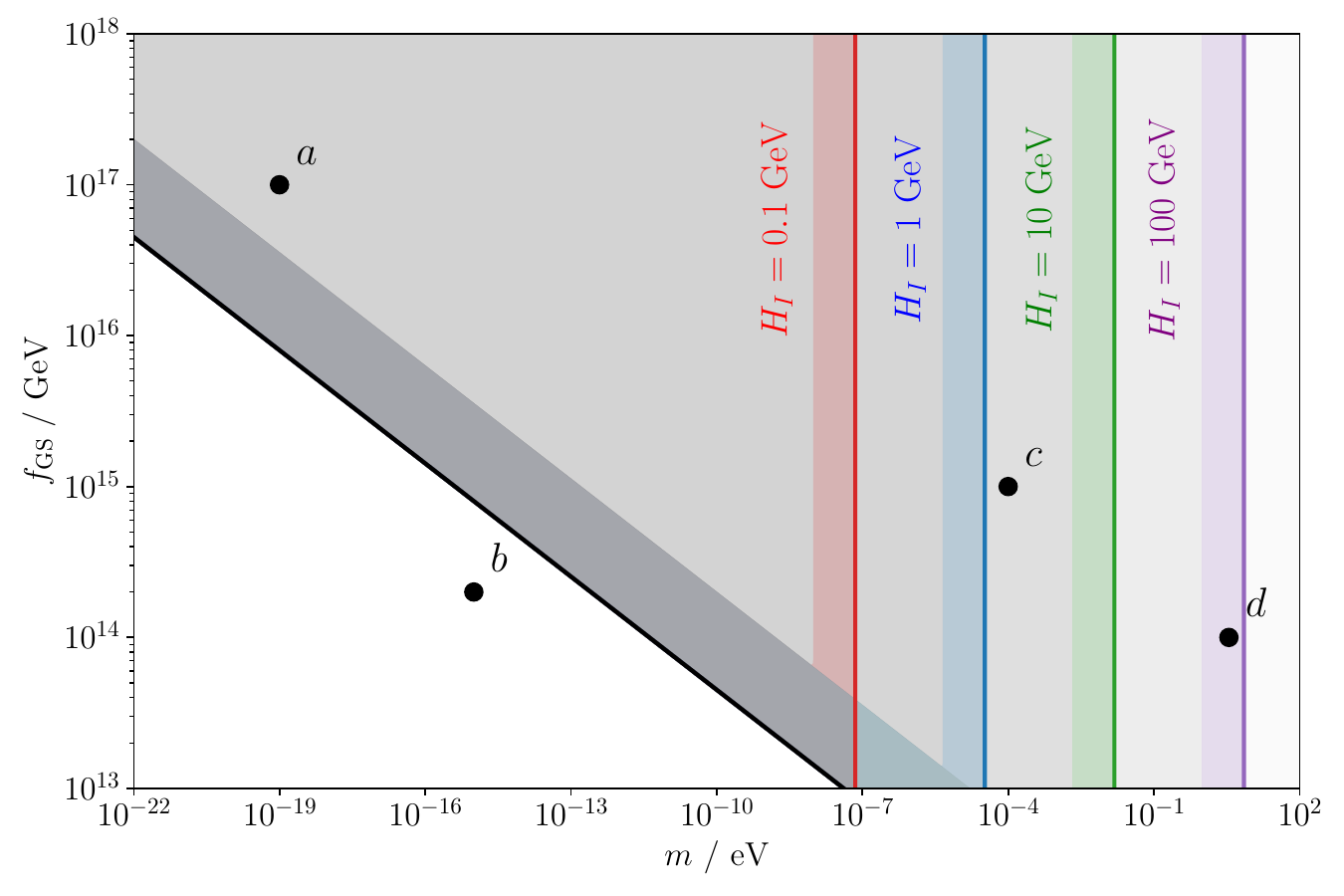}
    \caption{The parameter space of axions that contribute to the anthropic probability $\mathcal P$ in \cref{eqn:prob-theta-bounds}. The shaded region shows the part of the $f_{\rm GS}$ vs. $m$ parameter space that satisfies the criterion in \cref{eqn:anthropic-condition} and thus contributes to suppressing the anthropic probability. For a given set of axion parameters and inflationary scale, one estimates the anthropic probability by counting the number of axions that lie within this region. This number is then used to evaluate the probability in \cref{eqn:bookkeeping-result}. In the case of high-scale inflation (such that all axions have initial condition distributions), all of parameter space to the right of the black line is the anthropically relevant region; $a$, $c$, and $d$ are relevant axions, while $b$ is irrelevant. Example cases of low-scale inflation are illustrated by the colored vertical cutoffs: in these cases, axions in the parameter space to the right of the corresponding vertical line no longer count towards the anthropic probability. For example, if $H_I=10$ GeV, then the anthropic region is the space between the black and green lines; $d$ is no longer in the anthropic region, while $a$ and $c$ would still correspond to relevant axions. The colored shaded regions and the dark gray shaded area are transition regions (see text around \cref{fig:1-2-axion-probs}); an axion at point $d$ would therefore contribute a partial suppression to the anthropic probability if $H_I=100$ GeV. We conservatively consider such marginal axions to be relevant when computing the probability analytically. Note, relevant axions are exactly those which constitute the anthropic plateau discussed in \cref{sec:results} and visible in \cref{fig:relic_abundance,fig:relic_signatures}.
    }
    \label{fig:anthropic-parameter-space}
\end{figure}

Because the intermediate region is not sharply defined, we adopt a simple conservative criterion and classify an axion as relevant if $c_i' > \zeta_{\rm obs}$ (and irrelevant otherwise). In terms of the physical parameters, this condition is
\begin{equation}
\label{eqn:anthropic-condition}
    \left(\frac{m_i}{H_{\text{eq}}}\right)^{1/2} \left(\frac{f_{\text{GS},i}}{M_{\text{pl}}}\right)^2 \min{\left( \pi^2, \frac {3H_I^4}{8\pi^2m_i^2f_{\text{GS},i}^2}\right)} \gtrsim \zeta_{\rm obs}\,.
\end{equation}
The parameter space where the condition of \cref{eqn:anthropic-condition} is satisfied is shown in \cref{fig:anthropic-parameter-space}. The shaded gray region indicates the values of $(m,f_{\rm GS})$ for which an axion is relevant in the sense of \cref{eqn:anthropic-condition}, i.e.\ it can naturally contribute $\zeta\gtrsim \zeta_{\rm obs}$ for typical initial conditions. The boundary scales as $m^{-1/2}$ and corresponds to the high-scale inflation regime in which the axion potential during inflation is negligible, $\Lambda^4=m^2 f_{\rm GS}^2 \ll H_I^4$. In this case the misalignment angle is approximately drawn from a flat distribution on $[-\pi,\pi]$, and the typical relic abundance follows the usual scaling $\Omega\propto m^{1/2} f_{\rm GS}^2$ [cf.\ \cref{eqn:axion-relic-abundance}].

The colored curves show how this criterion is modified at larger masses for finite $H_I$. Once $m$ is large enough that inflationary fluctuations populate only a narrow range of misalignment angles, the typical misalignment is set by the equilibrium width $\theta_0^2\sim 3H_I^4/(8\pi^2 m^2 f_{\rm GS}^2)$. Substituting this expression into the misalignment abundance cancels the explicit $f_{\rm GS}$-dependence, so the relevance condition becomes approximately a statement about $m$ alone (equivalently, $\Omega\propto H_I^4m^{-3/2}$ in this regime \cite{Ho:2019ayl}). As a result, the high-mass boundary of the relevant region becomes approximately vertical: for $m$ to the right of the corresponding cutoff, inflation is low enough that typical $\theta_{0,i}$ values are naturally small, so these axions underproduce and no longer satisfy \cref{eqn:anthropic-condition}. This $f_{\rm GS}$-independence of the large-mass boundary was also noted in \cite{Reig:2021ipa}.

Determining the anthropic probability for an axiverse with hierarchical masses and an arbitrary scale of inflation therefore amounts to computing \cref{eqn:bookkeeping-result}, but only counting the axions that lie within the anthropic region, as only the relevant axions will contribute:
\begin{equation}
\label{eqn:anthrop-prob-analytical}
    \mathcal P \sim \left(\frac{{\zeta_{\rm obs}}}{\zeta_{\rm max}}\right)^{(N_{\rm rel}-2)/2}\,,\qquad N_{\rm rel}\gg 1\,.
\end{equation}
The number of relevant axions $N_{\rm rel}$ can be approximated by
\begin{equation}
\label{eqn:n-relevant}
    N_{\rm rel} = \min{\left(N,\frac{\log{\left(m_{\rm max}/m_{\rm min}\right)}}{\langle \log (m_{i}/m_{i + 1})\rangle}\right)},
\end{equation}
where the boundary of the relevant axion region is determined by
\begin{equation}
    m_{\rm max} \sim\frac{H_{\rm eq}}{(8\pi^2\zeta_{\rm obs}/3)^{2/3}}\left(\frac{H_I^2}{H_{\rm eq}M_{\rm pl}}\right)^{4/3}\,,
\end{equation}
and
\begin{equation}
    m_{\rm min} \sim H_{\rm eq}\frac{\zeta_{\rm obs}^2}{\pi^2}\left(\frac{M_{\rm pl}}{f_{{\rm GS},i}}\right)^{4}\,.
    \label{eqn:m_min}
\end{equation}
The relevant axions, whose masses lie in this range, constitute an anthropic plateau in which all axions have comparable relic abundances that sum to the observed dark matter density. We discuss implications of this plateau in \cref{sec:results}.

\begin{figure}[t!]
    \centering
    \includegraphics[width=\columnwidth]{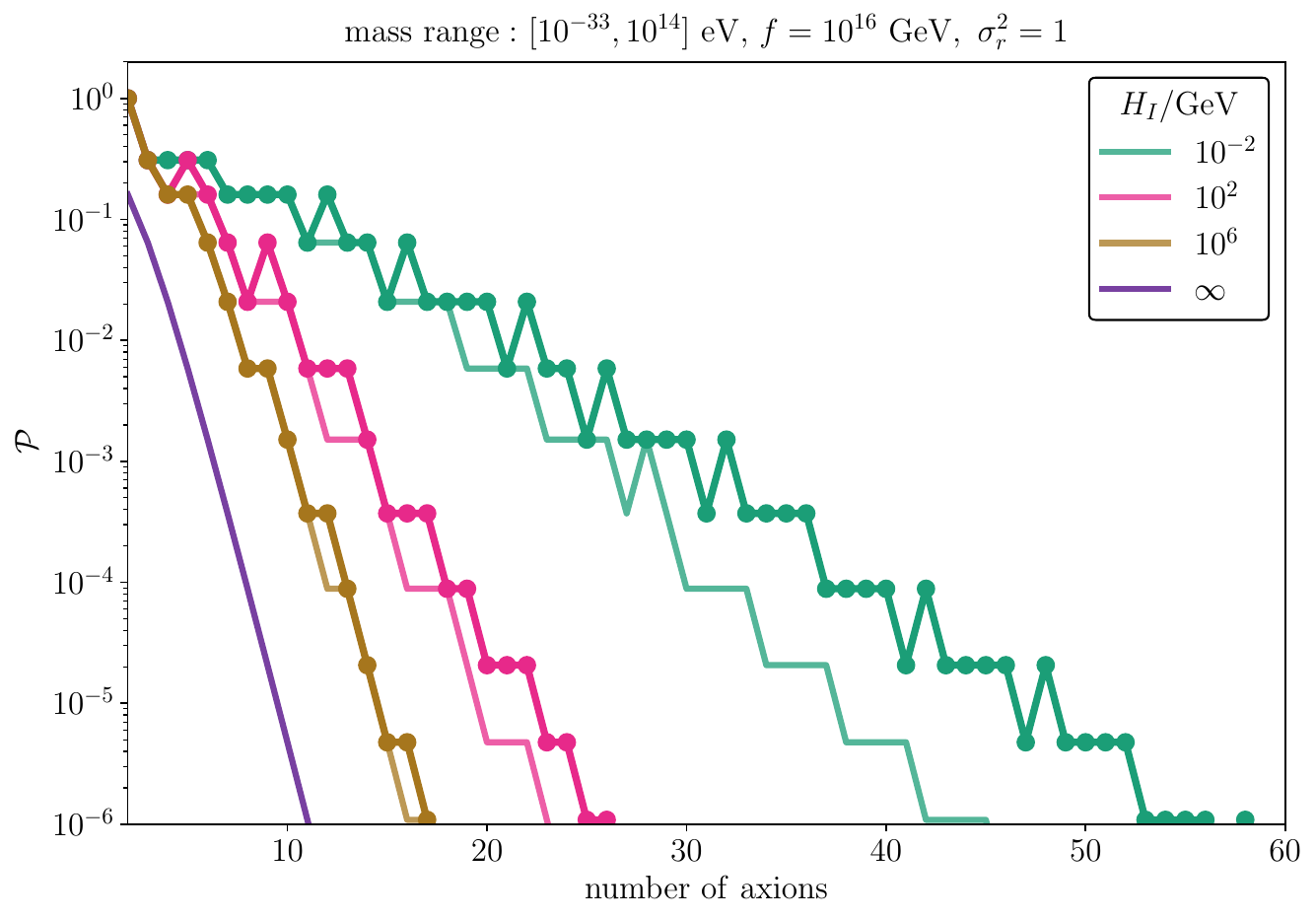}
    \caption{Anthropic probability determined by counting the relevant axions according to the criterion in \cref{eqn:anthropic-condition}. The purple line shows the result of Ref.~\cite{Arvanitaki:2010sy} reproduced in \cref{eqn:anthrop-prob-heavy}, where all axions are heavy enough, or the scale of inflation is high enough, that they satisfy \cref{eqn:anthropic-condition} and thus contribute to the anthropic probability. The other colors show the anthropic probability for different scales of inflation $H_I$. Each point corresponds to a scenario with a number of axions given by the horizontal axis, $\bm K=f^2\bm I$, and isotropic instanton charges drawn as $[\bm Q]_{ij}=[\bm r_i]_j\sim {\rm round}[{\cal N}(0,1)]$. The axion masses are taken to be equally spaced in the interval $10^{-33} \text{ eV} \leq m_i \leq 10^{14}$ eV, regardless of the number of axions; as a consequence, the density of axions per decade will increase as the number of axions does. The maximum number of axions per decade plotted is 1.7, which is still sparse enough for the hierarchical instanton regime to be valid. The un-dotted lines show the anthropic probability when axions are taken to be independent, while the dotted lines show the effect of the suppressed field ranges that arise from interactions. Because of axion mixing, the GS decay constants are suppressed by $\sqrt{N}$, and therefore the dotted lines show an enhancement of the anthropic probability relative to the independent case, growing as the number of axions increases. The non-monotonicity in the gold, pink, and green curves are binning artifacts due to the uniform mass spacing assumed in these examples, and would disappear if averaged over many draws of a mass spectrum distribution.}
    \label{fig:prob-vs-Na}
\end{figure}

\begin{figure}[t!]
    \centering
    \includegraphics[width=\columnwidth]{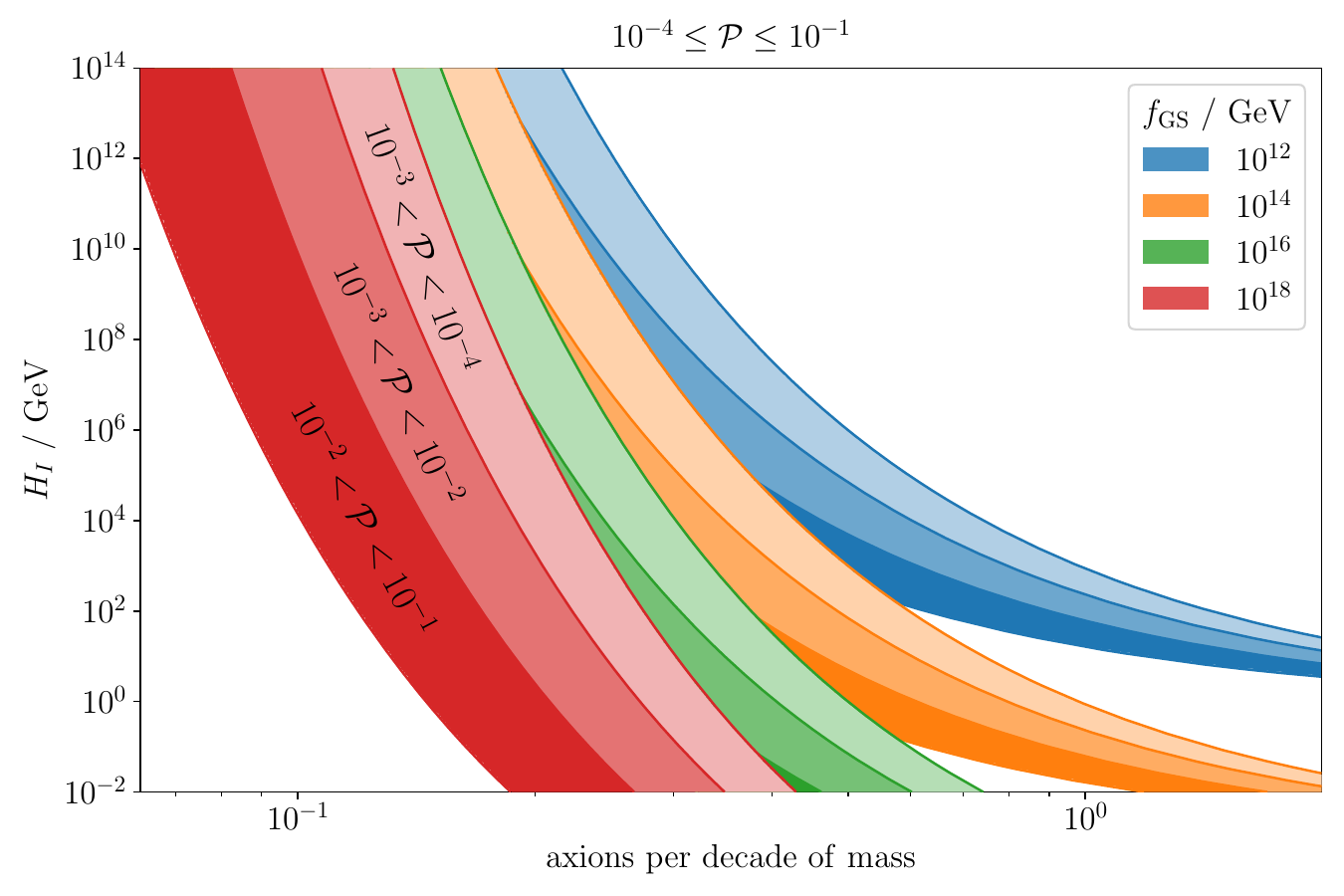}
    \caption{
    Contours of the anthropic probability as a function of the inflationary Hubble scale $H_I$ and the number of axions per decade of mass, for fixed $f_{\rm GS}$. Colored regions show the parameter space with $10^{-4}\leq \mathcal P\leq 10^{-1}$; darker colors indicate larger $\mathcal P$, with contours uniformly spaced in $\log \mathcal P$. As $f_{\rm GS}$ is decreased, larger values of $H_I$ become compatible with a fixed axion density per decade. Although values of $\mathcal P$ near unity can correspond to dynamically fine-tuned initial conditions, such regions are naturally interpreted as anthropically selected when observer formation itself requires the tuning; see the discussion in the text.
    }
    \label{fig:probHI-vs-ApD}
\end{figure}

The scaling of the anthropic probability with the number of relevant axions is illustrated in \cref{fig:prob-vs-Na} for different axiverse scenarios with hierarchical masses, for $N$ between 2 and 60, and for different choices of $H_I$. The mass range of the axions is constant across all scenarios: as the number of axions increases so does the number of axions per decade of mass. The thin solid lines show $\mathcal P$ as a function of $N$, for $N$ independent axions. On the other hand, the thick lines with points show $\mathcal P$ for $N$ axions with mixing, where all axions have a fundamental decay constant $f=10^{16}$ GeV (i.e.\ $\bm K = f^2 \bm I$ in the language of \cref{sec:Statistics}). There are two main results that are illustrated here. The first is that as $N$ increases, the effect of mixing among axions enhances the anthropic probability relative to the case where all axions are independent. This is due to the $\sqrt{N}$ suppression of the GS decay constants. The second is that lowering the scale of inflation also enhances the anthropic probability. In fact, the result of \cite{Arvanitaki:2010sy} shown in purple is the $H_I\rightarrow\infty $ limit of our result. 

\Cref{fig:probHI-vs-ApD} illustrates how the anthropic probability depends on $H_I$ and on the number of axions per decade of mass. The boundaries of the colorful bands show the contours of constant ${\cal P} = 10^{-4},10^{-3},10^{-2},10^{-1}$ for different choices of $f_{\rm GS}$. For high decay constants near the Planck scale, scenarios with ${\cal O}(0.1)$ axions per decade are favored by anthropics. As $f_{\rm GS}$ is taken smaller, scenarios with axions per decade above 1 become anthropically favored as long as the scale of inflation is also taken to be low.
The plot cuts off at $2$ axions per decade of mass, since the hierarchical axion mass assumption breaks down at modestly larger numbers of axions per decade of mass.

The enhancement of the anthropic probability for axiverse scenarios with coupled axions relative to the equivalent independent-axion axiverse scenario is visible in both \cref{fig:prob-vs-Na} and \cref{fig:probHI-vs-ApD}. Recall that the GS decay constants $f_{\text{GS},i}$ are responsible for this enhancement, as they are roughly suppressed by a factor of $\sqrt{N}$ when axions are coupled. We determine an analytic approximation for this relative enhancement by noting that the only dependence on $f_{\text{GS},i}$ of $\mathcal P$ is in $m_{\rm min}$ given in \cref{eqn:m_min}. If $f$ is the fundamental decay constant for an axiverse scenario with independent axions, then the equivalent scenario with coupled axions has $f_{\text{GS},i} \approx f/\sqrt{N}$. Making this substitution into \cref{eqn:m_min}, we express the anthropic probability for coupled axions, $\tilde{\mathcal P}$, as
\begin{equation}
    \tilde{\mathcal P} = \mathcal P \cdot N^{\log{\left(\zeta_{\rm max}/\zeta_{\rm obs}\right)}/{\langle \log (m_{i}/m_{i + 1})\rangle}},
\end{equation}
where $\mathcal P$ is the anthropic probability for an equivalent axiverse scenario but with independent axions.

\section{Results}
\label{sec:results}

\Cref{sec:Statistics,sec:anthropics} highlighted two distinct ways in which multi-axion theories can lead to qualitatively different conclusions from single-axion toy models, even under minimal (though still non-trivial) UV assumptions. In this section we illustrate these effects with Monte Carlo samples of axion parameters drawn in the fundamental basis according to the ensembles of \cref{sec:Statistics}, together with initial conditions appropriate to a long period of inflation as discussed in \cref{sec:anthropics}. Three features play a particularly prominent role: the relative enhancement of the QCD axion coupling, the emergence of an anthropic relic-abundance plateau, and a population of heavy axions which have suppressed abundances at low inflationary scales but whose decays can provide some of the most promising indirect-detection signatures.

This section is organized as follows. In \Cref{subsec:monte_carlo}, we outline the details of our Monte Carlo sampling. In \Cref{subsec:results_relic_abundances}, we summarize the relic-abundance scalings observed in the Monte Carlo sampling. In \Cref{subsec:results_detection}, we combine relic abundances with axion-matter couplings to assess detection prospects, and we give simple criteria for when heavy-axion decays are most relevant as signals or constraints. Finally, in \cref{subsec:measuring_N}, we note that if both the QCD axion and one axionlike particle are measured in the laboratory, they can provide a logarithmic estimate of the size of the axiverse. Taken together, these results provide qualitative intuition for observational expectations.

\begin{figure}
    \centering
    \includegraphics[width=\columnwidth]{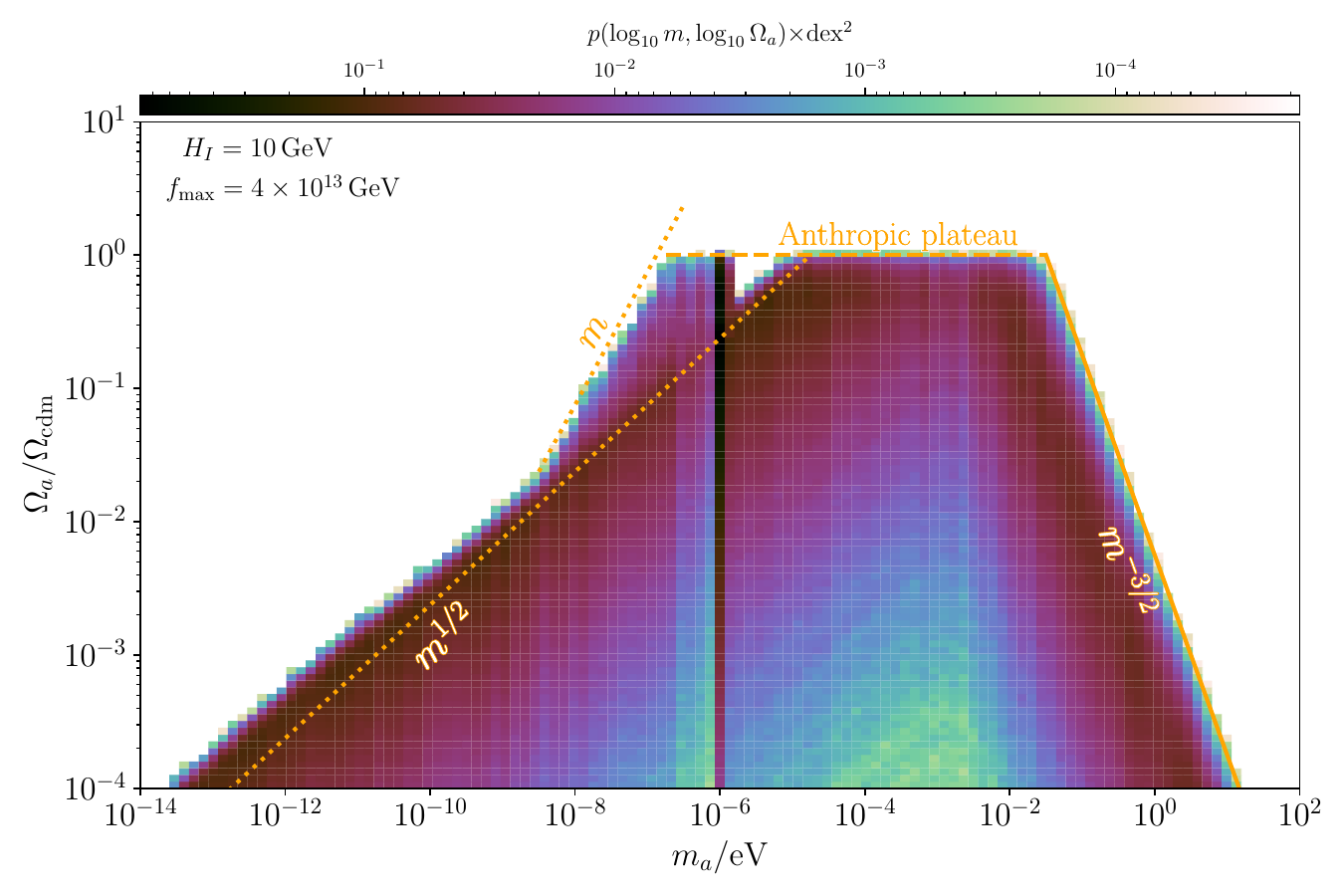}
    \caption{Example probability density $p(\log_{10}m,\log_{10}\Omega_a)$ inferred from a Monte Carlo scan over axion initial conditions, potentials, instanton charges, and kinetic terms, for a UV scale $f_{\rm max}=4\times 10^{13}\,{\rm GeV}$ and inflationary scale $H_I=10\,{\rm GeV}$ (see \cref{subsec:monte_carlo} for details and parameters). Orange lines highlight the expected scaling regimes. At high masses we observe $\Omega_a\propto m^{-3/2}$, consistent with suppressed misalignment when $\Lambda_i\gtrsim H_I$. Around $m\simeq10^{-1}\,{\rm eV}$ the ``anthropic plateau'' [\cref{eqn:anthropic_plateau}] is reached, where abundances are capped at $\Omega_a=\Omega_{\rm cdm}$; the probability density dips slightly near the plateau center because anthropically selected spectra tend to avoid axions that are likely to be overabundant. The anthropic probability is measured to be $ \mathcal{P} \sim 35$\% in this ensemble. At lower masses the light-axion scaling $\Omega_a\propto m^{1/2}$ emerges, with a mild upward deviation due to the slow growth of the GS decay constants toward lighter modes. Since the kinetic eigenvalue distribution is chosen to have only a mild hierarchy, the deviation from the $m^{1/2}$ power-law is also mild, though stronger behaviors are possible in general (see \cref{subsubsec:GS_relic_abundance_scaling}). This scaling is interrupted near the QCD axion, whose abundance is enhanced by delayed oscillation onset by a factor $(m_{\rm QCD}/H_{\rm osc})^{1/2}$, with $m_{\rm QCD}\sim10^{-6}\,{\rm eV}$ and $H_{\rm osc}\sim10^{-8}\,{\rm eV}$. Axions in the range $10^{-8}$--$10^{-6}\,{\rm eV}$ that undergo adiabatic level crossing with the QCD axion inherit energy density, scaling roughly as $\sqrt{m_i/m_{i+1}}$ relative to their lighter neighbor. Together with the $m^{1/2}$ trend, this produces an additional $m$-dependence for modes that cross the QCD axion, and extends the anthropic plateau to lower masses.}
    \label{fig:relic_abundance}
\end{figure}

\begin{figure}
    \centering
    \includegraphics[width=\textwidth]{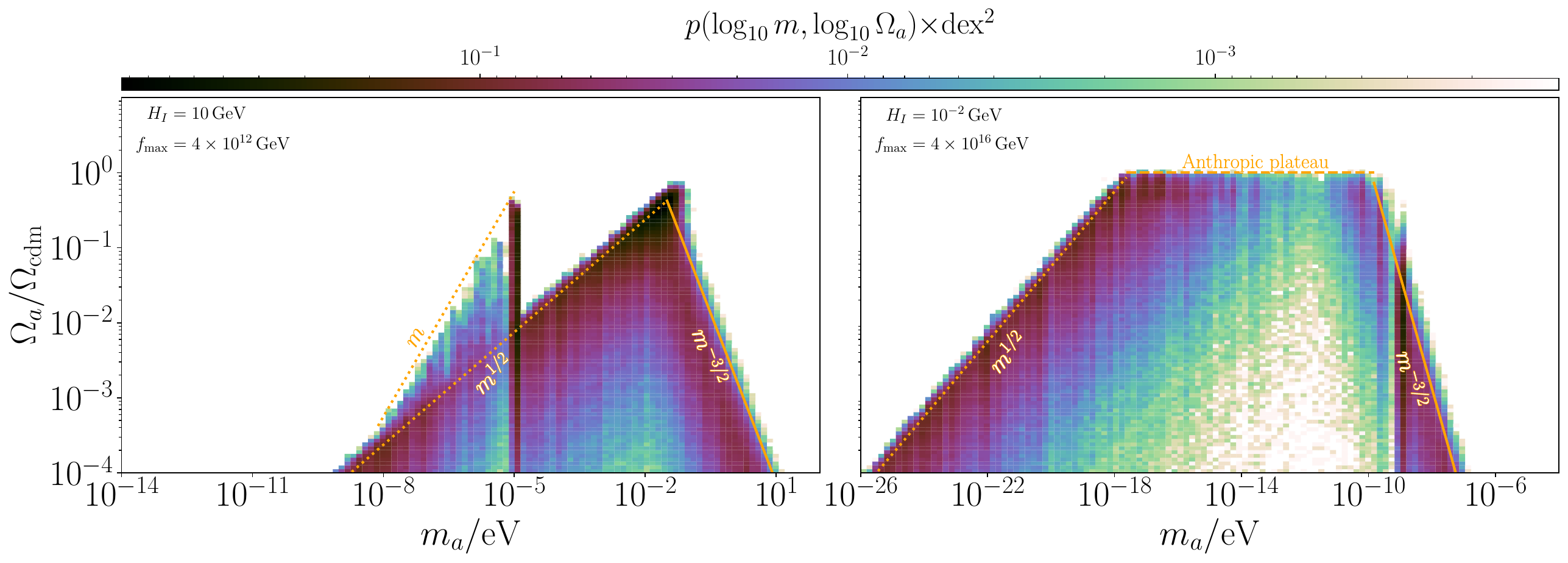}
    \caption{Example probability density $p(\log_{10}m,\log_{10}\Omega_a)$ inferred from the Monte Carlo scan described in \cref{subsec:monte_carlo} as in  \cref{fig:relic_abundance},for different values of the inflationary scale $H_I$ and maximum decay constant $f_{\rm max}$. (Left) $H_I = 10\,{\rm GeV}$ and $f_{\rm max} = 4\times 10^{12}\,{\rm GeV}$, yielding an anthropic probability ${\cal P}=99\%$. The smaller decay constants suppress the relic abundance, so only rare realizations exceed the observed dark matter density. As a result, the anthropic plateau is absent. (Right) $H_I = 10^{-2}\,{\rm GeV}$ and $f_{\rm max} = 4\times 10^{16}\,{\rm GeV}$, yielding ${\cal P}=25\%$. The large decay constants enhance the relic abundance, so a low inflationary scale is required to obtain even a modest anthropic probability. In this case the anthropic plateau is especially prominent. Here again, the population of axions is sparser near the plateau center, around $m\sim 10^{-12}\,{\rm eV}$, because anthropically selected spectra tend to avoid axions that are likely to be overabundant. Note that the two panels span very different axion mass ranges.}
    \label{fig:relic_abundance_extra}
\end{figure}

\subsection{Monte Carlo}
\label{subsec:monte_carlo}

We perform three Monte Carlo scans over axion initial conditions and over the parameters of the axion effective theory in the fundamental basis. Parameters of each scan are identical unless otherwise noted. We take an ensemble of $N=200$ axions with hierarchical instanton scales $\Lambda_i^4 = M_{\rm pl}^4 e^{-S_i}$, where the instanton actions are drawn independently from a uniform distribution $S_i\sim{\cal U}(0,614)$. We choose three benchmark values of the largest kinetic eigenvalue, $f_{\rm max}={4\times 10^{12} \ {\rm GeV}, \  4\times 10^{13} \ {\rm GeV}, \ 4\times 10^{16} \ {\rm GeV}}$, and draw the remaining kinetic eigenvalues from a broad distribution $(f_i/f_{\rm max})^2 \sim {\cal U}(0,1)$. The first two choices of $f_{\rm max}$ place the QCD axion in the classic axion dark matter window, with $m_{i_c}\approx 10^{-6} \ {\rm eV}$ and $m_{i_c}\approx 10^{-5} \ {\rm eV}$ respectively, while the third is representative of GUT-scale axions. We choose integer instanton charges by rounding unit-variance Gaussian draws, $[\bm r_i]_j \sim {\rm round}[{\cal N}(0,1)]$, and include a QCD sector so that one linear combination of axions plays the role of the QCD axion. For each of these three benchmark ensembles, we then choose the inflationary scale $H_I={10 \ {\rm GeV}, \ 10 \ {\rm GeV}, \ 10^{-2} \ {\rm GeV}}$ when drawing initial conditions from the Fokker-Planck equilibrium distribution \cref{eqn:inflationtheta0-dist}. These values are chosen to give reasonable anthropic probability for the corresponding choices of $f_{\rm max}$.

When imposing our simple anthropic selection, we find that a non-negligible fraction of realizations populate the observed range of dark matter to baryon ratios. We initially draw 3000 sets of axion parameters in the fundamental basis and 3000 sets of initial conditions for each of those parameter sets. For the parameter set $(f_{\rm max},H_I) = (4\times 10^{13}\,{\rm GeV},10\,{\rm GeV})$, among $7{,}792{,}179$ samples with total dark matter abundance below $100\times\Omega_b$, we find $1{,}061{,}501$ realizations (approximately $14\%$) satisfying $2.5 \Omega_b\le \Omega_{\rm cdm} \le 5 \Omega_b$, 
and we use this accepted subset to construct the relic-abundance histograms shown in \cref{fig:relic_abundance,fig:relic_signatures}. Weighted appropriately, the corresponding anthropic probability is ${\cal P}\approx 35\%$. Similarly, in \cref{fig:relic_abundance_extra,fig:relic_signatures_extra}, the left panels correspond to $(f_{\rm max},H_I) = (4\times 10^{12}\,{\rm GeV},10\,{\rm GeV})$, and among $169{,}853$ samples with total dark matter abundance below $100\times\Omega_b$, we find $168{,}369$ realizations (approximately $99\%$, and similarly for the anthropic probability ${\cal P}\approx 99\%$) satisfying $2.5 \Omega_b\le \Omega_{\rm cdm} \le 5 \Omega_b$.
The right panels correspond to $(f_{\rm max},H_I) = (4\times 10^{16}\,{\rm GeV},10^{-2}\,{\rm GeV})$, and among $1{,}579{,}083$ samples with total dark matter abundance below $100\times\Omega_b$, we find $88{,}313$ realizations (approximately $5.6\%$, corresponding to an anthropic probability ${\cal P}\approx25\%$) satisfying $2.5 \Omega_b\le \Omega_{\rm cdm} \le 5 \Omega_b$. 

\subsection{Relic abundances}
\label{subsec:results_relic_abundances}

The distributions of relic abundances obtained from the Monte Carlo sampling described in \cref{subsec:monte_carlo} are shown in \cref{fig:relic_abundance,fig:relic_abundance_extra}. Several distinct scaling regimes are evident. In this section we give simple analytic estimates for the envelope (often also the typical locus) of the distribution in each regime.

We begin at low masses, where inflationary fluctuations dominate over the axion potential and the  misalignment angles are approximately uniformly distributed. A sufficient condition is that the potential be subdominant during inflation,
\begin{align}
    \Lambda_i^4 \ll \frac{3H_I^4}{8\pi^2}\,,
\end{align}
so that the stationary distribution \cref{eqn:inflationtheta0-dist} is nearly flat. In this regime the relic abundance follows the usual misalignment scaling [up to an $\mathcal O(1)$ factor],
\begin{align}
\label{eqn:results_relic_low_mass}
    \Omega_i  \sim\ \left(\frac{m_i}{H_{\rm eq}}\right)^{1/2}\left(\frac{f_{{\rm GS},i}}{M_{\rm pl}}\right)^2\,.
\end{align}
We indicate the naive $\Omega\propto m^{1/2}$ behavior with the orange dotted line on the left of \cref{fig:relic_abundance}. However, it is not a perfect fit because $f_{{\rm GS},i}$ is itself a slowly varying function of mass rank: as discussed in \cref{subsec:field-ranges}, lighter modes typically have larger GS field ranges. This mild increase of $f_{{\rm GS},i}$ toward smaller masses shifts the typical relic abundance upward relative to the superficial $m^{1/2}$ expectation. In cases where the kinetic eigenvalues are more widely distributed, more extreme deviations from the $m^{1/2}$ power law are possible (see \cref{subsubsec:GS_relic_abundance_scaling}).

At masses above $\sim 10^{-8}\,\mathrm{eV}$, the envelope steepens from the naive $\Omega\propto m^{1/2}$ behavior toward an approximately linear scaling. This reflects mixing with the QCD axion: as the QCD potential turns on during the QCD crossover transition, the relevant mass eigenstates acquire a growing mass. In the adiabatic limit, the growing mass  enhances the axions' relic abundances~\cite{Li:2025cep,Daido:2015cba}.

To isolate the basic effect, consider first the QCD axion in isolation after the onset of oscillations about the minimum of its potential, when the comoving axion number density is approximately conserved. Let $m(a)$ denote the temperature-dependent axion mass and $m_0\equiv m(T=0)$ its vacuum value, and define $a_{\rm osc}$ by the condition $m(a_{\rm osc}) \approx 3H(a_{\rm osc})$. At the onset of oscillations, the number density is parametrically $n(a_{\rm osc}) \sim m(a_{\rm osc})f_a^2$, and number conservation implies
\begin{align}
    n(a) \approx m(a_{\rm osc})f_a^2\left(a_{\rm osc}/a\right)^3\,,\qquad a>a_{\rm osc}\,.
\end{align}
Now compare to a hypothetical axion with the same vacuum parameters $(m_0,f_a)$ but with a temperature-independent mass $m(a)\equiv m_0$. Oscillations would begin earlier at $a_{\rm osc}^{(0)}$ defined by $m_0\simeq 3H(a_{\rm osc}^{(0)})$, and
\begin{align}
    n^{(0)}(a)\sim m_0 f_a^2\left(a_{\rm osc}^{(0)}/a\right)^3\,,\qquad a > a_{\rm osc}^{(0)}\,.
\end{align}
Assuming the Universe is radiation dominated, the ratio of number densities at late times, once the QCD axion mass has stopped growing, gives
\begin{align}
    \frac{n(a)}{n^{(0)}(a)}\sim
    \frac{m(a_{\rm osc})}{m_0}\left(\frac{a_{\rm osc}}{a_{\rm osc}^{(0)}}\right)^3
    =\left(\frac{m_0}{m(a_{\rm osc})}\right)^{1/2}.
\end{align}
Thus, relative to the hypothetical constant-mass model, the QCD axion's relic abundance is enhanced by a factor $\sim \sqrt{m_0/m(a_{\rm osc})}$: the smaller mass at the onset of oscillations delays the onset of dilution due to Hubble friction, and this effect outweighs the smaller initial number density.

Moving now to the axiverse, multiple axions can have appreciable overlap with the QCD direction, depending on the sparsity of the fundamental instanton charge vectors. When the QCD topological susceptibility turns on, this overlap induces time-dependent mixing: the instantaneous mass eigenstates undergo a sequence of avoided level crossings in which one eigenvalue grows while the adjacent one remains constant~\cite{Li:2025cep}. In the adiabatic limit,\footnote{Ref.~\cite{Murai:2024nsp} finds that adiabatic conversion at the level crossing is favored when the zero-temperature QCD axion is heavier than the ALP and has a smaller decay constant. Since the GS construction typically assigns heavier modes slightly smaller effective decay constants, this condition may plausibly hold, though a numerical study would be needed if the decay constants are close to the adiabatic threshold.} the comoving occupation number of each instantaneous eigenstate is conserved, so whenever an eigenvalue grows from an initial value $m_{\rm in}$ to a final value $m_{\rm out}$, the relic abundance carried by that mode is enhanced by a factor $\sim \sqrt{m_{\rm out}/m_{\rm in}}$.

In particular, across an isolated avoided crossing between neighboring modes in the hierarchy, the net growth is set parametrically by the ratio of adjacent masses. Denoting by $n_i^{(0)}$ the relic number density the $i$th mode would have obtained in the absence of QCD-induced mixing (i.e.\ if its mass were held fixed), one finds
\begin{align}
    \frac{n_i(a)}{n_i^{(0)}(a)} \sim \left(\frac{m_{\rm out}}{m_{\rm in}}\right)^{1/2}
    \sim \left(\frac{m_{i}}{m_{i+1}}\right)^{1/2}\,,
\end{align}
where in the last step we have used that, in a hierarchical spectrum, the relevant avoided crossing typically interpolates between adjacent eigenvalues.

Returning to the relic abundance in \cref{fig:relic_abundance}, the net effect of QCD-induced mixing is to enhance the abundances of modes whose vacuum masses lie between the QCD axion mass at the onset of oscillations and the QCD axion’s zero-temperature mass, i.e.
\begin{align}
    m_{\rm QCD}(a_{\rm osc}) \lesssim\ m_i  \le m_{i_{c}} \equiv m_{\rm QCD}(T=0)\,.
\end{align}
For the benchmark parameters used in \cref{fig:relic_abundance}, this range corresponds roughly to $m_i\in[10^{-8}\,\mathrm{eV},\,10^{-6}\,\mathrm{eV}]$. The detailed enhancement depends on the density of the mass spectrum near the QCD scale: in a very dense spectrum one typically has $m_i/m_{i+1}\approx 1$, so each avoided crossing produces only a mild boost. The envelope, however, is set by the largest enhancement of $a_i$ compatible with the available hierarchy, which is approached when the level mixing is solely between $a_i$ and the QCD axion (i.e.\ there are no intermediate crossings). In that maximal case, one may estimate the relic abundance by multiplying the usual misalignment scaling by the adiabatic growth factor $\sim \sqrt{m_i/m_{\rm QCD}(a_{\rm osc})}$, giving
\begin{align}
\label{eqn:results_relic_QCD_mix}
    \Omega_i \ \sim\ 
    \left (\frac{m_i}{m_{\rm QCD}(a_{\rm osc})} \frac{m_i} {H_{\rm eq}}\right)^{1/2}
    \left(\frac{f_{{\rm GS},i}}{M_{\rm pl}}\right)^2\,,
    \qquad
    m_{\rm QCD}(a_{\rm osc}) \lesssim m_i \leq m_{i_{c}}\,.
\end{align}

Finally, we emphasize that we do not directly integrate the axion equations of motion in the evolving QCD potential, and therefore do not capture potential non-adiabatic effects in \cref{fig:relic_abundance,fig:relic_abundance_extra,fig:relic_signatures,fig:relic_signatures_extra}. For detailed treatments of level crossing across different regimes, see e.g.\ Refs.~\cite{Daido:2015bva,Daido:2015cba,Li:2024kdy,Li:2024okl,Li:2025cep,Cyncynates:2023esj,Kitajima:2014xla,Murai:2023xjn,Murai:2024nsp}.

For the benchmark parameters used here, the linear envelope from QCD-induced mixing does not persist up to the vacuum QCD axion mass. Instead, it is truncated by the anthropic plateau: once \cref{eqn:results_relic_QCD_mix} reaches $\Omega_i\sim \mathcal O(1)$, larger abundances are removed by anthropic selection, and the accepted realizations accumulate near $\Omega_i\simeq 1$.

Above the QCD axion mass, the envelope momentarily returns to the nominal $\Omega_i\propto m_i^{1/2}$ behavior, but it again terminates at the anthropic plateau. Parametrically, the plateau occupies the band
\begin{align}\label{eqn:anthropic_plateau}
    \Omega_i \approx 1\,,
    \qquad
    \left(\frac{M_{\rm pl}}{f_{{\rm GS},i}}\right)^{4}\lesssim
    \frac{m_i}{H_{\rm eq}}\lesssim \left(\frac{H_I^2}{H_{\rm eq}M_{\rm pl}}\right)^{4/3}\,.
\end{align}
The lower boundary is set by the misalignment estimate \cref{eqn:results_relic_low_mass}: demanding $\Omega_i\sim 1$ with $\theta_{0,i}=\mathcal O(1)$ gives $m_i/H_{\rm eq}\sim (M_{\rm pl}/f_{{\rm GS},i})^{4}$. The upper boundary is set by the suppression of initial misalignments at low inflationary scale. In this regime the typical initial displacement is bounded by inflationary fluctuations,
\begin{align}
    \theta_{0,i}\sim \frac{H_I^2}{m_i f_{{\rm GS},i}}\,,
\end{align}
so that the relic abundance scales as
\begin{align}
    \Omega_i\sim\left(\frac{m_i}{H_{\rm eq}}\right)^{1/2}\left(\frac{f_{{\rm GS},i}}{M_{\rm pl}}\right)^2\theta_{0,i}^2\sim\left(\frac{m_i}{H_{\rm eq}}\right)^{1/2}\frac{H_I^4}{m_i^2M_{\rm pl}^2}\propto m_i^{-3/2}\,.
\end{align}
Setting $\Omega_i\sim 1$ in this expression yields the characteristic cutoff scale for the anthropic plateau,
\begin{align}
    m_i \sim H_{\rm eq}\left(\frac{H_I^2}{H_{\rm eq}M_{\rm pl}}\right)^{4/3},
\end{align}
beyond which the abundances fall rapidly with increasing mass.

Finally, \cref{fig:relic_abundance}, and to a greater extent the right panel of \cref{fig:relic_abundance_extra}, shows depletion of axion density near the center of the anthropic plateau. This feature is a consequence of how the ensemble is generated: we scan not only over initial misalignments, but also over axion parameters. In the absence of any anthropic selection, the middle of the plateau corresponds to regions of parameter space where the axion relic abundance would be comparatively large, and therefore those parameters are more likely to overshoot the observed dark matter density. As a result, among the realizations that satisfy the anthropic cut, the most typical parameter draws tend to avoid masses that would otherwise populate the center of the plateau.

For the benchmark shown in \cref{fig:relic_abundance}, the fraction of Monte Carlo realizations that produce anthropically viable dark matter abundances is $\sim 14\%$ (while ${\cal P}\approx 35\%$), so the resulting depletion is mild. As the anthropic probability decreases, as in the benchmark in the right-hand panel of \cref{fig:relic_abundance_extra} where this fraction is $\sim 5.6\%$ (${\cal P}\approx25\%$), the accepted subset becomes increasingly dominated by rare realizations with atypical mass distributions---in particular, spectra that are unusually sparse near the plateau region. Thus, while modes that lie on the plateau generically have $\mathcal O(1)$ relic abundance, they also tend to be statistically rarer within the underlying parameter ensemble.

\subsubsection*{Freeze-in}
Axion couplings to Standard Model operators inevitably lead to some thermal production from the bath. This typically occurs through UV-sensitive freeze-in processes whose yield grows with the maximum temperature of the radiation era, and can therefore constrain cosmologies with high reheating temperature~\cite{Salvio:2013iaa,Elahi:2014fsa,Baumann:2016wac,Bernal:2019mhf}.For the low inflationary scale adopted in our benchmarks, $H_I=10\,\mathrm{GeV}$ and $10^{-2}\,{\rm GeV}$, we find that existing freeze-in bounds do not constrain the parameter space relevant for \cref{fig:relic_abundance,fig:relic_abundance_extra,fig:relic_signatures,fig:relic_signatures_extra}. A detailed effective-field-theory treatment of axiverse freeze-in and the associated constraints from contributions to $N_{\rm eff}$ can be found in Ref.~\cite{Dessert:2025yvk}.

\subsection{Relic detection prospects}
\label{subsec:results_detection}

\begin{figure}
    \centering
    \includegraphics[width=\columnwidth]{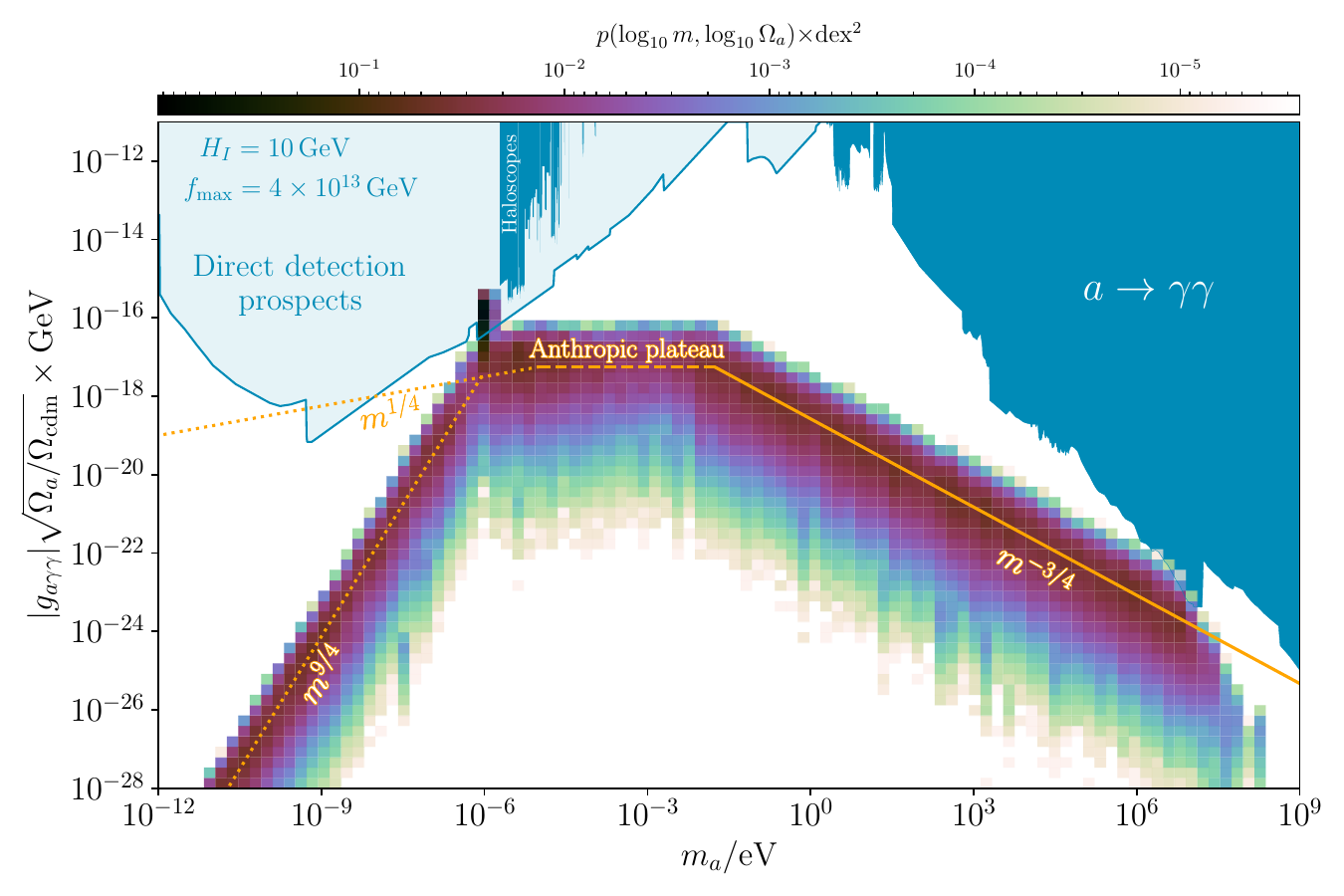}
    \caption{Detection prospects for the example ensemble shown in \cref{fig:relic_abundance} (see \cref{subsec:monte_carlo} for details). We assume the axion-photon coupling is entirely induced by the axions' couplings to QCD, and take the photon coupling in the fundamental basis to be related to the gluon coupling by $g_{a\gamma\gamma}=g_{agg}\,\alpha_{\rm EM}/2\pi$. The vertical axis shows the abundance-rescaled coupling $g_{a\gamma\gamma}\sqrt{\Omega_a/\Omega_{\rm cdm}}$. Indirect searches for decays $a\to\gamma\gamma$~\cite{Porras-Bedmar:2024uql,Liu:2023nct,Capozzi:2023xie,Cyr:2024sbd,Fong:2024qeq,Dekker:2021bos,Sun:2023acy,Nakayama:2022jza,Carenza:2023qxh,Todarello:2024qci,Calore:2022pks,Janish:2023kvi,Saha:2025any,Wadekar:2021qae,Wang:2023imi,Todarello:2023hdk,Thorpe-Morgan:2020rwc,Yin:2024lla,Foster:2021ngm,Yin:2025xad,Dessert:2023vyl} appear at high masses (solid blue), while existing~\cite{Ouellet:2018beu,Salemi:2021gck,Pandey:2024dcd,ADMX:2009iij,ADMX:2018gho,ADMX:2019uok,ADMX:2021nhd,ADMX:2024xbv,ADMX:2025vom,ADMX:2018ogs,ADMX:2021mio,Crisosto:2019fcj,Devlin:2021fpq,Lee:2020cfj,Jeong:2020cwz,CAPP:2020utb,Lee:2022mnc,Yoon:2022gzp,Kim:2022hmg,Yi:2022fmn,Yang:2023yry,Kim:2023vpo,CAPP:2024dtx,Bae:2024kmy,Adair:2022rtw,Oshima:2023csb,Nishizawa:2025xka,GigaBREAD:2025lzq,Grenet:2021vbb,Brubaker:2016ktl,HAYSTAC:2018rwy,HAYSTAC:2020kwv,HAYSTAC:2023cam,HAYSTAC:2024jch,Heinze:2023nfb,MADMAX:2024sxs,McAllister:2017lkb,Quiskamp:2022pks,Quiskamp:2023ehr,Quiskamp:2024oet,Alesini:2019ajt,Alesini:2020vny,Alesini:2022lnp,QUAX:2023gop,QUAX:2024fut,CAST:2020rlf,Ahyoune:2024klt,DePanfilis:1987dk,Wuensch:1989sa,Gramolin:2020ict,Sulai:2023zqw,Arza:2021ekq,Friel:2024shg,TASEH:2022vvu,Hagmann:1990tj,Hagmann:1996qd,Thomson:2021zvq,Thomson:2023moc} (projected ~\cite{Liu:2018icu,Stern:2016bbw,Nagano:2019rbw,Lawson:2019brd,Ahyoune:2023gfw,BREAD:2021tpx,Aja:2022csb,DeMiguel:2023nmz,Obata:2018vvr,Heinze:2024bdc,DMRadio:2022pkf,Fan:2024mhm,Alesini:2023qed,Baryakhtar:2018doz,Beurthey:2020yuq,Berlin:2020vrk,Qiu:2025bbi,Bourhill:2022alm,Zhang:2021bpa,Koppell:2025dmt}) direct detection searches appear at lower masses in solid (transparent) blue. Because the signal scales as $\sqrt{\Omega_a}$, the power-law trends (orange lines) mirror those in \cref{fig:relic_abundance}, with exponents reduced by a factor of two. At the highest masses the spectrum is cut off, since axions that decay well before the present epoch do not yield an observable abundance today. The QCD axion is particularly notable: its direct-detection reach is enhanced by $\sqrt{N}$ relative to typical axions in the ensemble (see \cref{subsubsec:QCD_axion_coupling}). Since the photon interaction descends from the QCD coupling, the low-mass tail has an overall $m^{9/4}$ scaling [\cref{eqn:light_axion_QCD_coupling}]. If the photon couplings are instead independent of the QCD instanton charge vector (though see also e.g.\ Ref.~\cite{Reig:2025dqb}), direct-detection prospects scale as $m^{1/4}$ [\cref{eqn:results_relic_low_mass}]. Overall, the most promising targets are the QCD axion and heavy axions with lifetimes comparable to the age of the Universe. Note that constraints from stellar, supernova and black hole production that do not depend on the relic abundance do not appear on these axes; see text for discussion.
    }
    \label{fig:relic_signatures}
\end{figure}

\begin{figure}
    \centering
    \includegraphics[width=\textwidth]{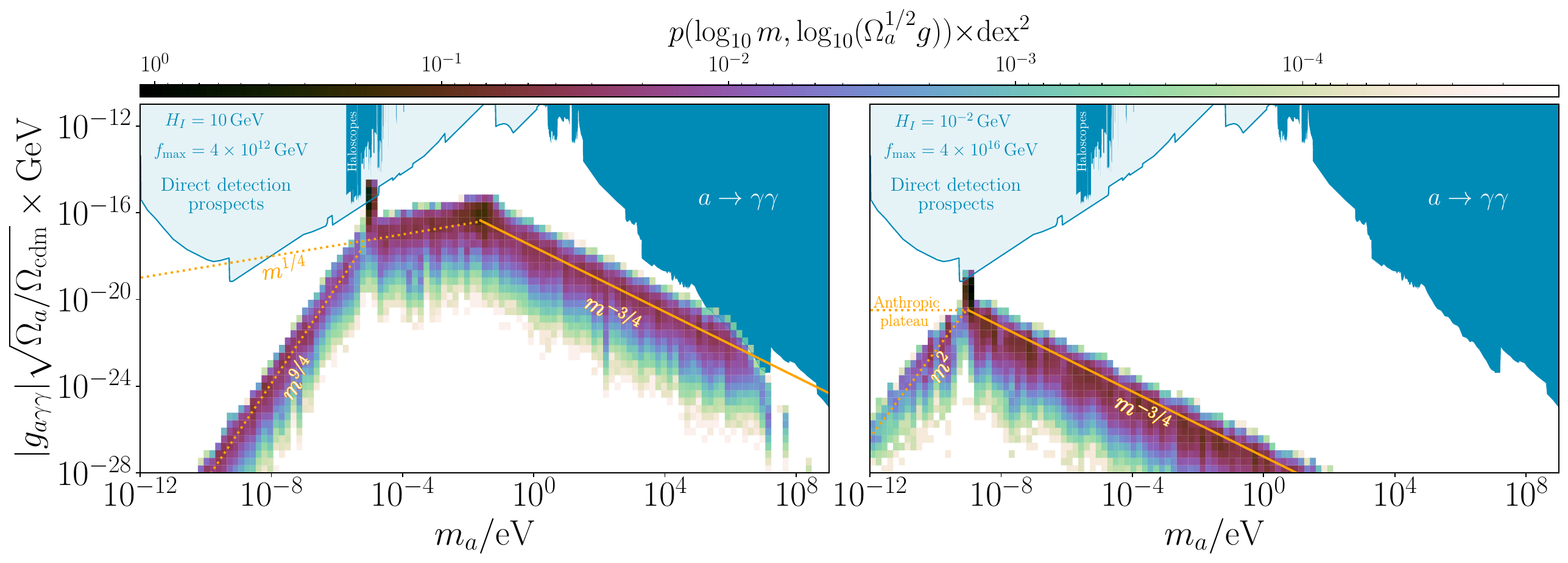}
    \caption{Detection prospects for the same benchmark ensembles shown in \cref{fig:relic_abundance_extra}, under the same assumptions as in \cref{fig:relic_signatures}. (Left) For $H_I = 10\,{\rm GeV}$ and $f_{\rm max} = 4\times 10^{12}\,{\rm GeV}$, the qualitative features are similar to those in \cref{fig:relic_signatures}, but the absence of an anthropic plateau removes the corresponding flattened region in parameter space. (Right) For $H_I = 10^{-2}\,{\rm GeV}$ and $f_{\rm max} = 4\times 10^{16}\,{\rm GeV}$, the same overall structure again appears, but now the anthropic plateau extends to masses below the QCD axion. As a result, the low-mass scaling is modified from $m^{9/4}$ to $m^2$.}
    \label{fig:relic_signatures_extra}
\end{figure}

\Cref{fig:relic_signatures,fig:relic_signatures_extra} summarize the direct and indirect detection targets for relic searches populated by our Monte Carlo ensemble, focusing on the axion-photon coupling. The vertical axis is the abundance-scaled photon coupling $g_{ a\gamma\gamma}\sqrt{\Omega_a/\Omega_{\rm cdm}}$, rather than $g_{a\gamma\gamma}$ itself. This choice is convenient because many searches, including haloscope-like direct detection and line searches for $a\to\gamma\gamma$ decays, have their reach set by the signal power/flux $\propto g_{a\gamma\gamma}^2\rho_a$. On these axes, subdominant components $(\Omega_a\ll \Omega_{\rm cdm})$ can be shown on the same footing as the dominant dark matter component. The populated region exhibits several distinct structures with clear origins. The first is the QCD axion, which stands out as the peak [near $10^{-6}\,{\rm eV}$ in \cref{fig:relic_signatures} and $10^{-5}\,{\rm eV}$ (left) and $10^{-9}\,{\rm eV}$ (right) in \cref{fig:relic_signatures_extra}] due to its unsuppressed coupling to matter. The second is the anthropic plateau, populated by axions that make up $O(1)$ of the dark matter abundance. The third is the heavy tail populated by axions at higher masses whose misalignment angles are inflation-suppressed. Finally, the fourth is the low-mass tail, whose slope depends on the origin of the photon coupling. In general, other realizations of this axiverse scenario will yield similar features, although details differ. We discuss each feature in detail below, and highlight their most promising experimental prospects in the context of the axiverse realizations depicted in \cref{fig:relic_signatures,fig:relic_signatures_extra}. 

The QCD axion stands out as a prominent feature above the bulk of the ensemble because its photon coupling is inherited from the same direction in field-space $\bm r_{i_{c}}$ that appears in the QCD potential.\footnote{Note that part of the reason the QCD axion is so heavily populated relative to the other points in the Monte Carlo ensemble is because we fix the scale of the QCD instanton, so its mass is overrepresented.} Consequently, it does not incur the typical $\sqrt{N}$ suppression of matter couplings found for generic axionlike particles in a large axiverse and falls approximately along the expected QCD axion band (see \cref{subsubsec:QCD_axion_coupling}). Away from the QCD direction, generic axions inherit both the sorting of the field ranges by the GS procedure, and the corresponding $N$-dimensional suppression of the couplings, so most of the parameter space lies parametrically below the QCD axion. Ongoing and proposed direct detection experiments (light blue shaded region in \cref{fig:relic_signatures,fig:relic_signatures_extra}) should be able to probe the QCD axion in these realizations of the axiverse.

The anthropic plateau visible in \Cref{fig:relic_abundance} maps into a corresponding band in \Cref{fig:relic_signatures}. Although the plateau contains most of the dark matter abundance, it is not always optimal as a direct detection target because it typically lies to the right of the QCD axion, where the product of abundance and coupling is relatively suppressed. Nevertheless, observing such a plateau would be particularly striking, enabling inference of the number of axions in the ensemble by comparing the signal strength to that of the QCD axion. In addition, if axionlike particles couple to matter independently of the QCD instanton direction (as, for example, in non-GUT theories~\cite{Agrawal:2022lsp}), observing axions lighter than the QCD axion could likewise yield an estimate of $N$. We discuss the statistics of inferring $N$ in both of these manners in \cref{subsec:measuring_N}.

At high masses, the ensemble develops a heavy tail populated by axions with inflation-suppressed misalignment angles. Using the long inflation prior for $\Lambda_i\gg H_I$ (see \cref{sec:anthropics} and \cref{eqn:pdf-sigma}), one finds a characteristic scaling
\begin{align}
    \Omega_{\rm heavy}\sim \frac{3H_I^4}{8\pi^2 M_{\rm pl}^2 H_{\rm eq}^{1/2}m^{3/2}}\,,
\end{align}
up to ${\cal O}(1)$ factors. Combining this with $g_{a\gamma\gamma}\sim\alpha_{\rm EM}/(2\pi \sqrt{N}f_{\rm GS})$, [ i.e. setting the ${\cal O}(1)$ anomaly coefficients to 1] gives
\begin{align}\label{eqn:heavy_tail}
    g_{a\gamma\gamma}\sqrt{\frac{\Omega_{\rm heavy}}{\Omega_{\rm cdm}}}\sim \frac{\alpha_{\rm EM}}{2\pi \sqrt{N}f_{\rm GS}}\sqrt{\frac{3}{8\pi^2}}\frac{H_I^2}{(m^3 H_{\rm eq})^{1/4} M_{\rm pl}}\,,
\end{align}
which approximates the envelope of the heavy axions in \cref{fig:relic_signatures}.

At sufficiently large masses this population is cut off by decays. For the photon coupling above, the lifetime is
\begin{align}
\label{eqn:decay-rate}
    \Gamma_{a\to\gamma\gamma} = \frac{g_{a\gamma\gamma}^2 m^3}{64\pi}\approx \frac{\alpha_{\rm EM}^2 m^3}{256\pi^3 N f_{\rm GS}^2}\,,
\end{align}
(again up to an ${\cal O}(1)$ model-dependent anomaly coefficient). Setting $\Gamma_{a\to\gamma\gamma} \approx H_0$ yields the characteristic decay cutoff
\begin{align}
    m_{\rm decay} \sim 4\pi \left(\frac{4 NH_0 f_{\rm GS}^2}{\alpha_{\rm EM}^2}\right)^{1/3}\,.
\end{align}
Thus, for fixed $f_{\rm GS}$ the heavy tail terminates near $m_{\rm decay}$, and only axions with lifetime close to the age of the Universe can produce an observable flux today. Note that $m_{\rm decay}$ assumes decays occur today; earlier-epoch probes such as the CMB are sensitive to heavier axion masses~\cite{Capozzi:2023xie,Liu:2023nct,Sun:2023acy,Cyr:2024sbd,Porras-Bedmar:2024uql}, though the same qualitative conclusions we derive below hold.

We now compare the heavy-tail locus \cref{eqn:heavy_tail} to the decay constraints shown in solid blue on the right of \cref{fig:relic_signatures}. Constraints on the lifetime $\tau_{a\rightarrow\gamma\gamma}= \Gamma^{-1}_{a\rightarrow\gamma\gamma}$, such as those from heating of the Leo T dwarf galaxy \cite{Wadekar:2021qae} and from spectral distortions of the CMB \cite{Liu:2023nct}, scale as $m^{-3/2}$, according to \cref{eqn:decay-rate}. For masses $m>$ keV, decay constraints are derived from the non-observation of X-ray and $\gamma$-ray photons by telescopes such as XMM-Newton \cite{Foster:2021ngm}, NuSTAR \cite{Roach:2022lgo}, INTEGRAL \cite{Calore:2022pks} and Fermi-LAT \cite{Essig:2013goa}. The scaling of these constraints with $m$ deviates from the $m^{-3/2}$ power law. This is because the measured photon flux expected from decays goes as $\Phi \sim \Gamma_{a\rightarrow\gamma\gamma}D/m$, where $D$ is the integral of the axion energy density along the line of sight \cite{Lisanti:2017qoz}. For telescopes whose sensitivity does not depend on photon energy, this extra factor translates directly into decay constraints on $g_{a\gamma\gamma}$ that scale as $m^{-1}$ rather than $m^{-3/2}$. When the sensitivity does vary with photon energy, however, the constraints deviate further from this scaling in a way that depends on the energy dependence of the experimental parameters and backgrounds. To capture this range of behaviors, we parametrize decay constraints from telescopes as scaling with $m^{-3/2-\varepsilon}$, where $\varepsilon$ accounts for the sensitivity scaling when decay constraints are flux-limited rather than lifetime-limited. 

Therefore, decay constraints are well-approximated by the power law 
\begin{align}
\label{eqn:decay-constraints-gagg}
    g_{a\gamma\gamma}\sqrt{\frac{\Omega}{\Omega_{\rm cdm}}}\lesssim g_{\rm obs}\left(\frac{m}{{\rm eV}}\right)^{-3/2-\varepsilon}\,,
\end{align}
where we have empirically found that $g_{\rm obs} \sim 10^{-12}\,{\rm GeV}^{-1}$ and $\varepsilon\sim 0$ match existing decay limits with a few orders of magnitude across the parameter space $m \in[{\rm eV},{\rm TeV}]$. In detail, both $g_{\rm obs}$ and $\varepsilon$ differ between constraints, and depend on the specific experimental parameters and backgrounds. Because the populated heavy tail falls only as $m^{-3/4}$ while the decay reach typically strengthens as $m^{-3/2-\varepsilon}$ \cite{Essig:2013goa}, indirect searches tend to dominate when the axion spectrum extends into the region with $\Gamma_{a\to\gamma\gamma}\sim H_0$, though whether these populations are visible depends on the scale of inflation. Phrased as a constraint on the axion decay constant, we find
\begin{equation}
\begin{aligned}
\label{eqn:heavy_axion_decay_bound}
    f_{\rm GS}&\gtrsim \frac{3\times10^{12}\,{\rm GeV}}{\sqrt{N}}\left(\frac{H_I}{10\,{\rm GeV}}\right)^{4 }\left(\frac{g_{\rm obs}}{10^{-12}\,{\rm GeV}^{-1}}\right)^{-2}\\
    &\qquad\times\left[4\times 10^4\left(\frac{H_I}{10\,{\rm GeV}}\right)^{-2}\left(\frac{g_{\rm obs}}{10^{-12}\,{\rm GeV}^{-1}}\right)\right]^{\frac{8\varepsilon}{3-4\varepsilon}} . 
\end{aligned}
\end{equation}
In practice, this implies that even for modest values of $H_I$, decay constraints can be among the most powerful probes whenever the axion spectrum populates masses across many decades. Further, the rapid scaling with $g_{\rm obs}$ indicates that substantial gains in axion parameter sensitivity can be made with even modest improvements in detector sensitivity.

Finally, we can repeat this analysis for the direct detection searches (light blue shaded region in \cref{fig:relic_signatures}), which are among the most promising probes of the low-mass tail of the axion distribution. The experiments currently setting the leading bounds roughly in the $2-20 \ \mu$eV range include ADMX \cite{ADMX:2009iij,ADMX:2018gho,ADMX:2019uok,ADMX:2021nhd,ADMX:2024xbv,ADMX:2025vom,ADMX:2018ogs,ADMX:2021mio}, HAYSTAC \cite{HAYSTAC:2018rwy,HAYSTAC:2020kwv,HAYSTAC:2023cam,HAYSTAC:2024jch}, and CAPP \cite{CAPP:2020utb,CAPP:2024dtx}, all microwave cavity haloscopes. The theoretical sensitivity limits for these types of electromagnetic experimental searches were studied in detail in Refs. \cite{Lasenby:2019hfz,Chaudhuri:2018rqn,Chaudhuri:2019ntz,Chaudhuri:2021xjd}. The sensitivity of a static-field experiment depends on how the experimental length scale $L$ compares to the inverse of the axion mass being probed. In the regime where $m > L^{-1}$, an experiment operating at the Standard Quantum Limit (SQL) achieves a mass-averaged signal-to-noise ratio that scales as $\overline{\rm SNR}^2 \propto g_{a\gamma\gamma}^4/(m^3\Delta m)$, where $\Delta m$ is the axion mass window being probed. Once the experiment enters the quasi-static regime, $m < L^{-1}$, the sensitivity is penalized by an additional factor of $(mL)^2$. Proposed up-conversion experiments avoid this suppression \cite{Berlin:2020vrk}: rather than detecting the low-frequency axion signal directly, these experiments convert the signal to a higher, more readily detectable frequency, and $\overline{\rm SNR}^2 \propto g_{a\gamma\gamma}^4/(m\Delta m)$ across the quasi-static regime. These scalings are shown in Fig.~1 of \cite{Lasenby:2019hfz}.

With these mass scalings in mind, we can approximate the direct detection constraints as~\cite{Lasenby:2019hfz}:
\begin{equation}
 g_{a\gamma\gamma}\sqrt{\frac{\Omega}{\Omega_{\rm cdm}}}\lesssim g_{\rm exp}  \left( \frac{m}{\text{peV}}\right)^{x}
\end{equation}
where the  scaling with mass is $x=1$ for static-field experiments at the SQL when $m> L^{-1}$ and $x=1/2$ for up-conversion experiments when $m> L^{-1}$. The constant $g_{\rm exp}$ depends on the parameters of the experimental setup, such as the volume $V\sim L^3$, quality factor $Q$, integration time $t$, and magnetic field strength $B$. For a static-field experiment able to reach the SQL with $L= 1$ m, $B= 4$  T, $t=1$ year per $e$-fold of mass, and $Q =10^6$, Ref. \cite{Lasenby:2019hfz} finds $g_{\rm exp} \sim 10^{-23}$ GeV$^{-1}$. For an up-conversion experiment with isolated-linear-amplifier sensitivity and physical temperature of 1.5 K, mode quality factor of $10^{11}$, $L=1$ m and $t=1$ year per $e$-fold of mass, Ref. \cite{Lasenby:2019hfz} finds $g_{\rm exp} \sim 10^{-19}$ GeV$^{-1}$.

The low-mass tail of the axiverse distribution scales as $m^{9/4}$; 1/4 power of the mass comes from the naive scaling of $\sqrt{\Omega_a}$ with the mass, and 2 powers come from the suppression in the couplings of axions lighter than the QCD axion (see \cref{subsubsec:QCD_axion_coupling}). This steeper scaling means that light axions that inherit their couplings to photons via the QCD instanton charge vector always have suppressed signatures, unlikely to be detected by prospective direct detection experiments. However, when the photon couplings are independent of the QCD instanton charge vector, $g_{a\gamma\gamma}\sqrt{\Omega_a/\Omega_{\rm cdm}} \propto m^{1/4}$. The projected reach of DMRadio~\cite{DMRadio:2022pkf} indicates that it is a particularly promising probe of the light axions when the photon couplings are not inherited from a coupling to QCD.

To summarize, when the coupling to photons is induced by the axions' coupling to QCD, the most accessible regions of parameter space are the QCD axion, which may be probed via future direct detection experiments, as well as the heavy axions, which make up a small subcomponent of dark matter, but may be probed by astrophysical searches for rare axion decays. Future direct detection experiments are also promising avenues to detect lighter axions if photon couplings are generated independently from the coupling to QCD. 

One might also consider other couplings to matter of the form studied in \cref{subsubsec:axionlike_couplings}, for example to electrons. The axion-electron coupling is constrained at low masses by stellar cooling bounds and by loop-induced decays to photons. At high masses above the kinematic threshold $m\gg2m_e$, the coupling is constrained by decays to electron-positron pairs. However, its weaker mass scaling means that these decays are not generically good probes of the heavy tail of the axiverse distribution in this parameter space. The exception is when a QCD coupling is present, in which case photon decays dominate and the heavy tail becomes accessible.

\subsubsection*{Non-relic detection}
We note that bounds from direct production in stars, neutron stars, and supernovae ~\cite{Ayala:2014pea,Vinyoles:2015aba,Meyer:2020vzy,Dessert:2020lil,Li:2020pcn,Dolan:2021rya,Dessert:2021bkv,Caputo:2022mah,Dessert:2022yqq,DeRocco:2022jyq,Dolan:2022kul,Hoof:2022xbe,Beaufort:2023zuj,Diamond:2023scc,Muller:2023vjm,Nguyen:2023czp,Manzari:2024jns,Fiorillo:2025gnd,Ruz:2024gkl,Fiorillo:2025yzf,Benabou:2025jcv,Candon:2024eah} and black hole superradiance constraints~\cite{Arvanitaki:2014wva, Baryakhtar:2020gao, Ng:2020ruv,Hoof:2024quk, Witte:2024drg,Aswathi:2025nxa, Caputo:2025oap} do not rely on cosmological abundances and therefore do not appear in \cref{fig:relic_signatures}. While the former are typically too weak to constrain the range of SM couplings in the ensembles shown here, superradiance may be sensitive to ensembles that are sufficiently dense; see e.g.~\cite{Mehta:2021pwf} for discussion in explicit string realizations. For the parameters of \cref{fig:relic_signatures},   $f_{\rm GS}\sim 3\times 10^{12}$~GeV, which is on the boundary of black hole spin constraints~\cite{Baryakhtar:2020gao}. Note that the enhanced axion self-interactions due to lower $f_{\rm GS}$ will generally weaken the constraints from black hole spins~\cite{Baryakhtar:2020gao},  while making axion radiation signals  in laboratory experiments more promising~\cite{Baryakhtar:2020gao,Gavilan-Martin:2026}.

\subsection{Detection statistics and strategy}
\label{subsec:measuring_N}

The existence of an axiverse changes the interpretation of axion searches in two related ways. First, it may modify the target for the QCD axion itself: since the QCD axion may be only one component of the dark matter, the single-axion target in the $(m_a,|g_{a\gamma\gamma}|\sqrt{\Omega_a})$ plane can overestimate the expected signal by an ${\cal O}(1)$ factor. Second, the discovery of an axion would provide statistical information about the broader ensemble. In this section we first discuss the search strategy for the QCD axion, and then turn to axionlike particles, emphasizing what their couplings and signal strengths can reveal about the size and density of the axiverse.

The QCD axion is special for two reasons. Its coupling to matter is parametrically stronger than that of a generic axionlike particle in the ensemble, and its relic abundance is enhanced by the temperature dependence of the QCD instanton potential. As a result, the QCD axion is often anthropically relevant (\cref{sec:anthropics}), across much of the available parameter space, and therefore naturally accounts for an ${\cal O}(1)$ fraction of the dark matter. In the simplest case, it lies on the usual QCD axion line in the $(m_a,|g_{a\gamma\gamma}|\sqrt{\Omega_a})$ plane. In an axiverse, however, other axions may share the anthropic plateau, reducing the QCD axion fraction and pushing its direct-detection signal somewhat below the canonical single-QCD-axion dark-matter target.

This motivates a modest revision to the usual search strategy. An experiment that reaches, for example, the DFSZ line may still miss the QCD axion if the QCD axion is only a codominant component of the dark matter. We therefore argue that searches should extend somewhat deeper than the canonical single-QCD-axion dark matter target, reaching roughly an ${\cal O}(1)$ factor below the DFSZ line.

Because experiments are especially focused on the QCD axion, we view it as more likely than not that the first axion discovered would be the QCD axion. It is therefore useful to imagine an optimistic future in which the QCD axion has been found and ask where one should search next. If the QCD axion lies precisely on one of the canonical KSVZ or DFSZ lines, the discovery by itself teaches us relatively little about the rest of the axiverse. If it lies appreciably below these lines in the $(m_a,|g_{a\gamma\gamma}|\sqrt{\Omega_a})$ plane, this may instead suggest that the QCD axion is codominant with one or more additional axions.

Where those additional axions are likely to lie depends on the inflationary scale and on the typical GS decay constants. Heavier axions are more likely to reveal themselves through astrophysical or cosmological signatures, while lighter axions may be more amenable to direct detection, especially if their couplings arise only through mixing with QCD (see \cref{subsec:results_detection}). Because of dimensional suppression and the unknown size of the axiverse, the heavier axionlike particles may typically be ${\cal O}(10)$ dimmer than the QCD axion, and therefore require more sensitive searches. At the same time, their masses are \emph{a priori} unknown, so broad coverage is also necessary. Ultimately, whether one prioritizes broad coverage above or below the QCD axion mass depends on one's theoretical priors. The most important considerations in developing a strategy are whether the axiverse is likely to be densely populated, and whether grand unification is likely~\cite{Agrawal:2022lsp,Agrawal:2024ejr,Agrawal:2025rbr}.

Let us now imagine an even more distant future in which an axionlike particle has been discovered in addition to the QCD axion. Such a discovery would already provide significant information about the deep UV, and in particular would hint at the existence of compact dimensions. An important inference can be made from the observed signal strength. The key point is that the typical power deposited in a direct detection experiment depends on whether the particle is the QCD axion or an axionlike particle whose coupling to matter arises through mixing with the QCD axion. For the QCD axion, the coupling strength is fixed by its GS decay constant, $f_{{\cal O},i_c}=f_{{\rm GS},i_c}$ (\cref{subsubsec:QCD_axion_coupling}). For an axionlike particle, the coupling is drawn from a distribution set by the instanton charges, with a mean interaction strength suppressed by the size of the axiverse, $f_{{\cal O},i\neq i_c}\propto \sqrt{N}f_{{\rm GS},i\neq i_c}$ (\cref{subsubsec:axionlike_couplings}). Thus, the ratio of powers deposited by the QCD axion and by an axionlike particle contains information about the total number of axions.

The precision of this inference depends on the density of the axiverse, or equivalently on how many axions lie within the experimentally accessible mass range. If there are a few axions in the accessible window, then a measurement of the QCD axion together with one axionlike particle, presumed to be the brightest in that window, can be used to infer $N$ to within roughly an order of magnitude. We give the details of this statistical analysis in appendix \ref{app:axiverse-size}. In addition to providing an estimate of $N$, the mass range over which the search was conducted itself carries information about the density of the axiverse. Finding an axionlike particle in a narrow window would suggest a denser axiverse than finding one only after many decades of coverage. In appendix \ref{app:axiverse-density}, we perform a toy statistical study of this inference.

\section{Conclusion}
\label{sec:conclusion}

The string axiverse offers an experimentally accessible window onto physics at very high scales. Because axions are protected by approximate shift symmetries, they can remain parametrically light in the four-dimensional effective theory even when other Kaluza--Klein and string excitations decouple. In realistic compactifications, the number of such axions can be large and often grows with the topological complexity of the extra-dimensional manifold. This abundance naturally raises a question: if many light degrees of freedom are generic, why has no direct evidence appeared so far? Two simple reasons are likely. First, axions are feebly coupled. Second, their late-time abundances can be small depending on cosmological history. Combined with the enormous dimensionality of multi-axion effective theories, these features make it challenging to extract robust, model-independent expectations. The aim of this work has been to identify such expectations under a minimal, theory-motivated set of assumptions, and to translate them into a clear set of experimental targets.

Concretely, we imposed hierarchical axion potential scales, as suggested by the exponential sensitivity of instanton actions, so that the limit $\Lambda_i \gg \Lambda_{i+1}$ organizes the spectrum through a sequence of integrating out heavy modes. In this regime, the mass eigenstates track an orthogonalization of the dominant charge directions, which is well captured by the Gram-Schmidt basis. We then treated the remaining structural input statistically by modeling the instanton charge vectors as independent and identically distributed random variables, while allowing general kinetic structure and phases. Within this framework we developed analytic control over effective field ranges and couplings, and then incorporated inflationary initial-condition priors together with anthropic weighting to identify which regions of parameter space are plausibly populated.

In \cref{sec:Statistics}, we analyzed the physical couplings and effective field ranges of the mass eigenstates in the hierarchical instanton limit. Beyond the hierarchy in instanton scales, our key structural assumption was that the instanton charge vectors are i.i.d.\ as vectors (that is, the components of each vector obey independent distributions), while we remained agnostic about the kinetic matrix and phases. In \cref{subsec:field-ranges}, we found that the effective field ranges are accurately captured by the analytic estimate \cref{eqn:fGS-ESP}, which exploits the relationship between Gram-Schmidt orthogonalization and the sequential ratios of volumes of parallelepipeds spanned by the instanton charge vectors. We tested this expression across representative examples in \cref{fig:GS_isotropic,fig:GS_anisotropic} and found that \cref{eqn:fGS-ESP} reproduces the mean behavior of the effective field ranges with excellent accuracy. We then estimated the spread about the mean in \cref{eqn:fGS_variance_estimate}: this formula is exact in the isotropic limit of degenerate kinetic eigenvalues, and it continues to capture the variance well even when the kinetic eigenvalues are broadly distributed. The qualitative outcome is that heavy modes tend to have reduced field ranges. When kinetic eigenvalues are comparable, the heavy axions exhibit a characteristic $\sqrt{N}$ suppression of $f_{{\rm GS},i}$ relative to the independent-axion picture, while broad kinetic hierarchies induce additional suppression set by the spread of kinetic eigenvalues. As a result, the usual single-axion intuition $\Omega \propto m^{1/2}$ need not hold parametrically: in \cref{subsubsec:GS_relic_abundance_scaling} we show that $\Omega(m)$ can realize power laws with exponents anywhere below $1/2$.

In \cref{subsec:couplings}, we turned to couplings to matter. For a generic axionlike coupling direction unrelated to any instanton charge vector (\cref{subsubsec:axionlike_couplings}), we found that typical modes incur a $\sqrt{N}$ suppression in signal strength: the coupling decay constants are parametrically larger than the corresponding field-range (Gram-Schmidt) decay constants, reducing direct-detection prospects for most axions in large-$N$ ensembles. The notable exceptions lie at the edges of the spectrum: the couplings of the lightest, and in some cases the heaviest, modes remain unsuppressed. The QCD axion is special (\cref{subsubsec:QCD_axion_coupling}): because its coupling direction is aligned with a term in the potential, it avoids the generic $\sqrt{N}$ suppression and is parametrically more visible than a typical ALP drawn from the same ensemble.

We showed how cosmological selection further sharpens this picture in \cref{sec:anthropics}. We reviewed initial conditions after a long period of inflation in \cref{subsubsec:inf-priors-theta0} and the resulting anthropic weighting of axion energy densities in \cref{subsubsec:anthropic-weight}. We found a useful simplification: at fixed inflationary scale $H_I$, the anthropic probability is well captured by identifying the relevant axions---those with an ${\mathcal O}(1)$ chance to overclose the Universe given their microscopic parameters and the inflationary prior---and penalizing only these. This generalizes the usual multi-axion anthropic argument to arbitrary (long) inflationary scales and arbitrary axion parameters within the regime of validity of the effective theory. In \cref{subsec:relevant-axions}, we use this generalization to find that interactions among axions as well as long, low-scale inflation enhance the anthropic probability of axiverse scenarios relative to the equivalent independent axion scenario, as shown in \cref{fig:prob-vs-Na}. The existence of an axiverse would thus show a preference for a low scale of inflation. Another qualitative implication of such anthropic considerations is the emergence of an anthropic plateau. Over a range of masses, several axions can contribute comparably to the total dark matter abundance, reflecting the fact that selection acts primarily on the subset of axions capable of overproducing dark matter for the chosen $H_I$.

Finally, in \cref{sec:results}, we illustrated these effects in a concrete realization by Monte Carlo sampling axion effective theories drawn from the ensembles of \cref{sec:Statistics}, combined with initial conditions drawn according to \cref{sec:anthropics}. The Monte Carlo procedure is described in \cref{subsec:monte_carlo}. In \cref{subsec:results_relic_abundances}, we presented the resulting relic-abundance distributions (\cref{fig:relic_abundance,fig:relic_abundance_extra}) and interpreted their dominant features: anthropic plateaus (including one tied to QCD dynamics), enhanced abundances of axions with masses smaller than, but close to, the QCD axion due to level mixing, and the suppression of very heavy and very light axion relic abundances due to low-scale inflation and small potential energy density respectively. In \cref{subsec:results_detection}, we translated these relics into detection prospects in \cref{fig:relic_signatures,fig:relic_signatures_extra}, plotting the abundance-rescaled coupling so that dark-matter subcomponents appear on equal footing with dominant components. The populated regions of parameter space track our simple estimates closely, and we find excellent agreement. In \cref{subsec:measuring_N}, we considered the future possibility that the QCD axion and possibly an axionlike particle relic are detected in the lab. We show how the observed signal strengths can be used to estimate the size and density of the axiverse.

Taken together, these results point to a few especially robust observational lessons. First, the QCD axion remains the most prominent direct detection target. Even in a mixed, many-axion theory, its coupling does not incur the generic $1/\sqrt{N}$ suppression. Its relic abundance is also typically enhanced relative to neighboring states because the QCD potential turns on late, delaying the onset of oscillations. Consequently, it tends to lie close to the standard ``QCD axion line'' across a wide range of inflationary histories and decay constants. Anthropic selection can place it within a plateau where several axions share the total abundance, mildly suppressing direct detection prospects, but a discovery would motivate follow-up searches for more-weakly coupled axionlike particles that could yield important information about the axiverse ensemble. Second, heavy-axion subcomponents provide powerful probes of hierarchical axiverse realizations. Low-scale inflation suppresses their abundances, but a high-mass tail with lifetimes comparable to the age of the Universe can be especially promising for indirect searches. In particular, the decay criterion of \cref{eqn:heavy_axion_decay_bound} is extremely sensitive to the inflationary scale and improves rapidly with experimental sensitivity. This makes rare decays into X-rays or $\gamma$-rays a compelling avenue for expanding coverage of axion parameter space even when these modes individually are subdominant. Third, when photon couplings are independent of the QCD instanton charge vector, the detection prospects of light axions do not receive the suppression induced by mixing with the QCD axion. For such scenarios, current and planned direct detection experiments may probe the low-mass tail of certain axiverse realizations. 

By its nature, the axiverse spans a vast landscape of possibilities. Any survey must choose a direction, and the scope of what can be claimed as generic is ultimately limited. Nevertheless, we have aimed to cast a broad net while still retaining analytic control and the ability to make definite statements. Under fairly weak assumptions, a simple picture emerges: two complementary search strategies are particularly well motivated. The first is direct detection of the QCD axion. It benefits both from the usual enhancement of its relic abundance due to delayed oscillations, and from the absence of large-$N$ suppression in its effective coupling, since the same interaction that makes it visible also contributes to its mass. The second is indirect detection of heavy axionlike subcomponents whose relic abundances are irrelevant for structure formation but whose decay rates are small enough to allow survival to the present epoch.
With these targets in view, the path forward is clear, and the search space remains rich with discovery potential. 

\begin{acknowledgments}

We thank Mustafa Amin, Asimina Arvanitaki, Joshua Benabou, Kevin Borisiak, Cyril Creque-Sarbinowski, Savas Dimopolous, Joshua Foster, Fabian Hahner, Naomi Gendler, Mudit Jain, Amalia Madden, Liam McAllister, Matt McQuinn, Viraf Mehta, Mehrdad Mirbabayi, Jakob Moritz, Mario Reig, Gray Rybka, Benjamin Safdi, Murali M. Saravanan, Matthew Seeley, Olivier Simon, and Giovanni Villadoro for helpful discussions. The authors are supported by the U.\,S. Department of Energy Office of Science under Award Number DE-SC0024375 and the Department of Physics and College of Arts and Science at the University of Washington. D.\,C. also acknowledges the receipt of a grant from the Abdus Salam International Centre for Theoretical Physics (ICTP), Trieste, Italy, and the support of the Istituto Nazionale di Fisica Nucleare (INFN). M.B. and E.H. gratefully acknowledge the Pacific Postdoctoral Program at the Dark Universe Science Center, University of Washington, during which part of this work was carried out. The Pacific Postdoctoral Program is supported by a grant from the Simons Foundation (SFI-MPS-T-Institutes-00012000, ML). We are also grateful for the hospitality of Perimeter Institute, where part of this work was carried out. Research at Perimeter Institute is supported in part by the Government of Canada through the Department of Innovation, Science and Economic Development and by the Province of Ontario through the Ministry of Colleges and Universities. M.\,B. was also supported by a grant from the Simons Foundation (1034867, Dittrich). Part of this work was carried out during the workshop Prospects for the String Axiverse (BIRS 25w5384), and we thank the organizers and the Banff International Research Station for hospitality and a stimulating research environment. Axion parameter space limits and direct detection prospects were assembled from Ref.~\cite{AxionLimits} and the associated GitHub repository.

\end{acknowledgments}

\appendix

\section{General statistics}
\label{app:General_Statistics}
In the main text we illustrated the statistical organization of axion field ranges and couplings in a simple representative ensemble. In this appendix we give a more general version of the same argument. We use a slightly different notation, chosen to make clear which assumptions are required. In particular, we do not assume that the full instanton charge vectors are independently and identically distributed. Instead, we assume only that their orientations are i.i.d., while their magnitudes are fixed but otherwise arbitrary. This allows the instanton charge-vector lengths to be correlated with the instanton actions, as one expects in many geometric constructions, so that instantons of different action can preferentially appear at different charge-vector lengths.

Throughout this appendix we work in the hierarchical-instanton regime, so that the GS basis coincides with the mass basis up to corrections controlled by ratios of neighboring instanton scales. We write
\begin{align}
    r_i \equiv |\bm r_i|\,,\qquad \hat{\bm r}_i \equiv \bm r_i/r_i\,,\qquad \bm\rho_i \equiv \bm K^{-1/2}\bm r_i\,.
\end{align}
The quantities $r_i$ are held fixed, while the orientations $\hat{\bm r}_i$ are drawn independently from a common angular distribution. The effects of both the kinetic matrix and the angular distribution are captured by the canonically normalized angular second moment
\begin{align}
\label{eqn:app_Crho}
    \bm C_\rho \equiv \bm K^{-1/2}\langle \hat{\bm r}\hat{\bm r}^T\rangle_{\hat r}\bm K^{-1/2}\,,\qquad
    {\rm eigs}\{\bm C_\rho\}=\{\lambda_1^2,\ldots,\lambda_N^2\}\,.
\end{align}
Without loss of generality, we take the eigenvalues to be ordered from largest to smallest $\lambda_i^2\gg\lambda_{i + 1}^2$.

\subsection{Field ranges}
\label{app:general_field_ranges}

The GS decay constant of the $i$th axion is set by the length of the $i$th charge vector projected orthogonally to the span of the previous $i-1$ charge vectors,
\begin{align}
    f_{{\rm GS},i}^{-1}=|\bm\rho_{i\perp}|\,.
\end{align}
Equivalently, in terms of the Gram matrices $\bm G_i=\bm Q_i\bm Q_i^T$, where $\bm Q_i$ has rows $\bm\rho_1,\ldots,\bm\rho_i$,
\begin{align}
\label{eqn:app_gram_ratio}
    |\bm\rho_{i\perp}|^2=\frac{\det \bm G_i}{\det \bm G_{i-1}}\,.
\end{align}
This ratio-of-volumes form is useful because the expectation value of the Gram determinant depends only on the second moment \eqref{eqn:app_Crho}. Using Cauchy--Binet, one finds
\begin{align}
\label{eqn:app_detGi}
    \langle \det \bm G_i\rangle_{\hat r}
    = i!\,e_i(\lambda_1^2,\ldots,\lambda_N^2)\prod_{j=1}^i r_j^2\,,
\end{align}
where $e_i$ denotes the $i$th elementary symmetric polynomial. We then use the mean-field approximation
\begin{align}
\label{eqn:app_meanfield}
    \langle f_{{\rm GS},i}^{-2}\rangle_{\hat r}
    = \left\langle\frac{\det \bm G_i}{\det \bm G_{i-1}}\right\rangle_{\hat r}
    \approx
    \frac{\langle\det \bm G_i\rangle_{\hat r}}
         {\langle\det \bm G_{i-1}\rangle_{\hat r}}\,.
\end{align}
This gives the central result
\begin{align}
\label{eqn:app_fGS_general}
    \langle f_{{\rm GS},i}^{-2}\rangle_{\hat r}
    \approx
    i\,r_i^2\,
    \frac{e_i(\lambda_1^2,\ldots,\lambda_N^2)}
         {e_{i-1}(\lambda_1^2,\ldots,\lambda_N^2)}\,.
\end{align}
The approximation is exact in the isotropic limit and is also accurate in the strongly anisotropic limit. In the latter case the GS procedure effectively removes the most concentrated canonically normalized directions first, so the relevant projectors become approximately deterministic.

If the kinetic matrix and the charge-vector orientation distribution are both isotropic, then $\bm C_\rho=\bm I/(Nf^2)$ and
\begin{align}
\label{eqn:app_fGS_isotropic}
    \langle f_{{\rm GS},i}^{-2}\rangle_{\hat r} = \frac{r_i^2}{Nf^2}(N-i+1)\,,\qquad ({\rm isotropic})\,.
\end{align}
Thus the field range of the $i$th axion is suppressed by the square root of the number of lighter fields, up to the overall charge-vector length. For dense charge vectors with typical component variance $\sigma^2$, one has $r_i^2\simeq N\sigma^2$ and hence $f_{{\rm GS},i}^{-2}\sim \sigma^2(N-i+1)/f^2$. For sparse charge vectors with $\mathcal O(1)$ nonzero entries, $r_i^2\sim \mathcal O(1)$, so the lightest mode obeys
\begin{align}
\label{eqn:app_sparse_lightest}
    \langle f_{{\rm GS},N}^{-2}\rangle_{\hat r}\sim \frac{1}{Nf^2}\,,
\end{align}
reproducing the familiar $f_{{\rm GS},N}\sim \sqrt N f$ delocalization enhancement.

In the opposite limit, suppose that the eigenvalues of $\bm C_\rho$ are strongly hierarchical, $\lambda_i\gg \lambda_{i+1}$. Then
\begin{align}
\label{eqn:app_fGS_anisotropic}
    \langle f_{{\rm GS},i}^{-2}\rangle_{\hat r}\approx i r_i^2\lambda_i^2\,,\qquad ({\rm strongly\ anisotropic})\,.
\end{align}
If the charge-vector orientations are isotropic and the anisotropy comes only from the kinetic matrix, with kinetic eigenvalues $f_i\ll f_{i + 1}$, this reduces to
\begin{align}
\label{eqn:app_fGS_kinetic_anisotropy}
    \langle f_{{\rm GS},i}^{-2}\rangle_{\hat r}\approx i\frac{r_i^2}{Nf_i^2}\,.
\end{align}
In general, however, it is more invariant to regard the anisotropy as encoded directly in the spectrum of $\bm C_\rho$; it need not be assigned separately to the kinetic matrix or to the angular distribution.

For the Gaussian examples used in the main text, the spread around the mean can be modeled by a scaled $\chi^2$ distribution. A useful estimate for the effective number of degrees of freedom is
\begin{align}
\label{eqn:app_nueff}
    \nu_{{\rm eff},i}\approx\frac{\left(\sum_{k=i}^N\lambda_k^2\right)^2}{\sum_{k=i}^N\lambda_k^4}\,,
\end{align}
so that
\begin{align}
\label{eqn:app_fGS_variance}
    \frac{{\rm var}(f_{{\rm GS},i}^{-2})}{\langle f_{{\rm GS},i}^{-2}\rangle^2}\sim \frac{2}{\nu_{{\rm eff},i}}\,.
\end{align}
This estimate captures the self-averaging of isotropic ensembles and the order-one fluctuations of strongly anisotropic ensembles. It can fail for very sparse integer charge matrices, where shot noise dominates.

\subsection{Generic axionlike couplings}
\label{app:general_axionlike_couplings}

Now consider a Standard Model coupling written in the fundamental basis as
\begin{align}
\label{eqn:app_generic_coupling}
    {\cal L}_{\rm int}=\bm q^T\bm\theta\,{\cal O}_{\rm SM}\,.
\end{align}
The transformation to the GS basis can be written as
\begin{align}
\label{eqn:app_A_def}
    \bm a=\bm A^{-1}\bm\theta\,,\qquad
    \bm A=\bm R_K\bm D_K^{-1}\bm R_Q\,,
\end{align}
where $\bm R_K\bm D_K^{-1}$ canonically normalizes the fields and $\bm R_Q$ implements the GS procedure. The GS-basis couplings are
\begin{align}
\label{eqn:app_g_def}
    \bm g^T=\bm q^T\bm A\,,\qquad
    g_i\equiv f_{{\cal O},i}^{-1}\,.
\end{align}

The most motivated generic case is that the coupling-vector orientation $\hat{\bm q}$ is independent of the instanton charge-vector orientations, but drawn from the same angular distribution. This reflects the expectation that both instanton charges and matter couplings are shaped by the same underlying geometry, without requiring the coupling vector to coincide with an instanton that contributes appreciably to the potential. Writing $q\equiv |\bm q|$, the averaged coupling is then controlled by the same matrix $\bm C_\rho$ that controlled the GS field ranges:
\begin{align}
\label{eqn:app_coupling_average_same_distribution}
    \langle f_{{\cal O},i}^{-2}\rangle_{\hat q}= q^2\,[\bm R_Q^T\bm C_\rho\bm R_Q]_{ii}\,.
\end{align}
This form makes the main point transparent. When $\hat{\bm q}$ is drawn from the same distribution as the instanton charge-vector orientations, there is not a separate classification into ``kinetic anisotropy,'' ``charge-vector anisotropy,'' and ``coupling-vector anisotropy.''  The only invariant object entering both field ranges and typical couplings is $\bm C_\rho$. The problem is therefore characterized simply by whether $\bm C_\rho$ is close to isotropic or strongly anisotropic. More general coupling distributions, with $\hat{\bm q}$ drawn from a different angular distribution, can be obtained by replacing $\bm C_\rho$ in \eqref{eqn:app_coupling_average_same_distribution} with the corresponding canonically normalized coupling second moment; we do not need this more general case in the main text.

In the isotropic limit, \eqref{eqn:app_coupling_average_same_distribution} becomes
\begin{align}
\label{eqn:app_fO_isotropic}
    \langle f_{{\cal O},i}^{-2}\rangle_{\hat q}
    = \frac{q^2}{Nf^2}\,,\qquad ({\rm isotropic})\,.
\end{align}
Comparing with \eqref{eqn:app_fGS_isotropic},
\begin{align}
\label{eqn:app_signal_isotropic}
    \frac{\langle f_{{\cal O},i}^{-2}\rangle_{\hat q}}{\langle f_{{\rm GS},i}^{-2}\rangle_{\hat r}} = \frac{q^2}{r_i^2}\frac{1}{N-i+1}\,,\qquad ({\rm isotropic})\,.
\end{align}
Thus generic couplings do not inherit the same dimensional GS projection that enhances the field range of light axions. For mid-spectrum modes this corresponds to the usual $\sqrt N$ suppression of the coupling relative to the naive expectation based on the field range.

In the strongly anisotropic limit, the GS procedure sorts the eigenvalues of $\bm C_\rho$, giving
\begin{align}
\label{eqn:app_fO_anisotropic}
    \langle f_{{\cal O},i}^{-2}\rangle_{\hat q}
    \sim q^2\lambda_i^2\,,\qquad ({\rm strongly\ anisotropic})\,.
\end{align}
Comparing with \eqref{eqn:app_fGS_anisotropic},
\begin{align}
\label{eqn:app_signal_anisotropic}
    \frac{\langle f_{{\cal O},i}^{-2}\rangle_{\hat q}}
         {\langle f_{{\rm GS},i}^{-2}\rangle_{\hat r}}
    \sim \frac{q^2}{r_i^2}\frac{1}{i}\,,\qquad ({\rm strongly\ anisotropic})\,.
\end{align}
The ordering of the dimensional suppression is reversed relative to the isotropic case, but typical mid-spectrum axions are again suppressed by a factor of order $\sqrt N$ in amplitude.

\subsection{QCD axion couplings}
\label{app:general_QCD_axion_couplings}

The QCD axion is special because the same linear combination of axions that appears in the QCD anomaly also appears in the potential generated by the QCD topological susceptibility,
\begin{align}
    \frac{1}{32\pi^2}\bm q^T\bm\theta\,G\widetilde G\,,\qquad
    V\supset \chi_c\left(1-\cos\bm q^T\bm\theta\right)\,.
\end{align}
If the QCD term appears at rank $i_c$ in the hierarchy, then
\begin{align}
    \bm q=\bm r_{i_c}\,.
\end{align}
Using \eqref{eqn:app_g_def}, together with the GS identity
\begin{align}
\label{eqn:app_GS_length_A}
    f_{{\rm GS},i}^{-1}=|\bm\rho_{i\perp}|=[\bm r_i^T\bm A]_i\,,
\end{align}
one immediately obtains
\begin{align}
\label{eqn:app_QCD_decay_constant}
    f_{{\cal O},i_c}=f_{{\rm GS},i_c}\,.
\end{align}
Thus the QCD axion avoids the $\sqrt N$ suppression that affects generic axionlike couplings. Its coupling decay constant is its field range, as in the single-axion theory.

It is useful to normalize the off-diagonal couplings to the corresponding GS decay constants by defining
\begin{align}
\label{eqn:app_xi_def}
    [\bm\xi_i^T]_j f_{{\rm GS},j}^{-1}\equiv [\bm r_i^T\bm A]_j\,,\qquad [\bm\xi_i]_i=1\,.
\end{align}
For axions heavier than the QCD axion, $i<i_c$, the QCD charge vector is statistically equivalent to an independent vector drawn from the same angular distribution as the instanton charges. Therefore the same isotropic and anisotropic estimates above apply, giving
\begin{align}
\label{eqn:app_heavy_QCD_coupling}
    [\bm\xi_{i_c}]_{i<i_c}^2
    \sim \frac{r_{i_c}^2}{r_i^2}
    \left\{
    \begin{array}{cc}
    (N-i+1)^{-1} & {\rm (isotropic)}\\[3pt]
    i^{-1} & {\rm (strongly\ anisotropic)}
    \end{array}
    \right.\,.
\end{align}
This is the same simplification emphasized above: because the QCD coupling direction is one of the instanton charge directions, the same matrix $\bm C_\rho$ controls both the field ranges and the off-diagonal QCD couplings.

\subsection{Independent sources of anisotropy}
\label{app:independent_coupling_anisotropy}

In the discussion above, we assumed that the coupling-vector orientation $\hat{\bm q}$ is independent of the instanton charge-vector orientations $\hat{\bm r}_i$, but drawn from the same statistical distribution. This is well motivated if both the axion potential and the couplings to matter are shaped by the same underlying geometry, even when the coupling vector does not itself correspond to an instanton that contributes appreciably to the potential. In this case, the coupling statistics are controlled by the same canonically normalized angular second moment that controls the GS field ranges, ${\bm C}_\rho$ defined in \cref{eqn:app_Crho}.
Thus there is not a separate invariant notion of coupling anisotropy. The kinetic matrix and the angular distribution of the instanton charge vectors may each contribute to anisotropy, but only through the combined matrix $\bm C_\rho$. The relevant distinction is therefore simply between the isotropic limit, in which the eigenvalues of $\bm C_\rho$ are approximately degenerate, and the strongly anisotropic limit, in which they are hierarchical.

It is nevertheless useful to consider the more general possibility that $\hat{\bm q}$ is drawn from a different angular distribution than the instanton charge vectors. In that case, the coupling-vector orientations are described by their own canonically normalized second moment, \begin{align}
    \bm C_{\cal O}\equiv\bm K^{-1/2}\bm \langle \hat{\bm q}\hat{\bm q}^{T}\rangle_{\hat{\bm q}}\bm K^{-1/2}\,,
\end{align}
and the coupling decay constant second moment becomes
\begin{align}\label{eqn:general_coupling_decay_constant_average}
    \langle f_{{\cal O},i}^{-2}\rangle_{\hat{\bm q}}=q^2[\bm R_Q^T\bm C_{\cal O}\bm R_Q]_{ii}\,.
\end{align}
This form makes clear that while the GS rotation $\bm R_Q$ is still determined by the instanton charges, through $\bm C_\rho$, the coupling strengths are set by the projection of the independent coupling anisotropy $\bm C_{\cal O}$ onto the GS directions.

There are then several qualitatively distinct limits. If the coupling-vector distribution is isotropic in the fundamental basis, while the instanton sector is anisotropic, then $\bm R_Q$ sorts the GS field ranges according to the eigenvalues of $\bm C_\rho$, but the coupling vector does not preferentially select the same directions. Conversely, if the instanton sector is isotropic while the coupling-vector distribution is anisotropic, then the GS field ranges follow the usual isotropic dimensional scaling, but the coupling decay constants can be ordered by the preferred directions of $\bm C_{\cal O}$. Finally, if both sectors are anisotropic, the coupling statistics depend on the relative orientation between the eigenvectors of $\bm C_\rho$ and those of $\bm C_{\cal O}$.

This last case is the genuinely new possibility relative to the discussion in the main text. The GS procedure removes the directions preferred by $\bm C_\rho$ in descending order, while the couplings are enhanced along the
directions preferred by $\bm C_{\cal O}$. If these two sets of preferred directions are aligned, then the coupling decay constants inherit an ordering similar to the GS field ranges. If they are misaligned, however, there is no
universal rank ordering of the couplings: the largest coupling need not be associated with the heaviest axion, and the scaling of $f_{{\rm GS},i}/f_{{\cal O},i}$ depends on the relative orientation of the two anisotropies.

\subsection{Summary}
\label{app:general_statistics_summary}

The general treatment differs from the simplified main-text ensemble in two ways. First, only the orientations of the instanton charge vectors are assumed to be i.i.d.; their lengths may be fixed arbitrarily and may correlate with their instanton actions. Second, the kinetic matrix and the angular distribution enter through the single canonically normalized second moment $\bm C_\rho$. The main field-range result is \eqref{eqn:app_fGS_general}, with isotropic and strongly anisotropic limits given by \eqref{eqn:app_fGS_isotropic} and \eqref{eqn:app_fGS_anisotropic}. For generic axionlike couplings whose orientations share the same geometric distribution as the instanton charges, the typical couplings are governed by \eqref{eqn:app_coupling_average_same_distribution}; consequently there is simply an isotropic regime and an anisotropic regime, rather than several independent kinds of anisotropy. Finally, the QCD axion remains special because its coupling vector is itself an instanton charge vector, giving $f_{{\cal O},i_c}=f_{{\rm GS},i_c}$ and eliminating the generic $\sqrt N$ suppression.

\section{Beyond independent and identically distributed instanton charge directions}
\label{subsec:iidnt}

The discussion in the main text and preceding \cref{app:General_Statistics} focused on the case in which the charge vectors orientations $\hat {\bm r}_i$ are i.i.d. This is the case, for example, when the instanton actions are statistically independent of the orientation of the corresponding charge vectors. Then selecting the $N$ smallest instanton actions does not bias the distribution of the associated charge vectors, so these vectors are effectively drawn from a large reservoir according to a common distribution. In the large-reservoir limit, this is equivalent to sampling with replacement, and hence the selected charge vectors are approximately i.i.d.
However, full independence between $S_i$ and $\hat{\bm r}_i$ is not required for the arguments of \cref{subsec:field-ranges,subsec:couplings} to carry through. It is sufficient that the action hierarchy be only locally correlated with the charge vector orientations, in the sense that the rank of an instanton in the action hierarchy is not rigidly fixed once its charge vector orientation is specified. We characterize this by a mobility scale ${\mathfrak m}$, defined as the typical range in rank over which an instanton, whose corresponding charge vector has fixed orientation, may move in the hierarchy.

The results of \cref{subsec:field-ranges,subsec:couplings} then continue to apply within sufficiently small local windows of the hierarchy. A conservative estimate for the size of such a window may be obtained from an extreme example in which the available charge vectors are a subset of the standard basis vectors of ${\mathbb R}^N$,
\begin{align}
    \{\hat{\bm e}_{1},\dots,\hat{\bm e}_{\mathfrak m}\}\,,
\end{align}
where here we take ${\mathfrak m}\leq N$.\footnote{The limitation ${\mathfrak m} \leq N$ in this example is due to there only being $N$ standard basis vectors in $N$ dimensions: generically one may have ${\mathfrak m}> N$ as well.} We compare two ensembles of $n$ draws from this set. In the first, vectors are drawn without replacement, mimicking the assignment of distinct charge vectors to nearby ranks in the hierarchy. In the second, vectors are drawn with replacement. Denoting the corresponding Gram matrices by $\bm G$ and $\bm G'$, respectively, one has
\begin{align}
    [\bm G_n]_{ij} = \hat{\bm e}_i^{\,T}\hat{\bm e}_j = \delta_{ij}
\end{align}
for the without-replacement ensemble, while in the with-replacement ensemble $\bm G'$ becomes singular whenever the same basis vector is drawn more than once. Thus
\begin{align}
    \det \bm G_n' =
    \begin{cases}
        1 & {\rm if\,\,all\,\,draws\,\,are\,\,distinct}\,,\\[3pt]
        0 & {\rm otherwise}\,,
    \end{cases}
\end{align}
and therefore
\begin{align}
    \langle \det \bm G_n'\rangle_{\rm draws}
    = P({\rm all\,\,distinct})
    = \frac{{\mathfrak m}({\mathfrak m}-1)\cdots({\mathfrak m}-n+1)}{{\mathfrak m}^n}
    \approx \exp\left\{-\frac{n(n-1)}{2{\mathfrak m}}\right\} .
\end{align}
Thus, drawing with replacement approximates drawing without replacement provided
\begin{align}
    \frac{n(n-1)}{2{\mathfrak m}} \ll 1\,.
\end{align}
Equivalently, a reservoir of size ${\mathfrak m}$ can supply ${\cal O}(\sqrt{2{\mathfrak m}})$ effectively i.i.d.\ draws before correlations significantly modify the Gram determinant.

Motivated by this estimate, we return to the physical picture of the bottom-up axiverse, where we draw $N$ charge vector orientations $\hat{\bm r}_i$ from a much larger set of charge vectors. We then treat any local window of $\lesssim \sqrt{2{\mathfrak m}}$ adjacent ranks as approximately i.i.d.\ for the purposes of \cref{subsec:field-ranges,subsec:couplings}. The influence of lower-rank instantons is then encoded through the sequential GS projection, and the results derived above apply directly to the effective theory of such a local block of $\sqrt{2{\mathfrak m}}$ axions. If ${\mathfrak m}\gtrsim N^2/2$, then a window of size $N$ is already approximately i.i.d., reproducing the fully i.i.d.\ regime discussed above. On the other hand, if ${\mathfrak m}\lesssim N^2/2$, the i.i.d.\ approximation applies only locally, over windows of size ${\cal O}(\sqrt{2{\mathfrak m}})$ along the hierarchy.

\section{Axiverse statistics from detection statistics}

In this appendix, we perform a toy statistical analysis of the inference of the density and size of the axiverse, given the discovery of an axionlike particle in addition to the QCD axion. However, a number of confounding variables complicate these inferences. Several stem from the relic abundances, which are in principle independent of the microscopic parameters of the theory. 
As we have described in \cref{subsec:results_relic_abundances}, the relic abundances depend on the axion mass through power law scalings that depend on both the scale of inflation and the GS decay constants themselves. While these systematic scalings are important, we neglect them here since they are parametrically small as long as the mass window over which direct detection experiments are sensitive is not too large. We further assume that direct detection experiments all have the same detecting power, and that together they have covered some fixed range of mass parameter space.

Following the simplified treatment of \cref{sec:Statistics}, we assume that the axionlike couplings are $\chi^2$-distributed
\begin{align}
    \frac{f_{{\rm GS},i\neq i_c}^2}{f_{{\cal O},i\neq i_c}^2}\equiv C_i\,,
\end{align}
where $C$ has PDF
\begin{align}
    f_{C}(c) = \frac{1}{\sqrt{2\pi c}}e^{-c/2}\,.
\end{align}

For both the QCD axion and axionlike particles, the power deposited also depends on the initial misalignment angle. For the sake of argument, we assume the initial misalignment angles are uniformly distributed over the interval $\theta_i\sim {\cal U}(0,1)$ (negative misalignments are redundant for the purpose of this calculation). The power scales as the initial misalignment squared:
\begin{align}
    \theta_i^2(0)\equiv X_i\,,
\end{align}
and $X$ has the PDF:
\begin{align}
    f_X(x) = \frac{1}{2\sqrt{x}}\,.
\end{align}

Thus, the power deposited by the QCD axion and by an axionlike particle are described at the level of distributions by
\begin{align}
    P_{i_c} \sim X_{i_c}\,,\qquad P_i\sim C_iX_i\,.
\end{align}
The PDF for the QCD axion power is then just $f_X(x)$, while the PDF for observing a single axionlike particle with power $P_{i\neq i_c} = z$ is
\begin{align}\label{eqn:PDF_Z}
    f_Z(z) = \frac{E_1(z/2)}{2\sqrt{2\pi z}}\,,
\end{align}
where $E_1(z) = \int_z^{\infty}e^{-t}/t \ {\rm d}t$ is the exponential integral.

\subsection{Size of the axiverse}
\label{app:axiverse-size}

Now let us focus on the optimistic scenario that we discover one axionlike particle in addition to the QCD axion. Here we show how one can infer an estimate for the size of the axiverse from this discovery. The typical power deposited in a direct detection experiment depends on whether the axion is the QCD axion or simply obtains its coupling to matter through mixing with the QCD axion. In the former case, the coupling strength is entirely deterministic, i.e.\ $f_{{\cal O},i_c} = f_{{\rm GS},i_c}$, while in the latter, the coupling strength is selected from a distribution determined by the instanton charges with a mean interaction strength suppressed by $N$: $f_{{\cal O},i\neq i_c} \propto \sqrt{N}f_{{\rm GS},i\neq i_c}$. Thus, information about the total size of the distribution is contained in the ratio of powers deposited in a direct detection experiment between the QCD axion and an axionlike particle.

We now make the important prior assumption that if there are $M$ axions within our direct detection window $\Delta m$, the first we see will be the one which deposits the most power, i.e.\ the brightest axion. The PDF of the maximum over $M$ axionlike particle powers
\begin{align}
    W\equiv \max_{i\neq i_c \wedge m_i\in \Delta m} P_{i}
\end{align}
is then given by
\begin{align}
    f_{W}(w|M) = M F_Z(w)^{M - 1}f_Z(w)\,,
\end{align}
where $F_Z(z)$ is the CDF of $f_Z(z)$.

Thus, it is the ratio of the power deposited by the QCD axion to that deposited by the brightest axionlike particle,
\begin{align}
    R = \frac{X_{i_c}}{W}\,,
\end{align}
that controls the statistical uncertainty on the inferred value of $N$. The corresponding PDF is 
\begin{align}\label{eqn:f_R_PDF}
    f_R(r|M) = \frac{M}{4\sqrt{2\pi r}}\int_0^{1/r}{\rm d}z\,E_{1}(z/2) F_Z(z)^{M - 1}\,.
\end{align}
Since the observed power ratio is proportional to $N$, the statistics encoded in \cref{eqn:f_R_PDF} determine the precision with which the size of the axiverse may be inferred from a measurement of the QCD axion and the brightest additional axionlike particle.
Depending on the density of the axiverse, or equivalently, how broad a swath of mass parameter space experiments are sensitive to, we can obtain a more-precise estimate of $N$. The 50\% confidence interval on $\log_{10}N$ inferred from the measurement of the QCD axion and one additional axionlike particle for the first few $M$ is given in \cref{tbl:N_uncertainty}. Under the marginally optimistic assumption that there are a few axions within the experimentally accessible window, the measurement of the QCD axion and one axionlike particle, presumed to be the brightest, can be used to infer the value of $N$ within half an order of magnitude of the true value. 
However, the uncertainty plateaus even for a very dense axiverse (large $M$) because the intrinsic uncertainty in the QCD axion relic abundance cannot be averaged away.

\begin{table}
\centering
\caption{Uncertainty on $N$ given QCD axion and one ALP observation.}
\label{tbl:N_uncertainty}
\begin{tabular}{|l|l|l|l|l|l|}
\hline
$M$ axions in experimental window & 1          & 2          & 3          & 4          & 5          \\ \hline
$50\%$ confidence interval $\log_{10}N$ & $\pm 0.90$ & $\pm 0.67$ & $\pm 0.60$ & $\pm 0.57$ & $\pm 0.55$ \\ \hline
\end{tabular}
\end{table}

\subsection{Density of the axiverse}
\label{app:axiverse-density}

Observing just one axionlike particle already provides us a great deal of information. Let us assume that we have discovered one axionlike particle at 5$\sigma$ significance, but have not observed other axionlike particles with more than $2\sigma$ significance. If there are only a few axions within the observing window, it is plausible that only one of them had a significant upward fluctuation that allowed us to see it. On the other hand, there cannot be too many axions in the searched mass window or else the lack of an upward fluctuation would itself be statistically improbable. 

We can formalize this in terms of the probability distributions above. Suppose that there are $M$ axionlike particles within the observing window. First, we assume that our search was likely to have discovered an axion at high significance in the first place. Phrased in terms of the CDF $F_Z(z)$ corresponding to the PDF \cref{eqn:PDF_Z},
\begin{align}
    P({\rm at\,least\,one} > 5\sigma) = 1-F_Z(z_5)^M > P_0\,,
\end{align}
where $P_0$ is some ${\cal O}(1)$ probability threshold and $z_5$ is the $5\sigma$ detection threshold. Second, the probability that one axion is observed above $5\sigma$ while the remaining $M-1$ are observed below $2\sigma$ should also not be too small:
\begin{align}
    P(1{\,\rm above\,}5\sigma\wedge M-1{\,\rm below\,}2\sigma) = M F_Z(z_2)^{M-1}[1-F_Z(z_5)] > P_0\,,
\end{align}
where $z_{2}$ is the $2\sigma$ detection threshold. Assuming, schematically, that the SNR scales linearly with the deposited power, $z_5 = 2.5 z_2$. As a function of $M$, we can then identify the maximum possible $P_0$ consistent with both conditions by scanning over all possible values of $z_5$. The resulting pairs of $P_0$ and $M$ are listed in \cref{tbl:axions_in_window} for the first few $M$. From this table one concludes that ${\cal O}({\rm few})$ axions may be reasonably consistent with the observation of a single ALP in a given window. Given the breadth of the observing window, this suggests that the typical axion mass spacing is within an order-one factor of the experimental window. Though this suggests that additional axions may have been missed within the original search window, they are more likely than not to have fluctuated downward. Expanding the search window while maintaining comparable sensitivity therefore offers the best chance of detecting yet another axion.

\begin{table}
\centering
\caption{Threshold probability $P_0$ and maximum allowed number of axions within observing window $M$.}
\label{tbl:axions_in_window}
\begin{tabular}{|l|l|l|l|l|l|}
\hline
$P_0$ & 1.00 & 0.377 & 0.275 & 0.224 & 0.191 \\ \hline
Maximum $M$     & 1 & 2     & 3     & 4     & 5     \\ \hline
\end{tabular}
\end{table}

\bibliography{refs}

@article{Lisanti:2017qoz,
    author = "Lisanti, Mariangela and Mishra-Sharma, Siddharth and Rodd, Nicholas L. and Safdi, Benjamin R. and Wechsler, Risa H.",
    title = "{Mapping Extragalactic Dark Matter Annihilation with Galaxy Surveys: A Systematic Study of Stacked Group Searches}",
    eprint = "1709.00416",
    archivePrefix = "arXiv",
    primaryClass = "astro-ph.CO",
    reportNumber = "MIT-CTP-4930, PUPT-2533, MCTP-17-14",
    doi = "10.1103/PhysRevD.97.063005",
    journal = "Phys. Rev. D",
    volume = "97",
    number = "6",
    pages = "063005",
    year = "2018"
}

@article{Fiorillo:2025gnd,
    author = "Fiorillo, Damiano F. G. and Gil Muyor, {\'A}ngel and Janka, Hans-Thomas and Raffelt, Georg G. and Vitagliano, Edoardo",
    title = "{Axion-photon conversion in transient compact stars: Systematics, constraints, and opportunities}",
    eprint = "2509.13322",
    archivePrefix = "arXiv",
    primaryClass = "hep-ph",
    doi = "10.1088/1475-7516/2026/03/053",
    journal = "JCAP",
    volume = "03",
    pages = "053",
    year = "2026"
}

@article{Chaudhuri:2018rqn,
    author = "Chaudhuri, Saptarshi and Irwin, Kent and Graham, Peter W. and Mardon, Jeremy",
    title = "{Optimal Impedance Matching and Quantum Limits of Electromagnetic Axion and Hidden-Photon Dark Matter Searches}",
    eprint = "1803.01627",
    archivePrefix = "arXiv",
    primaryClass = "hep-ph",
    month = "3",
    year = "2018",
    journal= ""
}

@article{Chaudhuri:2019ntz,
    author = "Chaudhuri, Saptarshi and Irwin, Kent D. and Graham, Peter W. and Mardon, Jeremy",
    title = "{Optimal Electromagnetic Searches for Axion and Hidden-Photon Dark Matter}",
    eprint = "1904.05806",
    archivePrefix = "arXiv",
    primaryClass = "hep-ex",
    month = "4",
    year = "2019",
    journal = ""
}

@article{Chaudhuri:2021xjd,
    author = "Chaudhuri, Saptarshi",
    title = "{Impedance matching to axion dark matter: considerations of the photon-electron interaction}",
    eprint = "2105.02005",
    archivePrefix = "arXiv",
    primaryClass = "hep-ph",
    doi = "10.1088/1475-7516/2021/12/033",
    journal = "JCAP",
    volume = "12",
    number = "12",
    pages = "033",
    year = "2021",
    journal = ""
}

@article{Lasenby:2019hfz,
    author = "Lasenby, Robert",
    title = "{Parametrics of Electromagnetic Searches for Axion Dark Matter}",
    eprint = "1912.11467",
    archivePrefix = "arXiv",
    primaryClass = "hep-ph",
    doi = "10.1103/PhysRevD.103.075007",
    journal = "Phys. Rev. D",
    volume = "103",
    number = "7",
    pages = "075007",
    year = "2021"
}

@article{deGiorgi:2025ldc,
    author = "de Giorgi, Arturo and Jaeckel, Joerg and Monath, Sebastian and Takhistov, Volodymyr",
    title = "{Multiple Axions in Laboratory Experiments}",
    eprint = "2512.16837",
    archivePrefix = "arXiv",
    primaryClass = "hep-ph",
    reportNumber = "IPPP/25/92, KEK-QUP-2025-0025, KEK-TH-2780",
    month = "12",
    year = "2025",
    journal = ""
}

@article{Ho:2019ayl,
    author = "Ho, Shu-Yu and Takahashi, Fuminobu and Yin, Wen",
    title = "{Relaxing the Cosmological Moduli Problem by Low-scale Inflation}",
    eprint = "1901.01240",
    archivePrefix = "arXiv",
    primaryClass = "hep-ph",
    reportNumber = "TU-1080, IPMU18-0210",
    doi = "10.1007/JHEP04(2019)149",
    journal = "JHEP",
    volume = "04",
    pages = "149",
    year = "2019"
}

@article{Takahashi:2018tdu,
    author = "Takahashi, Fuminobu and Yin, Wen and Guth, Alan H.",
    title = "{QCD axion window and low-scale inflation}",
    eprint = "1805.08763",
    archivePrefix = "arXiv",
    primaryClass = "hep-ph",
    reportNumber = "MIT-CTP/5022, TU-1064, IPMU18-0088, MIT-CTP-5022",
    doi = "10.1103/PhysRevD.98.015042",
    journal = "Phys. Rev. D",
    volume = "98",
    number = "1",
    pages = "015042",
    year = "2018"
}

@article{Kim:1979if,
    author = "Kim, Jihn E.",
    title = "{Weak Interaction Singlet and Strong CP Invariance}",
    reportNumber = "UPR-0120T",
    doi = "10.1103/PhysRevLett.43.103",
    journal = "Phys. Rev. Lett.",
    volume = "43",
    pages = "103",
    year = "1979"
}

@article{Dine:1981rt,
    author = "Dine, Michael and Fischler, Willy and Srednicki, Mark",
    title = "{A Simple Solution to the Strong CP Problem with a Harmless Axion}",
    reportNumber = "Print-81-0320 (IAS,PRINCETON)",
    doi = "10.1016/0370-2693(81)90590-6",
    journal = "Phys. Lett. B",
    volume = "104",
    pages = "199--202",
    year = "1981"
}

@article{Zhitnitsky:1980tq,
    author = "Zhitnitsky, A. R.",
    title = "{On Possible Suppression of the Axion Hadron Interactions. (In Russian)}",
    journal = "Sov. J. Nucl. Phys.",
    volume = "31",
    pages = "260",
    year = "1980"
}

@article{Srednicki:1985xd,
    author = "Srednicki, Mark",
    title = "{Axion Couplings to Matter. 1. CP Conserving Parts}",
    reportNumber = "Print-85-0247 (UC,SANTA BARBARA)",
    doi = "10.1016/0550-3213(85)90054-9",
    journal = "Nucl. Phys. B",
    volume = "260",
    pages = "689--700",
    year = "1985"
}

@article{Georgi:1986df,
    author = "Georgi, Howard and Kaplan, David B. and Randall, Lisa",
    title = "{Manifesting the Invisible Axion at Low-energies}",
    reportNumber = "HUTP-86/A004",
    doi = "10.1016/0370-2693(86)90688-X",
    journal = "Phys. Lett. B",
    volume = "169",
    pages = "73--78",
    year = "1986"
}

@article{Sikivie:1983ip,
    author = "Sikivie, P.",
    editor = "Srednicki, M. A.",
    title = "{Experimental Tests of the Invisible Axion}",
    reportNumber = "PRINT-83-0597 (FLORIDA), UF-TP-83-13",
    doi = "10.1103/PhysRevLett.51.1415",
    journal = "Phys. Rev. Lett.",
    volume = "51",
    pages = "1415--1417",
    year = "1983",
    note = "[Erratum: Phys.Rev.Lett. 52, 695 (1984)]"
}

@article{Arvanitaki:2009fg,
    author = "Arvanitaki, Asimina and Dimopoulos, Savas and Dubovsky, Sergei and Kaloper, Nemanja and March-Russell, John",
    title = "{String Axiverse}",
    eprint = "0905.4720",
    archivePrefix = "arXiv",
    primaryClass = "hep-th",
    doi = "10.1103/PhysRevD.81.123530",
    journal = "Phys. Rev. D",
    volume = "81",
    pages = "123530",
    year = "2010"
}

@article{Arvanitaki:2010sy,
    author = "Arvanitaki, Asimina and Dubovsky, Sergei",
    title = "{Exploring the String Axiverse with Precision Black Hole Physics}",
    eprint = "1004.3558",
    archivePrefix = "arXiv",
    primaryClass = "hep-th",
    doi = "10.1103/PhysRevD.83.044026",
    journal = "Phys. Rev. D",
    volume = "83",
    pages = "044026",
    year = "2011"
}

@article{Arvanitaki:2009hb,
    author = "Arvanitaki, Asimina and Craig, Nathaniel and Dimopoulos, Savas and Dubovsky, Sergei and March-Russell, John",
    title = "{String Photini at the LHC}",
    eprint = "0909.5440",
    archivePrefix = "arXiv",
    primaryClass = "hep-ph",
    reportNumber = "OUTP-09-23P",
    doi = "10.1103/PhysRevD.81.075018",
    journal = "Phys. Rev. D",
    volume = "81",
    pages = "075018",
    year = "2010"
}

@article{Randall:1999ee,
    author = "Randall, Lisa and Sundrum, Raman",
    title = "{A Large mass hierarchy from a small extra dimension}",
    eprint = "hep-ph/9905221",
    archivePrefix = "arXiv",
    reportNumber = "MIT-CTP-2860, PUPT-1860, BUHEP-99-9",
    doi = "10.1103/PhysRevLett.83.3370",
    journal = "Phys. Rev. Lett.",
    volume = "83",
    pages = "3370--3373",
    year = "1999"
}

@article{Arkani-Hamed:1998jmv,
    author = "Arkani-Hamed, Nima and Dimopoulos, Savas and Dvali, G. R.",
    title = "{The Hierarchy problem and new dimensions at a millimeter}",
    eprint = "hep-ph/9803315",
    archivePrefix = "arXiv",
    reportNumber = "SLAC-PUB-7769, SU-ITP-98-13",
    doi = "10.1016/S0370-2693(98)00466-3",
    journal = "Phys. Lett. B",
    volume = "429",
    pages = "263--272",
    year = "1998"
}

@article{Damour:1994ya,
    author = "Damour, T. and Polyakov, Alexander M.",
    title = "{String theory and gravity}",
    eprint = "gr-qc/9411069",
    archivePrefix = "arXiv",
    reportNumber = "IHES-P-94-1",
    doi = "10.1007/BF02106709",
    journal = "Gen. Rel. Grav.",
    volume = "26",
    pages = "1171--1176",
    year = "1994"
}

@article{Taylor:1988nw,
    author = "Taylor, T. R. and Veneziano, G.",
    title = "{Dilaton Couplings at Large Distances}",
    reportNumber = "CERN-TH-5116-88, FERMILAB-PUB-88-089-T",
    doi = "10.1016/0370-2693(88)91290-7",
    journal = "Phys. Lett. B",
    volume = "213",
    pages = "450--454",
    year = "1988"
}

@article{Goodsell:2009xc,
    author = "Goodsell, Mark and Jaeckel, Joerg and Redondo, Javier and Ringwald, Andreas",
    title = "{Naturally Light Hidden Photons in LARGE Volume String Compactifications}",
    eprint = "0909.0515",
    archivePrefix = "arXiv",
    primaryClass = "hep-ph",
    reportNumber = "IPPP-09-63, DCPT-09-126, DESY-09-123",
    doi = "10.1088/1126-6708/2009/11/027",
    journal = "JHEP",
    volume = "11",
    pages = "027",
    year = "2009"
}

@article{Svrcek:2006yi,
	title        = {{Axions In String Theory}},
	author       = {Svrcek, Peter and Witten, Edward},
	year         = 2006,
	journal      = {JHEP},
	volume       = {06},
	pages        = {051},
	doi          = {10.1088/1126-6708/2006/06/051},
	eprint       = {hep-th/0605206},
	archiveprefix = {arXiv},
	reportnumber = {SLAC-PUB-11894}
}

@article{Freese:1990rb,
    author = "Freese, Katherine and Frieman, Joshua A. and Olinto, Angela V.",
    title = "{Natural Inflation with Pseudo - Nambu-Goldstone Bosons}",
    reportNumber = "FERMILAB-PUB-90-177-A",
    doi = "10.1103/PhysRevLett.65.3233",
    journal = "Phys. Rev. Lett.",
    volume = "65",
    pages = "3233--3236",
    year = "1990"
}

@article{Abbott:1982af,
	title        = {{A Cosmological Bound on the Invisible Axion}},
	author       = {Abbott, L. F. and Sikivie, P.},
	year         = 1983,
	journal      = {Phys. Lett. B},
	volume       = 120,
	pages        = {133--136},
	doi          = {10.1016/0370-2693(83)90638-X},
	editor       = {Srednicki, M. A.},
	reportnumber = {PRINT-82-0695 (BRANDEIS)}
}

@article{Dine:1982ah,
    author = "Dine, Michael and Fischler, Willy",
    editor = "Srednicki, M. A.",
    title = "{The Not So Harmless Axion}",
    reportNumber = "UPR-0201T",
    doi = "10.1016/0370-2693(83)90639-1",
    journal = "Phys. Lett. B",
    volume = "120",
    pages = "137--141",
    year = "1983"
}

@article{Preskill:1982cy,
	title        = {{Cosmology of the Invisible Axion}},
	author       = {Preskill, John and Wise, Mark B. and Wilczek, Frank},
	year         = 1983,
	journal      = {Phys. Lett.},
	volume       = {B120},
	pages        = {127--132},
	doi          = {10.1016/0370-2693(83)90637-8},
	reportnumber = {HUTP-82-A048, NSF-ITP-82-103}
}

@article{Peccei:1977hh,
    author = "Peccei, R. D. and Quinn, Helen R.",
    title = "{CP Conservation in the Presence of Instantons}",
    reportNumber = "ITP-568-STANFORD",
    doi = "10.1103/PhysRevLett.38.1440",
    journal = "Phys. Rev. Lett.",
    volume = "38",
    pages = "1440--1443",
    year = "1977"
}

@article{Wilczek:1977pj,
    author = "Wilczek, Frank",
    title = "{Problem of Strong  $P$  and  $T$  Invariance in the Presence of Instantons}",
    reportNumber = "Print-77-0939 (COLUMBIA)",
    doi = "10.1103/PhysRevLett.40.279",
    journal = "Phys. Rev. Lett.",
    volume = "40",
    pages = "279--282",
    year = "1978"
}

@article{Weinberg:1977ma,
    author = "Weinberg, Steven",
    title = "{A New Light Boson?}",
    reportNumber = "HUTP-77/A074",
    doi = "10.1103/PhysRevLett.40.223",
    journal = "Phys. Rev. Lett.",
    volume = "40",
    pages = "223--226",
    year = "1978"
}

@article{Freivogel:2008qc,
    author = "Freivogel, Ben",
    title = "{Anthropic Explanation of the Dark Matter Abundance}",
    eprint = "0810.0703",
    archivePrefix = "arXiv",
    primaryClass = "hep-th",
    doi = "10.1088/1475-7516/2010/03/021",
    journal = "JCAP",
    volume = "03",
    pages = "021",
    year = "2010"
}

@article{Petrossian-Byrne:2025mto,
    author = "Petrossian-Byrne, Rudin and Villadoro, Giovanni",
    title = "{Open string axiverse}",
    eprint = "2503.16387",
    archivePrefix = "arXiv",
    primaryClass = "hep-ph",
    doi = "10.1007/JHEP07(2025)049",
    journal = "JHEP",
    volume = "07",
    pages = "049",
    year = "2025"
}

@article{Starobinsky:1986fx,
    author = "Starobinsky, Alexei A.",
    title = "{STOCHASTIC DE SITTER (INFLATIONARY) STAGE IN THE EARLY UNIVERSE}",
    doi = "10.1007/3-540-16452-9_6",
    journal = "Lect. Notes Phys.",
    volume = "246",
    pages = "107--126",
    year = "1986"
}

@article{Dimopoulos:1988pw,
    author = "Dimopoulos, S. and Hall, L. J.",
    title = "{Inflation and Invisible Axions}",
    doi = "10.1103/PhysRevLett.60.1899",
    journal = "Phys. Rev. Lett.",
    volume = "60",
    pages = "1899--1901",
    year = "1988"
}

@article{Starobinsky:1994bd,
    author = "Starobinsky, Alexei A. and Yokoyama, Junichi",
    title = "{Equilibrium state of a selfinteracting scalar field in the De Sitter background}",
    eprint = "astro-ph/9407016",
    archivePrefix = "arXiv",
    reportNumber = "YITP-U-94-12",
    doi = "10.1103/PhysRevD.50.6357",
    journal = "Phys. Rev. D",
    volume = "50",
    pages = "6357--6368",
    year = "1994"
}

@article{Graham:2018jyp,
    author = "Graham, Peter W. and Scherlis, Adam",
    title = "{Stochastic axion scenario}",
    eprint = "1805.07362",
    archivePrefix = "arXiv",
    primaryClass = "hep-ph",
    doi = "10.1103/PhysRevD.98.035017",
    journal = "Phys. Rev. D",
    volume = "98",
    number = "3",
    pages = "035017",
    year = "2018"
}

@article{Weinberg:1987dv,
    author = "Weinberg, Steven",
    title = "{Anthropic Bound on the Cosmological Constant}",
    reportNumber = "UTTG-06-87",
    doi = "10.1103/PhysRevLett.59.2607",
    journal = "Phys. Rev. Lett.",
    volume = "59",
    pages = "2607",
    year = "1987"
}

@article{Planck:2018vyg,
    author = "Aghanim, N. and others",
    collaboration = "Planck",
    title = "{Planck 2018 results. VI. Cosmological parameters}",
    eprint = "1807.06209",
    archivePrefix = "arXiv",
    primaryClass = "astro-ph.CO",
    doi = "10.1051/0004-6361/201833910",
    journal = "Astron. Astrophys.",
    volume = "641",
    pages = "A6",
    year = "2020",
    note = "[Erratum: Astron.Astrophys. 652, C4 (2021)]"
}

@article{Reig:2021ipa,
   title={The stochastic axiverse},
   volume={2021},
   ISSN={1029-8479},
   url={http://dx.doi.org/10.1007/JHEP09(2021)207},
   DOI={10.1007/jhep09(2021)207},
   number={9},
   journal={Journal of High Energy Physics},
   publisher={Springer Science and Business Media LLC},
   author={Reig, Mario},
   year={2021},
   month=sep }

@article{Tegmark:2005dy,
    author = "Tegmark, Max and Aguirre, Anthony and Rees, Martin and Wilczek, Frank",
    title = "{Dimensionless constants, cosmology and other dark matters}",
    eprint = "astro-ph/0511774",
    archivePrefix = "arXiv",
    doi = "10.1103/PhysRevD.73.023505",
    journal = "Phys. Rev. D",
    volume = "73",
    pages = "023505",
    year = "2006"
}

@article{Bousso:2006ev,
    author = "Bousso, Raphael",
    title = "{Holographic probabilities in eternal inflation}",
    eprint = "hep-th/0605263",
    archivePrefix = "arXiv",
    reportNumber = "UCB-PTH-06-10, LBNL-60251",
    doi = "10.1103/PhysRevLett.97.191302",
    journal = "Phys. Rev. Lett.",
    volume = "97",
    pages = "191302",
    year = "2006"
}

@article{Bousso:2007kq,
    author = "Bousso, Raphael and Harnik, Roni and Kribs, Graham D. and Perez, Gilad",
    title = "{Predicting the Cosmological Constant from the Causal Entropic Principle}",
    eprint = "hep-th/0702115",
    archivePrefix = "arXiv",
    reportNumber = "SLAC-PUB-12353, YITP-SB-07-04",
    doi = "10.1103/PhysRevD.76.043513",
    journal = "Phys. Rev. D",
    volume = "76",
    pages = "043513",
    year = "2007"
}

@misc{AxionLimits,
  author       = {Ciaran O'Hare},
  title        = {cajohare/AxionLimits: AxionLimits},
  month        = jul,
  year         = 2020,
  publisher    = {Zenodo},
  version      = {v1.0},
  doi          = {10.5281/zenodo.3932430},
  howpublished = {\url{https://cajohare.github.io/AxionLimits/}}
}

@article{Ouellet:2018beu,
    author = "Ouellet, Jonathan L. and others",
    title = "{First Results from ABRACADABRA-10 cm: A Search for Sub-$\mu$eV Axion Dark Matter}",
    eprint = "1810.12257",
    archivePrefix = "arXiv",
    primaryClass = "hep-ex",
    doi = "10.1103/PhysRevLett.122.121802",
    journal = "Phys. Rev. Lett.",
    volume = "122",
    number = "12",
    pages = "121802",
    year = "2019"
}

@article{Salemi:2021gck,
    author = "Salemi, Chiara P. and others",
    title = "{Search for Low-Mass Axion Dark Matter with ABRACADABRA-10~cm}",
    eprint = "2102.06722",
    archivePrefix = "arXiv",
    primaryClass = "hep-ex",
    doi = "10.1103/PhysRevLett.127.081801",
    journal = "Phys. Rev. Lett.",
    volume = "127",
    number = "8",
    pages = "081801",
    year = "2021"
}

@article{Pandey:2024dcd,
    author = "Pandey, Swadha and Hall, Evan D. and Evans, Matthew",
    title = "{First Results from the Axion Dark-Matter Birefringent Cavity (ADBC) Experiment}",
    eprint = "2404.12517",
    archivePrefix = "arXiv",
    primaryClass = "hep-ex",
    doi = "10.1103/PhysRevLett.133.111003",
    journal = "Phys. Rev. Lett.",
    volume = "133",
    number = "11",
    pages = "111003",
    year = "2024"
}

@article{ADMX:2009iij,
    author = "Asztalos, S. J. and others",
    collaboration = "ADMX",
    title = "{A SQUID-based microwave cavity search for dark-matter axions}",
    eprint = "0910.5914",
    archivePrefix = "arXiv",
    primaryClass = "astro-ph.CO",
    doi = "10.1103/PhysRevLett.104.041301",
    journal = "Phys. Rev. Lett.",
    volume = "104",
    pages = "041301",
    year = "2010"
}

@article{ADMX:2018gho,
    author = "Du, N. and others",
    collaboration = "ADMX",
    title = "{A Search for Invisible Axion Dark Matter with the Axion Dark Matter Experiment}",
    eprint = "1804.05750",
    archivePrefix = "arXiv",
    primaryClass = "hep-ex",
    reportNumber = "FERMILAB-PUB-18-101-AD-AE",
    doi = "10.1103/PhysRevLett.120.151301",
    journal = "Phys. Rev. Lett.",
    volume = "120",
    number = "15",
    pages = "151301",
    year = "2018"
}

@article{ADMX:2019uok,
    author = "Braine, T. and others",
    collaboration = "ADMX",
    title = "{Extended Search for the Invisible Axion with the Axion Dark Matter Experiment}",
    eprint = "1910.08638",
    archivePrefix = "arXiv",
    primaryClass = "hep-ex",
    reportNumber = "FERMILAB-PUB-19-569-AD-AE-PPD",
    doi = "10.1103/PhysRevLett.124.101303",
    journal = "Phys. Rev. Lett.",
    volume = "124",
    number = "10",
    pages = "101303",
    year = "2020"
}

@article{ADMX:2021nhd,
    author = "Bartram, C. and others",
    collaboration = "ADMX",
    title = "{Search for Invisible Axion Dark Matter in the 3.3{\textendash}4.2{\,}{\,}{\ensuremath{\mu}}eV Mass Range}",
    eprint = "2110.06096",
    archivePrefix = "arXiv",
    primaryClass = "hep-ex",
    reportNumber = "FERMILAB-PUB-21-774-DI-PPD-SQMS",
    doi = "10.1103/PhysRevLett.127.261803",
    journal = "Phys. Rev. Lett.",
    volume = "127",
    number = "26",
    pages = "261803",
    year = "2021"
}

@article{ADMX:2024xbv,
    author = "Goodman, C. and others",
    collaboration = "ADMX",
    title = "{ADMX Axion Dark Matter Bounds around 3.3{\,}{\,}{\ensuremath{\mu}}eV with Dine-Fischler-Srednicki-Zhitnitsky Discovery Ability}",
    eprint = "2408.15227",
    archivePrefix = "arXiv",
    primaryClass = "hep-ex",
    reportNumber = "FERMILAB-PUB-24-0602-V",
    doi = "10.1103/PhysRevLett.134.111002",
    journal = "Phys. Rev. Lett.",
    volume = "134",
    number = "11",
    pages = "111002",
    year = "2025"
}

@article{ADMX:2025vom,
    author = "Carosi, G. and others",
    collaboration = "ADMX",
    title = "{Search for Axion Dark Matter from 1.1 to 1.3~GHz with ADMX}",
    eprint = "2504.07279",
    archivePrefix = "arXiv",
    primaryClass = "hep-ex",
    reportNumber = "FERMILAB-PUB-25-0298-PPD",
    doi = "10.1103/d7mg-6sqq",
    journal = "Phys. Rev. Lett.",
    volume = "135",
    number = "19",
    pages = "191001",
    year = "2025"
}

@article{ADMX:2018ogs,
    author = "Boutan, C. and others",
    collaboration = "ADMX",
    title = "{Piezoelectrically Tuned Multimode Cavity Search for Axion Dark Matter}",
    eprint = "1901.00920",
    archivePrefix = "arXiv",
    primaryClass = "hep-ex",
    reportNumber = "FERMILAB-PUB-18-702-AD-AE",
    doi = "10.1103/PhysRevLett.121.261302",
    journal = "Phys. Rev. Lett.",
    volume = "121",
    number = "26",
    pages = "261302",
    year = "2018"
}

@article{ADMX:2021mio,
    author = "Bartram, C. and others",
    collaboration = "ADMX",
    title = "{Dark matter axion search using a Josephson Traveling wave parametric amplifier}",
    eprint = "2110.10262",
    archivePrefix = "arXiv",
    primaryClass = "hep-ex",
    reportNumber = "FERMILAB-PUB-22-615-PPD-SQMS",
    doi = "10.1063/5.0122907",
    journal = "Rev. Sci. Instrum.",
    volume = "94",
    number = "4",
    pages = "044703",
    year = "2023"
}

@article{Crisosto:2019fcj,
    author = "Crisosto, N. and Sikivie, P. and Sullivan, N. S. and Tanner, D. B. and Yang, J. and Rybka, G.",
    title = "{ADMX SLIC: Results from a Superconducting $LC$ Circuit Investigating Cold Axions}",
    eprint = "1911.05772",
    archivePrefix = "arXiv",
    primaryClass = "astro-ph.CO",
    doi = "10.1103/PhysRevLett.124.241101",
    journal = "Phys. Rev. Lett.",
    volume = "124",
    number = "24",
    pages = "241101",
    year = "2020"
}

@article{Devlin:2021fpq,
    author = "Devlin, Jack A. and others",
    title = "{Constraints on the Coupling between Axionlike Dark Matter and Photons Using an Antiproton Superconducting Tuned Detection Circuit in a Cryogenic Penning Trap}",
    eprint = "2101.11290",
    archivePrefix = "arXiv",
    primaryClass = "astro-ph.CO",
    doi = "10.1103/PhysRevLett.126.041301",
    journal = "Phys. Rev. Lett.",
    volume = "126",
    number = "4",
    pages = "041301",
    year = "2021"
}

@article{Lee:2020cfj,
    author = "Lee, S. and Ahn, S. and Choi, J. and Ko, B. R. and Semertzidis, Y. K.",
    title = "{Axion Dark Matter Search around 6.7 $\mu$eV}",
    eprint = "2001.05102",
    archivePrefix = "arXiv",
    primaryClass = "hep-ex",
    doi = "10.1103/PhysRevLett.124.101802",
    journal = "Phys. Rev. Lett.",
    volume = "124",
    number = "10",
    pages = "101802",
    year = "2020"
}

@article{Jeong:2020cwz,
    author = "Jeong, Junu and Youn, SungWoo and Bae, Sungjae and Kim, Jihngeun and Seong, Taehyeon and Kim, Jihn E. and Semertzidis, Yannis K.",
    title = "{Search for Invisible Axion Dark Matter with a Multiple-Cell Haloscope}",
    eprint = "2008.10141",
    archivePrefix = "arXiv",
    primaryClass = "hep-ex",
    doi = "10.1103/PhysRevLett.125.221302",
    journal = "Phys. Rev. Lett.",
    volume = "125",
    number = "22",
    pages = "221302",
    year = "2020"
}

@article{CAPP:2020utb,
    author = "Kwon, Ohjoon and others",
    collaboration = "CAPP",
    title = "{First Results from an Axion Haloscope at CAPP around 10.7  $\mu$eV}",
    eprint = "2012.10764",
    archivePrefix = "arXiv",
    primaryClass = "hep-ex",
    doi = "10.1103/PhysRevLett.126.191802",
    journal = "Phys. Rev. Lett.",
    volume = "126",
    number = "19",
    pages = "191802",
    year = "2021"
}

@article{Lee:2022mnc,
    author = "Lee, Youngjae and Yang, Byeongsu and Yoon, Hojin and Ahn, Moohyun and Park, Heejun and Min, Byeonghun and Kim, DongLak and Yoo, Jonghee",
    title = "{Searching for Invisible Axion Dark Matter with an 18~T Magnet Haloscope}",
    eprint = "2206.08845",
    archivePrefix = "arXiv",
    primaryClass = "hep-ex",
    doi = "10.1103/PhysRevLett.128.241805",
    journal = "Phys. Rev. Lett.",
    volume = "128",
    number = "24",
    pages = "241805",
    year = "2022"
}

@article{Yoon:2022gzp,
    author = "Yoon, Hojin and Ahn, Moohyun and Yang, Byeongsu and Lee, Youngjae and Kim, DongLak and Park, Heejun and Min, Byeonghun and Yoo, Jonghee",
    title = "{Axion haloscope using an 18~T high temperature superconducting magnet}",
    eprint = "2206.12271",
    archivePrefix = "arXiv",
    primaryClass = "hep-ex",
    doi = "10.1103/PhysRevD.106.092007",
    journal = "Phys. Rev. D",
    volume = "106",
    number = "9",
    pages = "092007",
    year = "2022"
}

@article{Kim:2022hmg,
    author = "Kim, Jinsu and others",
    title = "{Near-Quantum-Noise Axion Dark Matter Search at CAPP around 9.5{\,}{\,}{\ensuremath{\mu}}eV}",
    eprint = "2207.13597",
    archivePrefix = "arXiv",
    primaryClass = "hep-ex",
    doi = "10.1103/PhysRevLett.130.091602",
    journal = "Phys. Rev. Lett.",
    volume = "130",
    number = "9",
    pages = "091602",
    year = "2023"
}

@article{Yi:2022fmn,
    author = "Yi, Andrew K. and others",
    title = "{Axion Dark Matter Search around 4.55{\,}{\,}{\ensuremath{\mu}}eV with Dine-Fischler-Srednicki-Zhitnitskii Sensitivity}",
    eprint = "2210.10961",
    archivePrefix = "arXiv",
    primaryClass = "hep-ex",
    doi = "10.1103/PhysRevLett.130.071002",
    journal = "Phys. Rev. Lett.",
    volume = "130",
    number = "7",
    pages = "071002",
    year = "2023"
}

@article{Yang:2023yry,
    author = "Yang, Byeongsu and Yoon, Hojin and Ahn, Moohyun and Lee, Youngjae and Yoo, Jonghee",
    title = "{Extended Axion Dark Matter Search Using the CAPP18T Haloscope}",
    eprint = "2308.09077",
    archivePrefix = "arXiv",
    primaryClass = "hep-ex",
    doi = "10.1103/PhysRevLett.131.081801",
    journal = "Phys. Rev. Lett.",
    volume = "131",
    number = "8",
    pages = "081801",
    year = "2023"
}

@article{Kim:2023vpo,
    author = "Kim, Younggeun and others",
    title = "{Experimental Search for Invisible Dark Matter Axions around 22{\,}{\,}{\ensuremath{\mu}}eV}",
    eprint = "2312.11003",
    archivePrefix = "arXiv",
    primaryClass = "hep-ex",
    doi = "10.1103/PhysRevLett.133.051802",
    journal = "Phys. Rev. Lett.",
    volume = "133",
    number = "5",
    pages = "051802",
    year = "2024"
}

@article{CAPP:2024dtx,
    author = "Ahn, Saebyeok and others",
    collaboration = "CAPP",
    title = "{Extensive Search for Axion Dark Matter over 1~GHz with CAPP{\textquoteright}S Main Axion Experiment}",
    eprint = "2402.12892",
    archivePrefix = "arXiv",
    primaryClass = "hep-ex",
    doi = "10.1103/PhysRevX.14.031023",
    journal = "Phys. Rev. X",
    volume = "14",
    number = "3",
    pages = "031023",
    year = "2024"
}

@article{Bae:2024kmy,
    author = "Bae, Sungjae and Jeong, Junu and Kim, Younggeun and Youn, SungWoo and Park, Heejun and Seong, Taehyeon and Oh, Seongjeong and Semertzidis, Yannis K.",
    title = "{Search for Dark Matter Axions with Tunable TM020 Mode}",
    eprint = "2403.13390",
    archivePrefix = "arXiv",
    primaryClass = "hep-ex",
    doi = "10.1103/PhysRevLett.133.211803",
    journal = "Phys. Rev. Lett.",
    volume = "133",
    number = "21",
    pages = "211803",
    year = "2024"
}

@article{Adair:2022rtw,
    author = "Adair, C. M. and others",
    title = "{Search for Dark Matter Axions with CAST-CAPP}",
    eprint = "2211.02902",
    archivePrefix = "arXiv",
    primaryClass = "hep-ex",
    doi = "10.1038/s41467-022-33913-6",
    journal = "Nature Commun.",
    volume = "13",
    number = "1",
    pages = "6180",
    year = "2022"
}

@article{Oshima:2023csb,
    author = "Oshima, Yuka and Fujimoto, Hiroki and Kume, Jun'ya and Morisaki, Soichiro and Nagano, Koji and Fujita, Tomohiro and Obata, Ippei and Nishizawa, Atsushi and Michimura, Yuta and Ando, Masaki",
    title = "{First results of axion dark matter search with DANCE}",
    eprint = "2303.03594",
    archivePrefix = "arXiv",
    primaryClass = "hep-ex",
    reportNumber = "RESCEU-4/23",
    doi = "10.1103/PhysRevD.108.072005",
    journal = "Phys. Rev. D",
    volume = "108",
    number = "7",
    pages = "072005",
    year = "2023"
}

@article{Nishizawa:2025xka,
    author = "Nishizawa, Atsushi and Taruya, Atsushi and Himemoto, Yoshiaki",
    title = "{Axion dark matter search from terrestrial magnetic fields at extremely low frequencies}",
    eprint = "2504.07559",
     journal= " ",
    archivePrefix = "arXiv",
    primaryClass = "hep-ph",
    reportNumber = "YITP-25-55",
    month = "4",
    year = "2025"
}

@article{GigaBREAD:2025lzq,
    author = "Hoshino, Gabe and others",
    collaboration = "GigaBREAD",
    title = "{First Axionlike Particle Results from a Broadband Search for Wavelike Dark Matter in the 44 to 52{\,}{\,}{\ensuremath{\mu}}eV Range with a Coaxial Dish Antenna}",
    eprint = "2501.17119",
    archivePrefix = "arXiv",
    primaryClass = "hep-ex",
    reportNumber = "FERMILAB-PUB-25-0076-PPD",
    doi = "10.1103/PhysRevLett.134.171002",
    journal = "Phys. Rev. Lett.",
    volume = "134",
    number = "17",
    pages = "171002",
    year = "2025"
}

@article{Grenet:2021vbb,
    author = "Grenet, Thierry and Ballou, Rafik and Basto, Quentin and Martineau, Killian and Perrier, Pierre and Pugnat, Pierre and Quevillon, J{\'e}r{\'e}mie and Roch, Nicolas and Smith, Christopher",
    title = "{The Grenoble Axion Haloscope platform (GrAHal): development plan and first results}",
    eprint = "2110.14406",
     journal= " ",
    archivePrefix = "arXiv",
    primaryClass = "hep-ex",
    month = "10",
    year = "2021"
}

@article{Brubaker:2016ktl,
    author = "Brubaker, B. M. and others",
    title = "{First results from a microwave cavity axion search at 24 $\mu$eV}",
    eprint = "1610.02580",
    archivePrefix = "arXiv",
    primaryClass = "astro-ph.CO",
    doi = "10.1103/PhysRevLett.118.061302",
    journal = "Phys. Rev. Lett.",
    volume = "118",
    number = "6",
    pages = "061302",
    year = "2017"
}

@article{HAYSTAC:2018rwy,
    author = "Zhong, L. and others",
    collaboration = "HAYSTAC",
    title = "{Results from phase 1 of the HAYSTAC microwave cavity axion experiment}",
    eprint = "1803.03690",
    archivePrefix = "arXiv",
    primaryClass = "hep-ex",
    doi = "10.1103/PhysRevD.97.092001",
    journal = "Phys. Rev. D",
    volume = "97",
    number = "9",
    pages = "092001",
    year = "2018"
}

@article{HAYSTAC:2020kwv,
    author = "Backes, K. M. and others",
    collaboration = "HAYSTAC",
    title = "{A quantum-enhanced search for dark matter axions}",
    eprint = "2008.01853",
    archivePrefix = "arXiv",
    primaryClass = "quant-ph",
    doi = "10.1038/s41586-021-03226-7",
    journal = "Nature",
    volume = "590",
    number = "7845",
    pages = "238--242",
    year = "2021"
}

@article{HAYSTAC:2023cam,
    author = "Jewell, M. J. and others",
    collaboration = "HAYSTAC",
    title = "{New results from HAYSTAC{\textquoteright}s phase II operation with a squeezed state receiver}",
    eprint = "2301.09721",
    archivePrefix = "arXiv",
    primaryClass = "hep-ex",
    doi = "10.1103/PhysRevD.107.072007",
    journal = "Phys. Rev. D",
    volume = "107",
    number = "7",
    pages = "072007",
    year = "2023"
}

@article{HAYSTAC:2024jch,
    author = "Bai, Xiran and others",
    collaboration = "HAYSTAC",
    title = "{Dark Matter Axion Search with HAYSTAC Phase II}",
    eprint = "2409.08998",
    archivePrefix = "arXiv",
    primaryClass = "hep-ex",
    doi = "10.1103/PhysRevLett.134.151006",
    journal = "Phys. Rev. Lett.",
    volume = "134",
    number = "15",
    pages = "151006",
    year = "2025"
}

@article{Heinze:2023nfb,
    author = "Heinze, Joscha and Gill, Alex and Dmitriev, Artemiy and Smetana, Jiri and Yan, Tianliang and Boyer, Vincent and Martynov, Denis and Evans, Matthew",
    title = "{First Results of the Laser-Interferometric Detector for Axions (LIDA)}",
    eprint = "2307.01365",
    archivePrefix = "arXiv",
    primaryClass = "astro-ph.CO",
    doi = "10.1103/PhysRevLett.132.191002",
    journal = "Phys. Rev. Lett.",
    volume = "132",
    number = "19",
    pages = "191002",
    year = "2024"
}

@article{MADMAX:2024sxs,
    author = "Garcia, B. Ary dos Santos and others",
    collaboration = "MADMAX",
    title = "{First Search for Axion Dark Matter with a MADMAX Prototype}",
    eprint = "2409.11777",
    archivePrefix = "arXiv",
    primaryClass = "hep-ex",
    reportNumber = "FERMILAB-PUB-24-0095-PPD",
    doi = "10.1103/c749-419q",
    journal = "Phys. Rev. Lett.",
    volume = "135",
    number = "4",
    pages = "041001",
    year = "2025"
}

@article{McAllister:2017lkb,
    author = "McAllister, Ben T. and Flower, Graeme and Kruger, Justin and Ivanov, Eugene N. and Goryachev, Maxim and Bourhill, Jeremy and Tobar, Michael E.",
    collaboration = "ORGAN",
    title = "{The ORGAN Experiment: An axion haloscope above 15 GHz}",
    eprint = "1706.00209",
    archivePrefix = "arXiv",
    primaryClass = "physics.ins-det",
    doi = "10.1016/j.dark.2017.09.010",
    journal = "Phys. Dark Univ.",
    volume = "18",
    pages = "67--72",
    year = "2017"
}

@article{Quiskamp:2022pks,
    author = "Quiskamp, Aaron P. and McAllister, Ben T. and Altin, Paul and Ivanov, Eugene N. and Goryachev, Maxim and Tobar, Michael E.",
    collaboration = "ORGAN",
    title = "{Direct search for dark matter axions excluding ALP cogenesis in the 63- to 67-{\ensuremath{\mu}}eV range with the ORGAN experiment}",
    eprint = "2203.12152",
    archivePrefix = "arXiv",
    primaryClass = "hep-ex",
    doi = "10.1126/sciadv.abq3765",
    journal = "Sci. Adv.",
    volume = "8",
    number = "27",
    pages = "abq3765",
    year = "2022"
}

@article{Quiskamp:2023ehr,
    author = "Quiskamp, Aaron and McAllister, Ben T. and Altin, Paul and Ivanov, Eugene N. and Goryachev, Maxim and Tobar, Michael E.",
    title = "{Exclusion of Axionlike-Particle Cogenesis Dark Matter in a Mass Window above 100{\,}{\,}{\ensuremath{\mu}}eV}",
    eprint = "2310.00904",
    archivePrefix = "arXiv",
    primaryClass = "hep-ex",
    doi = "10.1103/PhysRevLett.132.031601",
    journal = "Phys. Rev. Lett.",
    volume = "132",
    number = "3",
    pages = "031601",
    year = "2024"
}

@article{Quiskamp:2024oet,
    author = "Quiskamp, Aaron P. and Flower, Graeme R. and Samuels, Steven and McAllister, Ben T. and Altin, Paul and Ivanov, Eugene N. and Goryachev, Maxim and Tobar, Michael E.",
    collaboration = "ORGAN",
    title = "{Near-quantum-limited axion dark matter search with the ORGAN experiment around 26{\,}{\,}{\ensuremath{\mu}}eV}",
    eprint = "2407.18586",
    archivePrefix = "arXiv",
    primaryClass = "hep-ex",
    doi = "10.1103/PhysRevD.111.095007",
    journal = "Phys. Rev. D",
    volume = "111",
    number = "9",
    pages = "095007",
    year = "2025"
}

@article{Alesini:2019ajt,
    author = "Alesini, D. and others",
    title = "{Galactic axions search with a superconducting resonant cavity}",
    eprint = "1903.06547",
    archivePrefix = "arXiv",
    primaryClass = "physics.ins-det",
    doi = "10.1103/PhysRevD.99.101101",
    journal = "Phys. Rev. D",
    volume = "99",
    number = "10",
    pages = "101101",
    year = "2019"
}

@article{Alesini:2020vny,
    author = "Alesini, D. and others",
    title = "{Search for invisible axion dark matter of mass m$_a=43~\mu$eV with the QUAX--$a\gamma$ experiment}",
    eprint = "2012.09498",
    archivePrefix = "arXiv",
    primaryClass = "hep-ex",
    doi = "10.1103/PhysRevD.103.102004",
    journal = "Phys. Rev. D",
    volume = "103",
    number = "10",
    pages = "102004",
    year = "2021"
}

@article{Alesini:2022lnp,
    author = "Alesini, D. and others",
    title = "{Search for Galactic axions with a high-Q dielectric cavity}",
    eprint = "2208.12670",
    archivePrefix = "arXiv",
    primaryClass = "hep-ex",
    doi = "10.1103/PhysRevD.106.052007",
    journal = "Phys. Rev. D",
    volume = "106",
    number = "5",
    pages = "052007",
    year = "2022"
}

@article{QUAX:2023gop,
    author = "Di Vora, R. and others",
    collaboration = "QUAX",
    title = "{Search for galactic axions with a traveling wave parametric amplifier}",
    eprint = "2304.07505",
    archivePrefix = "arXiv",
    primaryClass = "hep-ex",
    reportNumber = "FERMILAB-PUB-23-229-SQMS-V",
    doi = "10.1103/PhysRevD.108.062005",
    journal = "Phys. Rev. D",
    volume = "108",
    number = "6",
    pages = "062005",
    year = "2023"
}

@article{QUAX:2024fut,
    author = "Rettaroli, A. and others",
    collaboration = "QUAX",
    title = "{Search for axion dark matter with the QUAX{\textendash}LNF tunable haloscope}",
    eprint = "2402.19063",
    archivePrefix = "arXiv",
    primaryClass = "physics.ins-det",
    reportNumber = "FERMILAB-PUB-24-0511-SQMS-V",
    doi = "10.1103/PhysRevD.110.022008",
    journal = "Phys. Rev. D",
    volume = "110",
    number = "2",
    pages = "022008",
    year = "2024"
}

@article{CAST:2020rlf,
    author = "Melc{\'o}n, A. {\'A}lvarez and others",
    collaboration = "CAST",
    title = "{First results of the CAST-RADES haloscope search for axions at 34.67 $\mu$eV}",
    eprint = "2104.13798",
    archivePrefix = "arXiv",
    primaryClass = "hep-ex",
    reportNumber = "CERN-EP-2021-070",
    doi = "10.1007/JHEP10(2021)075",
    journal = "JHEP",
    volume = "10",
    pages = "075",
    year = "2020"
}

@article{Ahyoune:2024klt,
    author = "Ahyoune, S. and others",
    title = "{RADES axion search results with a high-temperature superconducting cavity in an 11.7 T magnet}",
    eprint = "2403.07790",
    archivePrefix = "arXiv",
    primaryClass = "hep-ex",
    reportNumber = "CERN-EP-2024-076, MPP-2024-55",
    doi = "10.1007/JHEP04(2025)113",
    journal = "JHEP",
    volume = "04",
    pages = "113",
    year = "2025"
}

@article{DePanfilis:1987dk,
    author = "De Panfilis, S. and Melissinos, A. C. and Moskowitz, B. E. and Rogers, J. T. and Semertzidis, Y. K. and Wuensch, Walter and Halama, H. J. and Prodell, A. G. and Fowler, W. B. and Nezrick, F. A.",
    title = "{Limits on the Abundance and Coupling of Cosmic Axions at 4.5-Microev {\ensuremath{<}} m(a) {\ensuremath{<}} 5.0-Microev}",
    reportNumber = "UR-994, FERMILAB-PUB-87-046, ER13065-482",
    doi = "10.1103/PhysRevLett.59.839",
    journal = "Phys. Rev. Lett.",
    volume = "59",
    pages = "839",
    year = "1987"
}

@article{Wuensch:1989sa,
    author = "Wuensch, Walter and De Panfilis-Wuensch, S. and Semertzidis, Y. K. and Rogers, J. T. and Melissinos, A. C. and Halama, H. J. and Moskowitz, B. E. and Prodell, A. G. and Fowler, W. B. and Nezrick, F. A.",
    title = "{Results of a Laboratory Search for Cosmic Axions and Other Weakly Coupled Light Particles}",
    reportNumber = "FERMILAB-PUB-89-185-E, BNL-43010",
    doi = "10.1103/PhysRevD.40.3153",
    journal = "Phys. Rev. D",
    volume = "40",
    pages = "3153",
    year = "1989"
}

@article{Gramolin:2020ict,
    author = "Gramolin, Alexander V. and Aybas, Deniz and Johnson, Dorian and Adam, Janos and Sushkov, Alexander O.",
    title = "{Search for axion-like dark matter with ferromagnets}",
    eprint = "2003.03348",
    archivePrefix = "arXiv",
    primaryClass = "hep-ex",
    doi = "10.1038/s41567-020-1006-6",
    journal = "Nature Phys.",
    volume = "17",
    number = "1",
    pages = "79--84",
    year = "2021"
}

@article{Sulai:2023zqw,
    author = "Sulai, Ibrahim A. and others",
    title = "{Hunt for magnetic signatures of hidden-photon and axion dark matter in the wilderness}",
    eprint = "2306.11575",
    archivePrefix = "arXiv",
    primaryClass = "hep-ph",
    reportNumber = "FERMILAB-PUB-23-394-SQMS-V",
    doi = "10.1103/PhysRevD.108.096026",
    journal = "Phys. Rev. D",
    volume = "108",
    number = "9",
    pages = "096026",
    year = "2023"
}

@article{Arza:2021ekq,
    author = "Arza, Ariel and Fedderke, Michael A. and Graham, Peter W. and Kimball, Derek F. Jackson and Kalia, Saarik",
    title = "{Earth as a transducer for axion dark-matter detection}",
    eprint = "2112.09620",
    archivePrefix = "arXiv",
    primaryClass = "hep-ph",
    reportNumber = "FERMILAB-PUB-23-653-SQMS-V",
    doi = "10.1103/PhysRevD.105.095007",
    journal = "Phys. Rev. D",
    volume = "105",
    number = "9",
    pages = "095007",
    year = "2022"
}

@article{Friel:2024shg,
    author = "Friel, Matt and Gjerloev, Jesper W. and Kalia, Saarik and Zamora, Alvaro",
    title = "{Search for ultralight dark matter in the SuperMAG high-fidelity dataset}",
    eprint = "2408.16045",
    archivePrefix = "arXiv",
    primaryClass = "hep-ph",
    reportNumber = "FERMILAB-PUB-24-0546-SQMS",
    doi = "10.1103/PhysRevD.110.115036",
    journal = "Phys. Rev. D",
    volume = "110",
    number = "11",
    pages = "115036",
    year = "2024"
}

@article{TASEH:2022vvu,
    author = "Chang, Hsin and others",
    collaboration = "TASEH",
    title = "{First Results from the Taiwan Axion Search Experiment with a Haloscope at 19.6{\,}{\,}{\ensuremath{\mu}}eV}",
    eprint = "2205.05574",
    archivePrefix = "arXiv",
    primaryClass = "hep-ex",
    doi = "10.1103/PhysRevLett.129.111802",
    journal = "Phys. Rev. Lett.",
    volume = "129",
    number = "11",
    pages = "111802",
    year = "2022"
}

@article{Hagmann:1990tj,
    author = "Hagmann, C. and Sikivie, P. and Sullivan, N. S. and Tanner, D. B.",
    title = "{Results from a search for cosmic axions}",
    reportNumber = "PRINT-90-0420 (FLORIDA)",
    doi = "10.1103/PhysRevD.42.1297",
    journal = "Phys. Rev. D",
    volume = "42",
    pages = "1297--1300",
    year = "1990"
}

@article{Hagmann:1996qd,
    author = "Hagmann, C. and others",
    editor = "Cline, D. B.",
    title = "{First results from a second generation galactic axion experiment}",
    eprint = "astro-ph/9607022",
    archivePrefix = "arXiv",
    reportNumber = "UCRL-JC-124162",
    doi = "10.1016/S0920-5632(96)00516-6",
    journal = "Nucl. Phys. B Proc. Suppl.",
    volume = "51",
    pages = "209--212",
    year = "1996"
}

@article{Thomson:2021zvq,
    author = "Thomson, Catriona A. and McAllister, Ben T. and Goryachev, Maxim and Ivanov, Eugene N. and Tobar, Michael E.",
    title = "{Upconversion Loop Oscillator Axion Detection Experiment: A Precision Frequency Interferometric Axion Dark Matter Search with a Cylindrical Microwave Cavity}",
    eprint = "1912.07751",
    archivePrefix = "arXiv",
    primaryClass = "hep-ex",
    reportNumber = "Erratum: Phys. Rev. Lett. 127, 019901 (2021)",
    doi = "10.1103/PhysRevLett.127.019901",
    journal = "Phys. Rev. Lett.",
    volume = "126",
    number = "8",
    pages = "081803",
    year = "2021",
    note = "[Erratum: Phys.Rev.Lett. 127, 019901 (2021)]"
}

@article{Thomson:2023moc,
    author = "Thomson, Catriona A. and Goryachev, Maxim and McAllister, Ben T. and Ivanov, Eugene N. and Altin, Paul and Tobar, Michael E.",
    title = "{Searching for low-mass axions using resonant upconversion}",
    eprint = "2301.06778",
    archivePrefix = "arXiv",
    primaryClass = "hep-ex",
    doi = "10.1103/PhysRevD.107.112003",
    journal = "Phys. Rev. D",
    volume = "107",
    number = "11",
    pages = "112003",
    year = "2023"
}

@article{Liu:2018icu,
    author = "Liu, Hongwan and Elwood, Brodi D. and Evans, Matthew and Thaler, Jesse",
    title = "{Searching for Axion Dark Matter with Birefringent Cavities}",
    eprint = "1809.01656",
    archivePrefix = "arXiv",
    primaryClass = "hep-ph",
    reportNumber = "MIT-CTP/5048",
    doi = "10.1103/PhysRevD.100.023548",
    journal = "Phys. Rev. D",
    volume = "100",
    number = "2",
    pages = "023548",
    year = "2019"
}

@article{Stern:2016bbw,
    author = "Stern, I.",
    title = "{ADMX Status}",
    eprint = "1612.08296",
    archivePrefix = "arXiv",
    primaryClass = "physics.ins-det",
    doi = "10.22323/1.282.0198",
    journal = "PoS",
    volume = "ICHEP2016",
    pages = "198",
    year = "2016"
}

@article{Nagano:2019rbw,
    author = "Nagano, Koji and Fujita, Tomohiro and Michimura, Yuta and Obata, Ippei",
    title = "{Axion Dark Matter Search with Interferometric Gravitational Wave Detectors}",
    eprint = "1903.02017",
    archivePrefix = "arXiv",
    primaryClass = "hep-ph",
    doi = "10.1103/PhysRevLett.123.111301",
    journal = "Phys. Rev. Lett.",
    volume = "123",
    number = "11",
    pages = "111301",
    year = "2019"
}

@article{Lawson:2019brd,
    author = "Lawson, Matthew and Millar, Alexander J. and Pancaldi, Matteo and Vitagliano, Edoardo and Wilczek, Frank",
    title = "{Tunable axion plasma haloscopes}",
    eprint = "1904.11872",
    archivePrefix = "arXiv",
    primaryClass = "hep-ph",
    reportNumber = "NORDITA-2019-038, MIT-CTP-5116, MPP-2019-83",
    doi = "10.1103/PhysRevLett.123.141802",
    journal = "Phys. Rev. Lett.",
    volume = "123",
    number = "14",
    pages = "141802",
    year = "2019"
}

@article{Ahyoune:2023gfw,
    author = "Ahyoune, Saiyd and others",
    title = "{A Proposal for a Low-Frequency Axion Search in the 1{\textendash}2 $\mu$ eV Range and Below with the BabyIAXO Magnet}",
    eprint = "2306.17243",
    archivePrefix = "arXiv",
    primaryClass = "physics.ins-det",
    doi = "10.1002/andp.202300326",
    journal = "Annalen Phys.",
    volume = "535",
    number = "12",
    pages = "2300326",
    year = "2023"
}

@article{BREAD:2021tpx,
    author = "Liu, Jesse and others",
    collaboration = "BREAD",
    title = "{Broadband Solenoidal Haloscope for Terahertz Axion Detection}",
    eprint = "2111.12103",
    archivePrefix = "arXiv",
    primaryClass = "physics.ins-det",
    reportNumber = "FERMILAB-PUB-21-694-AD-PPD-TD",
    doi = "10.1103/PhysRevLett.128.131801",
    journal = "Phys. Rev. Lett.",
    volume = "128",
    number = "13",
    pages = "131801",
    year = "2022"
}

@article{Aja:2022csb,
    author = "Aja, Beatriz and others",
    title = "{The Canfranc Axion Detection Experiment (CADEx): search for axions at 90 GHz with Kinetic Inductance Detectors}",
    eprint = "2206.02980",
    archivePrefix = "arXiv",
    primaryClass = "hep-ex",
    doi = "10.1088/1475-7516/2022/11/044",
    journal = "JCAP",
    volume = "11",
    pages = "044",
    year = "2022"
}

@article{DeMiguel:2023nmz,
    author = "De Miguel, Javier and Hern{\'a}ndez-Cabrera, Juan F. and Hern{\'a}ndez-Su{\'a}rez, Elvio and Joven-{\'A}lvarez, Enrique and Otani, Chiko and Rubi{\~n}o-Mart{\'\i}n, J. Alberto",
    collaboration = "DALI",
    title = "{Discovery prospects with the Dark-photons {\&} Axion-like particles Interferometer}",
    eprint = "2303.03997",
    archivePrefix = "arXiv",
    primaryClass = "hep-ph",
    doi = "10.1103/PhysRevD.109.062002",
    journal = "Phys. Rev. D",
    volume = "109",
    number = "6",
    pages = "062002",
    year = "2024"
}

@article{Obata:2018vvr,
    author = "Obata, Ippei and Fujita, Tomohiro and Michimura, Yuta",
    title = "{Optical Ring Cavity Search for Axion Dark Matter}",
    eprint = "1805.11753",
    archivePrefix = "arXiv",
    primaryClass = "astro-ph.CO",
    doi = "10.1103/PhysRevLett.121.161301",
    journal = "Phys. Rev. Lett.",
    volume = "121",
    number = "16",
    pages = "161301",
    year = "2018"
}

@article{Heinze:2024bdc,
    author = "Heinze, Joscha and others",
    title = "{DarkGEO: a large-scale laser-interferometric axion detector}",
    eprint = "2401.11907",
    archivePrefix = "arXiv",
    primaryClass = "astro-ph.CO",
    doi = "10.1088/1367-2630/ad48ac",
    journal = "New J. Phys.",
    volume = "26",
    number = "5",
    pages = "055002",
    year = "2024"
}

@article{DMRadio:2022pkf,
    author = "Brouwer, L. and others",
    collaboration = "DMRadio",
    title = "{Projected sensitivity of DMRadio-m3: A search for the QCD axion below 1{\,}{\,}{\ensuremath{\mu}}eV}",
    eprint = "2204.13781",
    archivePrefix = "arXiv",
    primaryClass = "hep-ex",
    doi = "10.1103/PhysRevD.106.103008",
    journal = "Phys. Rev. D",
    volume = "106",
    number = "10",
    pages = "103008",
    year = "2022"
}

@article{Fan:2024mhm,
    author = "Fan, Xing and Gabrielse, Gerald and Graham, Peter W. and Ramani, Harikrishnan and Wong, Samuel S. Y. and Xiao, Yawen",
    title = "{Highly excited electron cyclotron for QCD axion and dark-photon detection}",
    eprint = "2410.05549",
    archivePrefix = "arXiv",
    primaryClass = "hep-ph",
    reportNumber = "FERMILAB-PUB-24-0893-SQMS-V",
    doi = "10.1103/PhysRevD.111.075022",
    journal = "Phys. Rev. D",
    volume = "111",
    number = "7",
    pages = "075022",
    year = "2025"
}

@article{Alesini:2023qed,
    author = "Alesini, David and others",
    title = "{The future search for low-frequency axions and new physics with the FLASH resonant cavity experiment at Frascati National Laboratories}",
    eprint = "2309.00351",
    archivePrefix = "arXiv",
    primaryClass = "physics.ins-det",
    reportNumber = "CA21106; CA21136",
    doi = "10.1016/j.dark.2023.101370",
    journal = "Phys. Dark Univ.",
    volume = "42",
    pages = "101370",
    year = "2023"
}

@article{Baryakhtar:2018doz,
    author = "Baryakhtar, Masha and Huang, Junwu and Lasenby, Robert",
    title = "{Axion and hidden photon dark matter detection with multilayer optical haloscopes}",
    eprint = "1803.11455",
    archivePrefix = "arXiv",
    primaryClass = "hep-ph",
    doi = "10.1103/PhysRevD.98.035006",
    journal = "Phys. Rev. D",
    volume = "98",
    number = "3",
    pages = "035006",
    year = "2018"
}

@article{Beurthey:2020yuq,
    author = "Beurthey, S. and others",
    title = "{MADMAX Status Report}",
    eprint = "2003.10894",
    archivePrefix = "arXiv",
    journal= " ",
    primaryClass = "physics.ins-det",
    month = "3",
    year = "2020"
}

@article{Berlin:2020vrk,
    author = "Berlin, Asher and D'Agnolo, Raffaele Tito and Ellis, Sebastian A. R. and Zhou, Kevin",
    title = "{Heterodyne broadband detection of axion dark matter}",
    eprint = "2007.15656",
    archivePrefix = "arXiv",
    primaryClass = "hep-ph",
    doi = "10.1103/PhysRevD.104.L111701",
    journal = "Phys. Rev. D",
    volume = "104",
    number = "11",
    pages = "L111701",
    year = "2021"
}

@article{Qiu:2025bbi,
    author = "Qiu, Jian-Xiang and others",
    title = "{Observation of the axion quasiparticle in 2D MnBi$_{2}$Te$_{4}$}",
    eprint = "2504.12572",
    archivePrefix = "arXiv",
    primaryClass = "cond-mat.mes-hall",
    doi = "10.1038/s41586-025-08862-x",
    journal = "Nature",
    volume = "641",
    number = "8061",
    pages = "62--69",
    year = "2025"
}

@article{Bourhill:2022alm,
    author = "Bourhill, J. F. and Paterson, E. C. I. and Goryachev, M. and Tobar, M. E.",
    title = "{Searching for ultralight axions with twisted cavity resonators of anyon rotational symmetry with bulk modes of nonzero helicity}",
    eprint = "2208.01640",
    archivePrefix = "arXiv",
    primaryClass = "hep-ph",
    doi = "10.1103/PhysRevD.108.052014",
    journal = "Phys. Rev. D",
    volume = "108",
    number = "5",
    pages = "052014",
    year = "2023"
}

@article{Zhang:2021bpa,
    author = "Zhang, Zhongyue and Horns, Dieter and Ghosh, Oindrila",
    title = "{Search for dark matter with an LC circuit}",
    eprint = "2111.04541",
    archivePrefix = "arXiv",
    primaryClass = "hep-ex",
    doi = "10.1103/PhysRevD.106.023003",
    journal = "Phys. Rev. D",
    volume = "106",
    number = "2",
    pages = "023003",
    year = "2022"
}

@article{Essig:2013goa,
    author = "Essig, Rouven and Kuflik, Eric and McDermott, Samuel D. and Volansky, Tomer and Zurek, Kathryn M.",
    title = "{Constraining Light Dark Matter with Diffuse X-Ray and Gamma-Ray Observations}",
    eprint = "1309.4091",
    archivePrefix = "arXiv",
    primaryClass = "hep-ph",
    reportNumber = "YITP-SB-29-13, FERMILAB-PUB-13-377-A-T, MCTP-13-27",
    doi = "10.1007/JHEP11(2013)193",
    journal = "JHEP",
    volume = "11",
    pages = "193",
    year = "2013"
}

@article{Roach:2022lgo,
    author = "Roach, Brandon M. and Rossland, Steven and Ng, Kenny C. Y. and Perez, Kerstin and Beacom, John F. and Grefenstette, Brian W. and Horiuchi, Shunsaku and Krivonos, Roman and Wik, Daniel R.",
    title = "{Long-exposure NuSTAR constraints on decaying dark matter in the Galactic halo}",
    eprint = "2207.04572",
    archivePrefix = "arXiv",
    primaryClass = "astro-ph.HE",
    doi = "10.1103/PhysRevD.107.023009",
    journal = "Phys. Rev. D",
    volume = "107",
    number = "2",
    pages = "023009",
    year = "2023"
}

@article{Candon:2024eah,
    author = "Cand{\'o}n, Francisco R. and Fiorillo, Damiano F. G. and Lucente, Giuseppe and Vitagliano, Edoardo and Vogel, Julia K.",
    title = "{NuSTAR Bounds on Radiatively Decaying Particles from M82}",
    eprint = "2412.03660",
    archivePrefix = "arXiv",
    primaryClass = "hep-ph",
    doi = "10.1103/PhysRevLett.134.171004",
    journal = "Phys. Rev. Lett.",
    volume = "134",
    number = "17",
    pages = "171004",
    year = "2025"
}

@article{Porras-Bedmar:2024uql,
    author = "Porras-Bedmar, S. and Meyer, M. and Horns, D.",
    title = "{Novel bounds on decaying axionlike particle dark matter from the cosmic background}",
    eprint = "2407.10618",
    archivePrefix = "arXiv",
    primaryClass = "astro-ph.CO",
    doi = "10.1103/PhysRevD.110.103501",
    journal = "Phys. Rev. D",
    volume = "110",
    pages = "103501",
    month = "7",
    year = "2024"
}

@article{Liu:2023nct,
    author = "Liu, Hongwan and Qin, Wenzer and Ridgway, Gregory W. and Slatyer, Tracy R.",
    title = "{Exotic energy injection in the early Universe. II. CMB spectral distortions and constraints on light dark matter}",
    eprint = "2303.07370",
    archivePrefix = "arXiv",
    primaryClass = "astro-ph.CO",
    reportNumber = "MIT-CTP/5524",
    doi = "10.1103/PhysRevD.108.043531",
    journal = "Phys. Rev. D",
    volume = "108",
    number = "4",
    pages = "043531",
    year = "2023"
}

@article{Capozzi:2023xie,
    author = "Capozzi, Francesco and Ferreira, Ricardo Z. and Lopez-Honorez, Laura and Mena, Olga",
    title = "{CMB and Lyman-{\ensuremath{\alpha}} constraints on dark matter decays to photons}",
    eprint = "2303.07426",
    archivePrefix = "arXiv",
    primaryClass = "astro-ph.CO",
    reportNumber = "ULB-TH/23-03",
    doi = "10.1088/1475-7516/2023/06/060",
    journal = "JCAP",
    volume = "06",
    pages = "060",
    year = "2023"
}

@article{Cyr:2024sbd,
    author = "Cyr, Bryce and Chluba, Jens and Manoj, Pranav Bharadwaj Gangrekalve",
    title = "{Revisiting Constraints on Resonant Axion-Photon Conversions from CMB Spectral Distortions}",
    eprint = "2411.13701",
    journal= " ",
    archivePrefix = "arXiv",
    primaryClass = "astro-ph.CO",
    month = "11",
    year = "2024"
}

@article{Fong:2024qeq,
    author = "Fong, Chingam and Ng, Kenny C. Y. and Liu, Qishan",
    title = "{Searching for particle dark matter with eROSITA early data}",
    eprint = "2401.16747",
    archivePrefix = "arXiv",
    primaryClass = "hep-ph",
    doi = "10.1103/f2dq-hq2j",
    journal = "Phys. Rev. D",
    volume = "112",
    number = "8",
    pages = "083052",
    year = "2025"
}

@article{Dekker:2021bos,
    author = "Dekker, Ariane and Peerbooms, Ebo and Zimmer, Fabian and Ng, Kenny C. Y. and Ando, Shin'ichiro",
    title = "{Searches for sterile neutrinos and axionlike particles from the Galactic halo with eROSITA}",
    eprint = "2103.13241",
    archivePrefix = "arXiv",
    primaryClass = "astro-ph.HE",
    doi = "10.1103/PhysRevD.104.023021",
    journal = "Phys. Rev. D",
    volume = "104",
    number = "2",
    pages = "023021",
    year = "2021"
}

@article{Sun:2023acy,
    author = "Sun, Yitian and Foster, Joshua W. and Liu, Hongwan and Mu{\~n}oz, Julian B. and Slatyer, Tracy R.",
    title = "{Inhomogeneous energy injection in the 21-cm power spectrum: Sensitivity to dark matter decay}",
    eprint = "2312.11608",
    archivePrefix = "arXiv",
    primaryClass = "hep-ph",
    reportNumber = "MIT-CTP/5657, FERMILAB-PUB-23-0816-T-V",
    doi = "10.1103/PhysRevD.111.043015",
    journal = "Phys. Rev. D",
    volume = "111",
    number = "4",
    pages = "043015",
    year = "2025"
}

@article{Nakayama:2022jza,
    author = "Nakayama, Kazunori and Yin, Wen",
    title = "{Anisotropic cosmic optical background bound for decaying dark matter in light of the LORRI anomaly}",
    eprint = "2205.01079",
    archivePrefix = "arXiv",
    primaryClass = "hep-ph",
    reportNumber = "TU-1154",
    doi = "10.1103/PhysRevD.106.103505",
    journal = "Phys. Rev. D",
    volume = "106",
    number = "10",
    pages = "103505",
    year = "2022"
}

@article{Carenza:2023qxh,
    author = "Carenza, Pierluca and Lucente, Giuseppe and Vitagliano, Edoardo",
    title = "{Probing the blue axion with cosmic optical background anisotropies}",
    eprint = "2301.06560",
    archivePrefix = "arXiv",
    primaryClass = "hep-ph",
    doi = "10.1103/PhysRevD.107.083032",
    journal = "Phys. Rev. D",
    volume = "107",
    number = "8",
    pages = "083032",
    year = "2023"
}

@article{Todarello:2024qci,
    author = "Todarello, Elisa and Regis, Marco",
    title = "{Bounds on axions-like particles shining in the ultra-violet}",
    eprint = "2412.02543",
    archivePrefix = "arXiv",
    primaryClass = "hep-ph",
    doi = "10.1088/1475-7516/2025/05/070",
    journal = "JCAP",
    volume = "05",
    pages = "070",
    year = "2025"
}

@article{Calore:2022pks,
    author = "Calore, Francesca and Dekker, Ariane and Serpico, Pasquale Dario and Siegert, Thomas",
    title = "{Constraints on light decaying dark matter candidates from 16~yr of INTEGRAL/SPI observations}",
    eprint = "2209.06299",
    archivePrefix = "arXiv",
    primaryClass = "hep-ph",
    doi = "10.1093/mnras/stad457",
    journal = "Mon. Not. Roy. Astron. Soc.",
    volume = "520",
    number = "3",
    pages = "4167--4172",
    year = "2023",
    note = "[Erratum: Mon.Not.Roy.Astron.Soc. 538, 132 (2025)]"
}

@article{Janish:2023kvi,
    author = "Janish, Ryan and Pinetti, Elena",
    title = "{Hunting Dark Matter Lines in the Infrared Background with the James Webb Space Telescope}",
    eprint = "2310.15395",
    archivePrefix = "arXiv",
    primaryClass = "hep-ph",
    reportNumber = "FERMILAB-PUB-23-633-T",
    doi = "10.1103/PhysRevLett.134.071002",
    journal = "Phys. Rev. Lett.",
    volume = "134",
    number = "7",
    pages = "071002",
    year = "2025"
}

@article{Saha:2025any,
    author = "Saha, Akash Kumar and Bouri, Subhadip and Das, Anirban and Dubey, Abhishek and Laha, Ranjan",
    title = "{Shedding Infrared Light on QCD Axion and ALP Dark Matter with JWST}",
    eprint = "2503.14582",
    journal= " ",
    archivePrefix = "arXiv",
    primaryClass = "hep-ph",
    month = "3",
    year = "2025"
}

@article{Wadekar:2021qae,
    author = "Wadekar, Digvijay and Wang, Zihui",
    title = "{Strong constraints on decay and annihilation of dark matter from heating of gas-rich dwarf galaxies}",
    eprint = "2111.08025",
    archivePrefix = "arXiv",
    primaryClass = "hep-ph",
    doi = "10.1103/PhysRevD.106.075007",
    journal = "Phys. Rev. D",
    volume = "106",
    number = "7",
    pages = "075007",
    year = "2022"
}

@article{Wang:2023imi,
    author = "Wang, Hanyue and others",
    title = "{Spectroscopic search for optical emission lines from dark matter decay}",
    eprint = "2311.05476",
    archivePrefix = "arXiv",
    primaryClass = "astro-ph.CO",
    doi = "10.1103/PhysRevD.110.103007",
    journal = "Phys. Rev. D",
    volume = "110",
    number = "10",
    pages = "103007",
    year = "2024"
}

@article{Todarello:2023hdk,
    author = "Todarello, Elisa and Regis, Marco and Reynoso-Cordova, Javier and Taoso, Marco and Vaz, Daniel and Brinchmann, Jarle and Steinmetz, Matthias and Zoutendijke, Sebastiaan L.",
    title = "{Robust bounds on ALP dark matter from dwarf spheroidal galaxies in the optical MUSE-Faint survey}",
    eprint = "2307.07403",
    archivePrefix = "arXiv",
    primaryClass = "astro-ph.CO",
    doi = "10.1088/1475-7516/2024/05/043",
    journal = "JCAP",
    volume = "05",
    pages = "043",
    year = "2024"
}

@article{Thorpe-Morgan:2020rwc,
    author = {Thorpe-Morgan, Charles and Malyshev, Denys and Santangelo, Andrea and Jochum, Josef and J{\"a}ger, Barbara and Sasaki, Manami and Saeedi, Sara},
    title = "{THESEUS insights into axionlike particles, dark photon, and sterile neutrino dark matter}",
    eprint = "2008.08306",
    archivePrefix = "arXiv",
    primaryClass = "astro-ph.HE",
    doi = "10.1103/PhysRevD.102.123003",
    journal = "Phys. Rev. D",
    volume = "102",
    number = "12",
    pages = "123003",
    year = "2020"
}

@article{Yin:2024lla,
    author = "Yin, Wen and others",
    title = "{First Result for Dark Matter Search by WINERED}",
    eprint = "2402.07976",
    archivePrefix = "arXiv",
    primaryClass = "astro-ph.CO",
    reportNumber = "TU-1220",
    doi = "10.1103/PhysRevLett.134.051004",
    journal = "Phys. Rev. Lett.",
    volume = "134",
    number = "5",
    pages = "051004",
    year = "2025"
}

@article{Foster:2021ngm,
    author = "Foster, Joshua W. and Kongsore, Marius and Dessert, Christopher and Park, Yujin and Rodd, Nicholas L. and Cranmer, Kyle and Safdi, Benjamin R.",
    title = "{Deep Search for Decaying Dark Matter with XMM-Newton Blank-Sky Observations}",
    eprint = "2102.02207",
    archivePrefix = "arXiv",
    primaryClass = "astro-ph.CO",
    reportNumber = "LCTP-21-05",
    doi = "10.1103/PhysRevLett.127.051101",
    journal = "Phys. Rev. Lett.",
    volume = "127",
    number = "5",
    pages = "051101",
    year = "2021"
}

@article{Yin:2025xad,
    author = "Yin, Wen and Fujita, Yutaka and Ezoe, Yuichiro and Ishisaki, Yoshitaka",
    title = "{Double Narrow-Line Signatures of Dark Matter Decay and New Constraints from XRISM Observations}",
    eprint = "2503.04726",
    journal= " ",
    archivePrefix = "arXiv",
    primaryClass = "hep-ph",
    month = "3",
    year = "2025"
}

@article{Dessert:2023vyl,
    author = "Dessert, Christopher and Ning, Orion and Rodd, Nicholas L. and Safdi, Benjamin R.",
    title = "{Resurrecting Hitomi for Decaying Dark Matter and Forecasting Leading Sensitivity for XRISM}",
    eprint = "2305.17160",
    archivePrefix = "arXiv",
    primaryClass = "astro-ph.CO",
    reportNumber = "CERN-TH-2023-088",
    doi = "10.1103/PhysRevLett.132.211002",
    journal = "Phys. Rev. Lett.",
    volume = "132",
    number = "21",
    pages = "211002",
    year = "2024"
}

@article{Ayala:2014pea,
    author = "Ayala, Adrian and Dom{\'\i}nguez, Inma and Giannotti, Maurizio and Mirizzi, Alessandro and Straniero, Oscar",
    title = "{Revisiting the bound on axion-photon coupling from Globular Clusters}",
    eprint = "1406.6053",
    archivePrefix = "arXiv",
    primaryClass = "astro-ph.SR",
    doi = "10.1103/PhysRevLett.113.191302",
    journal = "Phys. Rev. Lett.",
    volume = "113",
    number = "19",
    pages = "191302",
    year = "2014"
}

@article{Vinyoles:2015aba,
    author = "Vinyoles, N{\'u}ria and Serenelli, Aldo and Villante, Francesco L. and Basu, Sarbani and Redondo, Javier and Isern, Jordi",
    title = "{New axion and hidden photon constraints from a solar data global fit}",
    eprint = "1501.01639",
    archivePrefix = "arXiv",
    primaryClass = "astro-ph.SR",
    doi = "10.1088/1475-7516/2015/10/015",
    journal = "JCAP",
    volume = "10",
    pages = "015",
    year = "2015"
}

@article{Meyer:2020vzy,
    author = "Meyer, Manuel and Petrushevska, Tanja",
    title = "{Search for Axionlike-Particle-Induced Prompt $\gamma$-Ray Emission from Extragalactic Core-Collapse Supernovae with the $Fermi$ Large Area Telescope}",
    eprint = "2006.06722",
    archivePrefix = "arXiv",
    primaryClass = "astro-ph.HE",
    doi = "10.1103/PhysRevLett.124.231101",
    journal = "Phys. Rev. Lett.",
    volume = "124",
    number = "23",
    pages = "231101",
    year = "2020",
    note = "[Erratum: Phys.Rev.Lett. 125, 119901 (2020)]"
}

@article{Dessert:2020lil,
    author = "Dessert, Christopher and Foster, Joshua W. and Safdi, Benjamin R.",
    title = "{X-ray Searches for Axions from Super Star Clusters}",
    eprint = "2008.03305",
    archivePrefix = "arXiv",
    primaryClass = "hep-ph",
    doi = "10.1103/PhysRevLett.125.261102",
    journal = "Phys. Rev. Lett.",
    volume = "125",
    number = "26",
    pages = "261102",
    year = "2020"
}

@article{Li:2020pcn,
    author = "Li, Hai-Jun and Guo, Jun-Guang and Bi, Xiao-Jun and Lin, Su-Jie and Yin, Peng-Fei",
    title = "{Limits on axion-like particles from Mrk 421 with 4.5-year period observations by ARGO-YBJ and Fermi-LAT}",
    eprint = "2008.09464",
    archivePrefix = "arXiv",
    primaryClass = "astro-ph.HE",
    doi = "10.1103/PhysRevD.103.083003",
    journal = "Phys. Rev. D",
    volume = "103",
    number = "8",
    pages = "083003",
    year = "2021"
}

@article{Dolan:2021rya,
    author = "Dolan, Matthew J. and Hiskens, Frederick J. and Volkas, Raymond R.",
    title = "{Constraining axion-like particles using the white dwarf initial-final mass relation}",
    eprint = "2102.00379",
    archivePrefix = "arXiv",
    primaryClass = "hep-ph",
    doi = "10.1088/1475-7516/2021/09/010",
    journal = "JCAP",
    volume = "09",
    pages = "010",
    year = "2021"
}

@article{Dessert:2021bkv,
    author = "Dessert, Christopher and Long, Andrew J. and Safdi, Benjamin R.",
    title = "{No Evidence for Axions from Chandra Observation of the Magnetic White Dwarf RE J0317-853}",
    eprint = "2104.12772",
    archivePrefix = "arXiv",
    primaryClass = "hep-ph",
    doi = "10.1103/PhysRevLett.128.071102",
    journal = "Phys. Rev. Lett.",
    volume = "128",
    number = "7",
    pages = "071102",
    year = "2022"
}

@article{Caputo:2022mah,
    author = "Caputo, Andrea and Janka, Hans-Thomas and Raffelt, Georg and Vitagliano, Edoardo",
    title = "{Low-Energy Supernovae Severely Constrain Radiative Particle Decays}",
    eprint = "2201.09890",
    archivePrefix = "arXiv",
    primaryClass = "astro-ph.HE",
    doi = "10.1103/PhysRevLett.128.221103",
    journal = "Phys. Rev. Lett.",
    volume = "128",
    number = "22",
    pages = "221103",
    year = "2022"
}

@article{Dessert:2022yqq,
    author = "Dessert, Christopher and Dunsky, David and Safdi, Benjamin R.",
    title = "{Upper limit on the axion-photon coupling from magnetic white dwarf polarization}",
    eprint = "2203.04319",
    archivePrefix = "arXiv",
    primaryClass = "hep-ph",
    doi = "10.1103/PhysRevD.105.103034",
    journal = "Phys. Rev. D",
    volume = "105",
    number = "10",
    pages = "103034",
    year = "2022"
}

@article{DeRocco:2022jyq,
    author = "DeRocco, William and Wegsman, Shalma and Grefenstette, Brian and Huang, Junwu and Van Tilburg, Ken",
    title = "{First Indirect Detection Constraints on Axions in the Solar Basin}",
    eprint = "2205.05700",
    archivePrefix = "arXiv",
    primaryClass = "hep-ph",
    doi = "10.1103/PhysRevLett.129.101101",
    journal = "Phys. Rev. Lett.",
    volume = "129",
    number = "10",
    pages = "101101",
    year = "2022"
}

@article{Dolan:2022kul,
    author = "Dolan, Matthew J. and Hiskens, Frederick J. and Volkas, Raymond R.",
    title = "{Advancing globular cluster constraints on the axion-photon coupling}",
    eprint = "2207.03102",
    archivePrefix = "arXiv",
    primaryClass = "hep-ph",
    doi = "10.1088/1475-7516/2022/10/096",
    journal = "JCAP",
    volume = "10",
    pages = "096",
    year = "2022"
}

@article{Hoof:2022xbe,
    author = "Hoof, Sebastian and Schulz, Lena",
    title = "{Updated constraints on axion-like particles from temporal information in supernova SN1987A gamma-ray data}",
    eprint = "2212.09764",
    archivePrefix = "arXiv",
    primaryClass = "hep-ph",
    reportNumber = "TTP22-072",
    doi = "10.1088/1475-7516/2023/03/054",
    journal = "JCAP",
    volume = "03",
    pages = "054",
    year = "2023"
}

@article{Beaufort:2023zuj,
    author = "Beaufort, Cyprien and Bastero-Gil, Mar and Luce, Tiffany and Santos, Daniel",
    title = "{New solar x-ray constraints on keV axionlike particles}",
    eprint = "2303.06968",
    archivePrefix = "arXiv",
    primaryClass = "hep-ph",
    doi = "10.1103/PhysRevD.108.L081302",
    journal = "Phys. Rev. D",
    volume = "108",
    number = "8",
    pages = "L081302",
    year = "2023"
}

@article{Diamond:2023scc,
    author = "Diamond, Melissa and Fiorillo, Damiano F. G. and Marques-Tavares, Gustavo and Vitagliano, Edoardo",
    title = "{Axion-sourced fireballs from supernovae}",
    eprint = "2303.11395",
    archivePrefix = "arXiv",
    primaryClass = "hep-ph",
    doi = "10.1103/PhysRevD.107.103029",
    journal = "Phys. Rev. D",
    volume = "107",
    number = "10",
    pages = "103029",
    year = "2023",
    note = "[Erratum: Phys.Rev.D 108, 049902 (2023)]"
}

@article{Muller:2023vjm,
    author = {M{\"u}ller, Eike and Calore, Francesca and Carenza, Pierluca and Eckner, Christopher and Marsh, M. C. David},
    title = "{Investigating the gamma-ray burst from decaying MeV-scale axion-like particles produced in supernova explosions}",
    eprint = "2304.01060",
    archivePrefix = "arXiv",
    primaryClass = "astro-ph.HE",
    doi = "10.1088/1475-7516/2023/07/056",
    journal = "JCAP",
    volume = "07",
    pages = "056",
    year = "2023"
}

@article{Nguyen:2023czp,
    author = "Nguyen, Ngan H. and Tanin, Erwin H. and Kamionkowski, Marc",
    title = "{Spectra of axions emitted from main sequence stars}",
    eprint = "2307.11216",
    archivePrefix = "arXiv",
    primaryClass = "hep-ph",
    doi = "10.1088/1475-7516/2023/11/091",
    journal = "JCAP",
    volume = "11",
    pages = "091",
    year = "2023"
}

@article{Manzari:2024jns,
    author = "Manzari, Claudio Andrea and Park, Yujin and Safdi, Benjamin R. and Savoray, Inbar",
    title = "{Supernova Axions Convert to Gamma Rays in Magnetic Fields of Progenitor Stars}",
    eprint = "2405.19393",
    archivePrefix = "arXiv",
    primaryClass = "hep-ph",
    doi = "10.1103/PhysRevLett.133.211002",
    journal = "Phys. Rev. Lett.",
    volume = "133",
    number = "21",
    pages = "211002",
    year = "2024"
}

@article{Ruz:2024gkl,
    author = "Ruz, J. and others",
    title = "{NuSTAR as an Axion Helioscope}",
    eprint = "2407.03828",
    archivePrefix = "arXiv",
    primaryClass = "astro-ph.CO",
    doi = "10.1103/18sn-hxtb",
    journal = "Phys. Rev. Lett.",
    volume = "135",
    number = "14",
    pages = "141001",
    year = "2025"
}

@article{Fiorillo:2025yzf,
    author = "Fiorillo, Damiano F. G. and Pitik, Tetyana and Vitagliano, Edoardo",
    title = "{Energy Transfer by Feebly Interacting Particles in Supernovae: The Trapping Regime}",
    eprint = "2503.13653",
    archivePrefix = "arXiv",
    primaryClass = "hep-ph",
    doi = "10.1103/cz94-dqxt",
    journal = "Phys. Rev. Lett.",
    volume = "135",
    number = "7",
    pages = "071005",
    year = "2025"
}

@preprint{Benabou:2025jcv,
    author = "Benabou, Joshua N. and Dessert, Christopher and Patra, Kishore C. and Brink, Thomas G. and Zheng, WeiKang and Filippenko, Alexei V. and Safdi, Benjamin R.",
    title = "{Search for Axions in Magnetic White Dwarf Polarization at Lick and Keck Observatories}",
    eprint = "2504.12377",
    archivePrefix = "arXiv",
    primaryClass = "hep-ph",
    month = "4",
    year = "2025"
}

@article{Arvanitaki:2014wva,
    author = "Arvanitaki, Asimina and Baryakhtar, Masha and Huang, Xinlu",
    title = "{Discovering the QCD Axion with Black Holes and Gravitational Waves}",
    eprint = "1411.2263",
    archivePrefix = "arXiv",
    primaryClass = "hep-ph",
    doi = "10.1103/PhysRevD.91.084011",
    journal = "Phys. Rev. D",
    volume = "91",
    number = "8",
    pages = "084011",
    year = "2015"
}

@article{Baryakhtar:2020gao,
    author = "Baryakhtar, Masha and Galanis, Marios and Lasenby, Robert and Simon, Olivier",
    title = "{Black hole superradiance of self-interacting scalar fields}",
    eprint = "2011.11646",
    archivePrefix = "arXiv",
    primaryClass = "hep-ph",
    doi = "10.1103/PhysRevD.103.095019",
    journal = "Phys. Rev. D",
    volume = "103",
    number = "9",
    pages = "095019",
    year = "2021"
}

@article{Koppell:2025dmt,
    author = "Koppell, Stewart and Bittencourt, Otavio D. A. R. and Paul, Dip Joti and Huang, Junwu and Baryakhtar, Masha and Berggren, Karl K.",
    title = "{Dark Matter Haloscope with a Disordered Dielectric Absorber}",
    eprint = "2506.00115",
    journal ="",
    archivePrefix = "arXiv",
    primaryClass = "hep-ph",
    month = "5",
    year = "2025"
}

@article{Hoof:2024quk,
    author = "Hoof, Sebastian and Marsh, David J. E. and Sisk-Reyn{\'e}s, J{\'u}lia and Matthews, James H. and Reynolds, Christopher",
    title = "{Getting more out of black hole superradiance: a statistically rigorous approach to ultralight boson constraints from black hole spin measurements}",
    eprint = "2406.10337",
    archivePrefix = "arXiv",
    primaryClass = "hep-ph",
    doi = "10.1093/mnras/staf1564",
    journal = "Mon. Not. Roy. Astron. Soc.",
    volume = "546",
    number = "2",
    pages = "staf1564",
    year = "2026"
}

@article{Caputo:2025oap,
    author = "Caputo, Andrea and Franciolini, Gabriele and Witte, Samuel J.",
    title = "{Superradiance Constraints from GW231123}",
    eprint = "2507.21788",
    journal="",
    archivePrefix = "arXiv",
    primaryClass = "hep-ph",
    reportNumber = "CERN-TH-2025-147, DESY-25-110",
    month = "7",
    year = "2025"
}

@article{Witte:2024drg,
    author = "Witte, Samuel J. and Mummery, Andrew",
    title = "{Stepping up superradiance constraints on axions}",
    eprint = "2412.03655",
    archivePrefix = "arXiv",
    primaryClass = "hep-ph",
    doi = "10.1103/PhysRevD.111.083044",
    journal = "Phys. Rev. D",
    volume = "111",
    number = "8",
    pages = "083044",
    year = "2025"
}

@article{Aswathi:2025nxa,
    author = "Aswathi, P. S. and East, William E. and Siemonsen, Nils and Sun, Ling and Jones, Dana",
    title = "{Ultralight boson constraints from gravitational wave observations of spinning binary black holes}",
    eprint = "2507.20979",
    archivePrefix = "arXiv",
    primaryClass = "gr-qc",
    doi = "10.1103/n5hr-zljn",
    journal = "Phys. Rev. D",
    volume = "112",
    number = "12",
    pages = "123048",
    year = "2025"
}

@article{Ng:2020ruv,
    author = "Ng, Ken K. Y. and Vitale, Salvatore and Hannuksela, Otto A. and Li, Tjonnie G. F.",
    title = "{Constraints on Ultralight Scalar Bosons within Black Hole Spin Measurements from the LIGO-Virgo GWTC-2}",
    eprint = "2011.06010",
    archivePrefix = "arXiv",
    primaryClass = "gr-qc",
    doi = "10.1103/PhysRevLett.126.151102",
    journal = "Phys. Rev. Lett.",
    volume = "126",
    number = "15",
    pages = "151102",
    year = "2021"
}

@article{Liddle:1998jc,
    author = "Liddle, Andrew R. and Mazumdar, Anupam and Schunck, Franz E.",
    title = "{Assisted inflation}",
    eprint = "astro-ph/9804177",
    archivePrefix = "arXiv",
    reportNumber = "SUSSEX-AST-98-4-3",
    doi = "10.1103/PhysRevD.58.061301",
    journal = "Phys. Rev. D",
    volume = "58",
    pages = "061301",
    year = "1998"
}

@article{Dimopoulos:2005ac,
    author = "Dimopoulos, S. and Kachru, S. and McGreevy, J. and Wacker, Jay G.",
    title = "{N-flation}",
    eprint = "hep-th/0507205",
    archivePrefix = "arXiv",
    reportNumber = "SLAC-PUB-11016, SU-ITP-05-08",
    doi = "10.1088/1475-7516/2008/08/003",
    journal = "JCAP",
    volume = "08",
    pages = "003",
    year = "2008"
}

@article{Kim:2004rp,
    author = "Kim, Jihn E. and Nilles, Hans Peter and Peloso, Marco",
    title = "{Completing natural inflation}",
    eprint = "hep-ph/0409138",
    archivePrefix = "arXiv",
    doi = "10.1088/1475-7516/2005/01/005",
    journal = "JCAP",
    volume = "01",
    pages = "005",
    year = "2005"
}

@article{Bachlechner:2017hsj,
    author = "Bachlechner, Thomas C. and Eckerle, Kate and Janssen, Oliver and Kleban, Matthew",
    title = "{Systematics of Aligned Axions}",
    eprint = "1709.01080",
    archivePrefix = "arXiv",
    primaryClass = "hep-th",
    doi = "10.1007/JHEP11(2017)036",
    journal = "JHEP",
    volume = "11",
    pages = "036",
    year = "2017"
}

@article{Bachlechner:2018gew,
    author = "Bachlechner, Thomas C. and Eckerle, Kate and Janssen, Oliver and Kleban, Matthew",
    title = "{Axion Landscape Cosmology}",
    eprint = "1810.02822",
    archivePrefix = "arXiv",
    primaryClass = "hep-th",
    doi = "10.1088/1475-7516/2019/09/062",
    journal = "JCAP",
    volume = "09",
    pages = "062",
    year = "2019"
}

@article{Bachlechner:2017zpb,
    author = "Bachlechner, Thomas C. and Eckerle, Kate and Janssen, Oliver and Kleban, Matthew",
    title = "{Multiple-axion framework}",
    eprint = "1703.00453",
    archivePrefix = "arXiv",
    primaryClass = "hep-th",
    doi = "10.1103/PhysRevD.98.061301",
    journal = "Phys. Rev. D",
    volume = "98",
    number = "6",
    pages = "061301",
    year = "2018"
}

@article{Bachlechner:2019vcb,
    author = "Bachlechner, Thomas C. and Eckerle, Kate and Janssen, Oliver and Kleban, Matthew",
    title = "{The Axidental Universe}",
    eprint = "1902.05952",
    archivePrefix = "arXiv",
    primaryClass = "hep-th",
    doi = "10.1088/1475-7516/2025/03/050",
    journal = "JCAP",
    volume = "03",
    pages = "050",
    year = "2025"
}

@article{Bachlechner:2014hsa,
    author = "Bachlechner, Thomas C. and Dias, Mafalda and Frazer, Jonathan and McAllister, Liam",
    title = "{Chaotic inflation with kinetic alignment of axion fields}",
    eprint = "1404.7496",
    archivePrefix = "arXiv",
    primaryClass = "hep-th",
    doi = "10.1103/PhysRevD.91.023520",
    journal = "Phys. Rev. D",
    volume = "91",
    number = "2",
    pages = "023520",
    year = "2015"
}

@article{Kim:2007bc,
    author = "Kim, Soo A and Liddle, Andrew R.",
    title = "{Nflation: observable predictions from the random matrix mass spectrum}",
    eprint = "0707.1982",
    archivePrefix = "arXiv",
    primaryClass = "astro-ph",
    doi = "10.1103/PhysRevD.76.063515",
    journal = "Phys. Rev. D",
    volume = "76",
    pages = "063515",
    year = "2007"
}

@article{Stott:2017hvl,
    author = "Stott, Matthew J. and Marsh, David J. E. and Pongkitivanichkul, Chakrit and Price, Layne C. and Acharya, Bobby S.",
    title = "{Spectrum of the axion dark sector}",
    eprint = "1706.03236",
    archivePrefix = "arXiv",
    primaryClass = "astro-ph.CO",
    doi = "10.1103/PhysRevD.96.083510",
    journal = "Phys. Rev. D",
    volume = "96",
    number = "8",
    pages = "083510",
    year = "2017"
}

@article{Gendler:2023hwg,
    author = "Gendler, Naomi and Janssen, Oliver and Kleban, Matthew and La Madrid, Joan and Mehta, Viraf M.",
    title = "{Axion minima in string theory}",
    eprint = "2309.01831",
    archivePrefix = "arXiv",
    primaryClass = "hep-th",
    doi = "10.1007/JHEP02(2025)134",
    journal = "JHEP",
    volume = "02",
    pages = "134",
    year = "2025"
}

@article{GrillidiCortona:2015jxo,
    author = "Grilli di Cortona, Giovanni and Hardy, Edward and Pardo Vega, Javier and Villadoro, Giovanni",
    title = "{The QCD axion, precisely}",
    eprint = "1511.02867",
    archivePrefix = "arXiv",
    primaryClass = "hep-ph",
    doi = "10.1007/JHEP01(2016)034",
    journal = "JHEP",
    volume = "01",
    pages = "034",
    year = "2016"
}

@article{Li:2025cep,
    author = "Li, Hai-Jun and Zhou, Yu-Feng",
    title = "{Axion Mixing in the String Axiverse}",
    eprint = "2504.10170",
    archivePrefix = "arXiv",
    journal = " ",
    primaryClass = "hep-th",
    reportNumber = "ITP-CAS-25-088",
    month = "4",
    year = "2025"
}

@article{Demirtas:2018akl,
    author = "Demirtas, Mehmet and Long, Cody and McAllister, Liam and Stillman, Mike",
    title = "{The Kreuzer-Skarke Axiverse}",
    eprint = "1808.01282",
    archivePrefix = "arXiv",
    primaryClass = "hep-th",
    doi = "10.1007/JHEP04(2020)138",
    journal = "JHEP",
    volume = "04",
    pages = "138",
    year = "2020"
}

@article{Mehta:2021pwf,
    author = "Mehta, Viraf M. and Demirtas, Mehmet and Long, Cody and Marsh, David J. E. and McAllister, Liam and Stott, Matthew J.",
    title = "{Superradiance in string theory}",
    eprint = "2103.06812",
    archivePrefix = "arXiv",
    primaryClass = "hep-th",
    doi = "10.1088/1475-7516/2021/07/033",
    journal = "JCAP",
    volume = "07",
    pages = "033",
    year = "2021"
}

@article{Gendler:2023kjt,
    author = "Gendler, Naomi and Marsh, David J. E. and McAllister, Liam and Moritz, Jakob",
    title = "{Glimmers from the axiverse}",
    eprint = "2309.13145",
    archivePrefix = "arXiv",
    primaryClass = "hep-th",
    reportNumber = "KCL-PH-TH/2023-49",
    doi = "10.1088/1475-7516/2024/09/071",
    journal = "JCAP",
    volume = "09",
    pages = "071",
    year = "2024"
}

@article{Demirtas:2021gsq,
    author = "Demirtas, Mehmet and Gendler, Naomi and Long, Cody and McAllister, Liam and Moritz, Jakob",
    title = "{PQ axiverse}",
    eprint = "2112.04503",
    archivePrefix = "arXiv",
    primaryClass = "hep-th",
    doi = "10.1007/JHEP06(2023)092",
    journal = "JHEP",
    volume = "06",
    pages = "092",
    year = "2023"
}

@preprint{Demirtas:2022hqf,
    author = "Demirtas, Mehmet and Rios-Tascon, Andres and McAllister, Liam",
    title = "{CYTools: A Software Package for Analyzing Calabi-Yau Manifolds}",
    eprint = "2211.03823",
    archivePrefix = "arXiv",
    primaryClass = "hep-th",
    month = "11",
    year = "2022"
}

@article{Sheridan:2024vtt,
    author = "Sheridan, Elijah and Carta, Federico and Gendler, Naomi and Jain, Mudit and Marsh, David J. E. and McAllister, Liam and Righi, Nicole and Rogers, Keir K. and Schachner, Andreas",
    title = "{Fuzzy axions and associated relics}",
    eprint = "2412.12012",
    archivePrefix = "arXiv",
    primaryClass = "hep-th",
    reportNumber = "KCL-PH-TH/2024-75, KCL-PH-TH/2024-75",
    doi = "10.1007/JHEP09(2025)016",
    journal = "JHEP",
    volume = "09",
    pages = "016",
    year = "2025"
}

@article{Gendler:2024adn,
    author = "Gendler, Naomi and Marsh, David J. E.",
    title = "{Possible Implications of QCD Axion Dark Matter Constraints from Helioscopes and Haloscopes for the String Theory Landscape}",
    eprint = "2407.07143",
    archivePrefix = "arXiv",
    primaryClass = "hep-th",
    doi = "10.1103/PhysRevLett.134.081602",
    journal = "Phys. Rev. Lett.",
    volume = "134",
    number = "8",
    pages = "081602",
    year = "2025"
}

@article{Cheng:2025ggf,
    author = "Cheng, Junyi and Gendler, Naomi",
    title = "{Universality in the axiverse}",
    eprint = "2507.12516",
    archivePrefix = "arXiv",
    primaryClass = "hep-th",
    doi = "10.1007/JHEP11(2025)012",
    journal = "JHEP",
    volume = "11",
    pages = "012",
    year = "2025"
}

@preprint{Fallon:2025lvn,
    author = "Fallon, Sebastian Vander Ploeg and Halverson, James and McAllister, Liam and Zhu, Yunhao",
    title = "{F-theory Axiverse}",
    eprint = "2511.20458",
    archivePrefix = "arXiv",
    primaryClass = "hep-th",
    month = "11",
    year = "2025"
}

@article{Halverson:2019cmy,
    author = "Halverson, James and Long, Cody and Nelson, Brent and Salinas, Gustavo",
    title = "{Towards string theory expectations for photon couplings to axionlike particles}",
    eprint = "1909.05257",
    archivePrefix = "arXiv",
    primaryClass = "hep-th",
    doi = "10.1103/PhysRevD.100.106010",
    journal = "Phys. Rev. D",
    volume = "100",
    number = "10",
    pages = "106010",
    year = "2019"
}

@article{Halverson:2019kna,
    author = "Halverson, James and Long, Cody and Nelson, Brent and Salinas, Gustavo",
    title = "{Axion reheating in the string landscape}",
    eprint = "1903.04495",
    archivePrefix = "arXiv",
    primaryClass = "hep-th",
    doi = "10.1103/PhysRevD.99.086014",
    journal = "Phys. Rev. D",
    volume = "99",
    number = "8",
    pages = "086014",
    year = "2019"
}

@article{Cicoli:2012sz,
    author = "Cicoli, Michele and Goodsell, Mark and Ringwald, Andreas",
    title = "{The type IIB string axiverse and its low-energy phenomenology}",
    eprint = "1206.0819",
    archivePrefix = "arXiv",
    primaryClass = "hep-th",
    reportNumber = "DESY-12-058, CERN-PH-TH-2012-153",
    doi = "10.1007/JHEP10(2012)146",
    journal = "JHEP",
    volume = "10",
    pages = "146",
    year = "2012"
}

@article{Dienes:2021woi,
    author = "Dienes, Keith R. and Heurtier, Lucien and Huang, Fei and Kim, Doojin and Tait, Tim M. P. and Thomas, Brooks",
    title = "{Stasis in an expanding universe: A recipe for stable mixed-component cosmological eras}",
    eprint = "2111.04753",
    archivePrefix = "arXiv",
    primaryClass = "astro-ph.CO",
    reportNumber = "IPPP/21/47, UCI-HEP-TR-2021-28, MI-HET-767",
    doi = "10.1103/PhysRevD.105.023530",
    journal = "Phys. Rev. D",
    volume = "105",
    number = "2",
    pages = "023530",
    year = "2022"
}

@article{Dienes:2023ziv,
    author = "Dienes, Keith R. and Heurtier, Lucien and Huang, Fei and Tait, Tim M. P. and Thomas, Brooks",
    title = "{Stasis, Stasis, Triple Stasis}",
    eprint = "2309.10345",
    archivePrefix = "arXiv",
    primaryClass = "astro-ph.CO",
    doi = "10.1103/PhysRevD.109.083508",
    journal = "Phys. Rev. D",
    volume = "109",
    number = "8",
    pages = "083508",
    year = "2024"
}

@article{Dienes:2024wnu,
    author = "Dienes, Keith R. and Heurtier, Lucien and Huang, Fei and Tait, Tim M. P. and Thomas, Brooks",
    title = "{Cosmological stasis from dynamical scalars: Tracking solutions and the possibility of a stasis-induced inflation}",
    eprint = "2406.06830",
    archivePrefix = "arXiv",
    primaryClass = "astro-ph.CO",
    reportNumber = "KCL-PH-TH/2024-23, UCI-HEP-TR-2024-09",
    doi = "10.1103/PhysRevD.110.123514",
    journal = "Phys. Rev. D",
    volume = "110",
    number = "12",
    pages = "123514",
    year = "2024"
}

@preprint{Dienes:2025tox,
    author = "Dienes, Keith R. and Heurtier, Lucien and Hoover, Daniel and Huang, Fei and Paulsen, Anna and Thomas, Brooks",
    title = "{Spotting Stasis in Cosmological Perturbations}",
    eprint = "2503.19959",
    archivePrefix = "arXiv",
    primaryClass = "astro-ph.CO",
    month = "3",
    year = "2025"
}

@article{Halverson:2024oir,
    author = "Halverson, James and Pandya, Sneh",
    title = "{Generality and persistence of cosmological stasis}",
    eprint = "2408.00835",
    archivePrefix = "arXiv",
    primaryClass = "astro-ph.CO",
    doi = "10.1103/PhysRevD.110.075041",
    journal = "Phys. Rev. D",
    volume = "110",
    number = "7",
    pages = "075041",
    year = "2024"
}

@article{Dienes:2011ja,
    author = "Dienes, Keith R. and Thomas, Brooks",
    title = "{Dynamical Dark Matter: I. Theoretical Overview}",
    eprint = "1106.4546",
    archivePrefix = "arXiv",
    primaryClass = "hep-ph",
    doi = "10.1103/PhysRevD.85.083523",
    journal = "Phys. Rev. D",
    volume = "85",
    pages = "083523",
    year = "2012"
}

@article{Dienes:2011sa,
    author = "Dienes, Keith R. and Thomas, Brooks",
    title = "{Dynamical Dark Matter: II. An Explicit Model}",
    eprint = "1107.0721",
    archivePrefix = "arXiv",
    primaryClass = "hep-ph",
    doi = "10.1103/PhysRevD.85.083524",
    journal = "Phys. Rev. D",
    volume = "85",
    pages = "083524",
    year = "2012"
}

@article{Dienes:2012jb,
    author = "Dienes, Keith R. and Thomas, Brooks",
    title = "{Phenomenological Constraints on Axion Models of Dynamical Dark Matter}",
    eprint = "1203.1923",
    archivePrefix = "arXiv",
    primaryClass = "hep-ph",
    reportNumber = "UH511-1189-12",
    doi = "10.1103/PhysRevD.86.055013",
    journal = "Phys. Rev. D",
    volume = "86",
    pages = "055013",
    year = "2012"
}

@article{Co:2019wyp,
    author = "Co, Raymond T. and Harigaya, Keisuke",
    title = "{Axiogenesis}",
    eprint = "1910.02080",
    archivePrefix = "arXiv",
    primaryClass = "hep-ph",
    reportNumber = "LCTP-19-27",
    doi = "10.1103/PhysRevLett.124.111602",
    journal = "Phys. Rev. Lett.",
    volume = "124",
    number = "11",
    pages = "111602",
    year = "2020"
}

@preprint{Asadi:2025cvm,
    author = "Asadi, Pouya and Cyncynates, David and Gori, Stefania",
    title = "{Axiverse Baryogenesis}",
    eprint = "2511.15794",
    archivePrefix = "arXiv",
    primaryClass = "hep-ph",
    month = "11",
    year = "2025"
}

@article{Agrawal:2019lkr,
    author = "Agrawal, Prateek and Hook, Anson and Huang, Junwu",
    title = "{A CMB Millikan experiment with cosmic axiverse strings}",
    eprint = "1912.02823",
    archivePrefix = "arXiv",
    primaryClass = "astro-ph.CO",
    doi = "10.1007/JHEP07(2020)138",
    journal = "JHEP",
    volume = "07",
    pages = "138",
    year = "2020"
}

@article{Agrawal:2020euj,
    author = "Agrawal, Prateek and Hook, Anson and Huang, Junwu and Marques-Tavares, Gustavo",
    title = "{Axion string signatures: a cosmological plasma collider}",
    eprint = "2010.15848",
    archivePrefix = "arXiv",
    primaryClass = "hep-ph",
    doi = "10.1007/JHEP01(2022)103",
    journal = "JHEP",
    volume = "01",
    pages = "103",
    year = "2022"
}

@article{Daido:2015bva,
    author = "Daido, Ryuji and Kitajima, Naoya and Takahashi, Fuminobu",
    title = "{Domain Wall Formation from Level Crossing in the Axiverse}",
    eprint = "1505.07670",
    archivePrefix = "arXiv",
    primaryClass = "hep-ph",
    reportNumber = "TU-995, IPMU15-0076",
    doi = "10.1103/PhysRevD.92.063512",
    journal = "Phys. Rev. D",
    volume = "92",
    number = "6",
    pages = "063512",
    year = "2015"
}

@article{Daido:2015cba,
    author = "Daido, Ryuji and Kitajima, Naoya and Takahashi, Fuminobu",
    title = "{Level crossing between the QCD axion and an axionlike particle}",
    eprint = "1510.06675",
    archivePrefix = "arXiv",
    primaryClass = "hep-ph",
    reportNumber = "TU-1007, APCTP-PRE2015-026, IPMU15-0181",
    doi = "10.1103/PhysRevD.93.075027",
    journal = "Phys. Rev. D",
    volume = "93",
    number = "7",
    pages = "075027",
    year = "2016"
}

@article{Kitajima:2014xla,
    author = "Kitajima, Naoya and Takahashi, Fuminobu",
    title = "{Resonant conversions of QCD axions into hidden axions and suppressed isocurvature perturbations}",
    eprint = "1411.2011",
    archivePrefix = "arXiv",
    primaryClass = "hep-ph",
    reportNumber = "TU-985, IPMU14-0334",
    doi = "10.1088/1475-7516/2015/01/032",
    journal = "JCAP",
    volume = "01",
    pages = "032",
    year = "2015"
}

@article{Murai:2023xjn,
    author = "Murai, Kai and Takahashi, Fuminobu and Yin, Wen",
    title = "{QCD axion: A unique player in the axiverse with mixings}",
    eprint = "2305.18677",
    archivePrefix = "arXiv",
    primaryClass = "hep-ph",
    reportNumber = "TU-1190",
    doi = "10.1103/PhysRevD.108.036020",
    journal = "Phys. Rev. D",
    volume = "108",
    number = "3",
    pages = "036020",
    year = "2023"
}

@article{Murai:2024nsp,
    author = "Murai, Kai and Narita, Yuma and Takahashi, Fuminobu and Yin, Wen",
    title = "{QCD axion dark matter from level crossing with refined adiabatic condition}",
    eprint = "2412.10232",
    archivePrefix = "arXiv",
    primaryClass = "hep-ph",
    reportNumber = "TU-1252",
    doi = "10.1007/JHEP04(2025)124",
    journal = "JHEP",
    volume = "04",
    pages = "124",
    year = "2025"
}

@article{Li:2024okl,
    author = "Li, Hai-Jun and Zhou, Yu-Feng",
    title = "{Mass mixing between QCD axions}",
    eprint = "2408.00267",
    archivePrefix = "arXiv",
    primaryClass = "hep-ph",
    reportNumber = "ITP-CAS-24-166, ITP-24-166",
    doi = "10.1088/1674-1137/aded01",
    journal = "Chin. Phys. C",
    volume = "49",
    number = "11",
    pages = "115101",
    year = "2025"
}

@article{Li:2024kdy,
    author = "Li, Hai-Jun",
    title = "{On the temperature effects in QCD axion mass mixing}",
    eprint = "2410.00377",
    archivePrefix = "arXiv",
    primaryClass = "hep-ph",
    reportNumber = "ITP-CAS-24-226, ITP-24-226",
    doi = "10.1016/j.aop.2025.170141",
    journal = "Annals Phys.",
    volume = "480",
    pages = "170141",
    year = "2025"
}

@article{Cyncynates:2023esj,
    author = "Cyncynates, David and Thompson, Jedidiah O.",
    title = "{Heavy QCD axion dark matter from avoided level crossing}",
    eprint = "2306.04678",
    archivePrefix = "arXiv",
    primaryClass = "hep-ph",
    doi = "10.1103/PhysRevD.108.L091703",
    journal = "Phys. Rev. D",
    volume = "108",
    number = "9",
    pages = "L091703",
    year = "2023"
}

@article{Nakagawa:2020eeg,
    author = "Nakagawa, Shota and Takahashi, Fuminobu and Yin, Wen",
    title = "{Stochastic Axion Dark Matter in Axion Landscape}",
    eprint = "2002.12195",
    archivePrefix = "arXiv",
    primaryClass = "hep-ph",
    reportNumber = "TU-1097",
    doi = "10.1088/1475-7516/2020/05/004",
    journal = "JCAP",
    volume = "05",
    pages = "004",
    year = "2020"
}

@preprint{Katewongveerachart:2026ovj,
    author = "Katewongveerachart, Supakorn and Marsh, David J. E.",
    title = "{Cosmological Dynamics of Multi-Axion Quintessence}",
    eprint = "2602.00820",
    archivePrefix = "arXiv",
    primaryClass = "astro-ph.CO",
    month = "1",
    year = "2026"
}

@article{Kamionkowski:2014zda,
    author = "Kamionkowski, Marc and Pradler, Josef and Walker, Devin G. E.",
    title = "{Dark energy from the string axiverse}",
    eprint = "1409.0549",
    archivePrefix = "arXiv",
    primaryClass = "hep-ph",
    reportNumber = "SLAC-PUB-16085",
    doi = "10.1103/PhysRevLett.113.251302",
    journal = "Phys. Rev. Lett.",
    volume = "113",
    number = "25",
    pages = "251302",
    year = "2014"
}

@article{Karwal:2016vyq,
    author = "Karwal, Tanvi and Kamionkowski, Marc",
    title = "{Dark energy at early times, the Hubble parameter, and the string axiverse}",
    eprint = "1608.01309",
    archivePrefix = "arXiv",
    primaryClass = "astro-ph.CO",
    doi = "10.1103/PhysRevD.94.103523",
    journal = "Phys. Rev. D",
    volume = "94",
    number = "10",
    pages = "103523",
    year = "2016"
}

@article{Emami:2016mrt,
    author = "Emami, Razieh and Grin, Daniel and Pradler, Josef and Raccanelli, Alvise and Kamionkowski, Marc",
    title = "{Cosmological tests of an axiverse-inspired quintessence field}",
    eprint = "1603.04851",
    archivePrefix = "arXiv",
    primaryClass = "astro-ph.CO",
    doi = "10.1103/PhysRevD.93.123005",
    journal = "Phys. Rev. D",
    volume = "93",
    number = "12",
    pages = "123005",
    year = "2016"
}

@article{Dessert:2025yvk,
    author = "Dessert, Christopher and Kumar, Soubhik and Ruderman, Joshua T.",
    title = "{Freezing-in the Axiverse}",
    eprint = "2511.09631",
    journal = " ",
    archivePrefix = "arXiv",
    primaryClass = "hep-ph",
    month = "11",
    year = "2025"
}

@article{Acharya:2015zfk,
    author = "Acharya, Bobby Samir and Pongkitivanichkul, Chakrit",
    title = "{The Axiverse induced Dark Radiation Problem}",
    eprint = "1512.07907",
    archivePrefix = "arXiv",
    primaryClass = "hep-ph",
    doi = "10.1007/JHEP04(2016)009",
    journal = "JHEP",
    volume = "04",
    pages = "009",
    year = "2016"
}

@article{Bousso:2000xa,
    author = "Bousso, Raphael and Polchinski, Joseph",
    title = "{Quantization of four form fluxes and dynamical neutralization of the cosmological constant}",
    eprint = "hep-th/0004134",
    archivePrefix = "arXiv",
    reportNumber = "SU-ITP-00-12, NSF-ITP-00-40",
    doi = "10.1088/1126-6708/2000/06/006",
    journal = "JHEP",
    volume = "06",
    pages = "006",
    year = "2000"
}

@inproceedings{Polchinski:2006gy,
    author = "Polchinski, Joseph",
    title = "{The Cosmological Constant and the String Landscape}",
    booktitle = "{23rd Solvay Conference in Physics: The Quantum Structure of Space and Time}",
    eprint = "hep-th/0603249",
    archivePrefix = "arXiv",
    pages = "216--236",
    month = "3",
    year = "2006"
}

@article{Halverson:2018cio,
    author = "Halverson, James and Ruehle, Fabian",
    title = "{Computational Complexity of Vacua and Near-Vacua in Field and String Theory}",
    eprint = "1809.08279",
    archivePrefix = "arXiv",
    primaryClass = "hep-th",
    doi = "10.1103/PhysRevD.99.046015",
    journal = "Phys. Rev. D",
    volume = "99",
    number = "4",
    pages = "046015",
    year = "2019"
}

@article{Agrawal:2024ejr,
    author = "Agrawal, Prateek and Nee, Michael and Reig, Mario",
    title = "{Axion couplings in heterotic string theory}",
    eprint = "2410.03820",
    archivePrefix = "arXiv",
    primaryClass = "hep-ph",
    doi = "10.1007/JHEP02(2025)188",
    journal = "JHEP",
    volume = "02",
    pages = "188",
    year = "2025"
}

@article{Reig:2025dqb,
    author = "Reig, Mario and Weigand, Timo",
    title = "{Testing the heterotic string with the axion-photon coupling}",
    eprint = "2509.08042",
    archivePrefix = "arXiv",
    primaryClass = "hep-th",
    doi = "10.1007/JHEP01(2026)006",
    journal = "JHEP",
    volume = "01",
    pages = "006",
    year = "2026"
}

@article{Agrawal:2022lsp,
    author = "Agrawal, Prateek and Nee, Michael and Reig, Mario",
    title = "{Axion couplings in grand unified theories}",
    eprint = "2206.07053",
    archivePrefix = "arXiv",
    primaryClass = "hep-ph",
    doi = "10.1007/JHEP10(2022)141",
    journal = "JHEP",
    volume = "10",
    pages = "141",
    year = "2022"
}

@article{Agrawal:2025rbr,
    author = "Agrawal, Prateek and Nee, Michael and Reig, Mario",
    title = "{Axion couplings in Orbifold GUTs}",
    eprint = "2511.21830",
    archivePrefix = "arXiv",
    primaryClass = "hep-ph",
    month = "11",
    year = "2025",
    journal = ""
}

@article{Baumann:2016wac,
    author = "Baumann, Daniel and Green, Daniel and Wallisch, Benjamin",
    title = "{New Target for Cosmic Axion Searches}",
    eprint = "1604.08614",
    archivePrefix = "arXiv",
    primaryClass = "astro-ph.CO",
    doi = "10.1103/PhysRevLett.117.171301",
    journal = "Phys. Rev. Lett.",
    volume = "117",
    number = "17",
    pages = "171301",
    year = "2016"
}

@article{Elahi:2014fsa,
    author = "Elahi, Fatemeh and Kolda, Christopher and Unwin, James",
    title = "{UltraViolet Freeze-in}",
    eprint = "1410.6157",
    archivePrefix = "arXiv",
    primaryClass = "hep-ph",
    doi = "10.1007/JHEP03(2015)048",
    journal = "JHEP",
    volume = "03",
    pages = "048",
    year = "2015"
}

@article{Bernal:2019mhf,
    author = "Bernal, Nicol{\'a}s and Elahi, Fatemeh and Maldonado, Carlos and Unwin, James",
    title = "{Ultraviolet Freeze-in and Non-Standard Cosmologies}",
    eprint = "1909.07992",
    archivePrefix = "arXiv",
    primaryClass = "hep-ph",
    reportNumber = "PI/UAN-2019-654FT",
    doi = "10.1088/1475-7516/2019/11/026",
    journal = "JCAP",
    volume = "11",
    pages = "026",
    year = "2019"
}

@article{Salvio:2013iaa,
    author = "Salvio, Alberto and Strumia, Alessandro and Xue, Wei",
    title = "{Thermal axion production}",
    eprint = "1310.6982",
    archivePrefix = "arXiv",
    primaryClass = "hep-ph",
    reportNumber = "FTUAM-13-29, IFT-UAM-CSIC-13-113",
    doi = "10.1088/1475-7516/2014/01/011",
    journal = "JCAP",
    volume = "01",
    pages = "011",
    year = "2014"
}

@article{Visinelli:2009zm,
    author = "Visinelli, Luca and Gondolo, Paolo",
    title = "{Dark Matter Axions Revisited}",
    eprint = "0903.4377",
    archivePrefix = "arXiv",
    primaryClass = "astro-ph.CO",
    doi = "10.1103/PhysRevD.80.035024",
    journal = "Phys. Rev. D",
    volume = "80",
    pages = "035024",
    year = "2009"
}

@article{Cyncynates:2021xzw,
    author = "Cyncynates, David and Giurgica-Tiron, Tudor and Simon, Olivier and Thompson, Jedidiah O.",
    title = "{Resonant nonlinear pairs in the axiverse and their late-time direct and astrophysical signatures}",
    eprint = "2109.09755",
    archivePrefix = "arXiv",
    primaryClass = "hep-ph",
    doi = "10.1103/PhysRevD.105.055005",
    journal = "Phys. Rev. D",
    volume = "105",
    number = "5",
    pages = "055005",
    year = "2022"
}

@article{Cyncynates:2022wlq,
    author = "Cyncynates, David and Simon, Olivier and Thompson, Jedidiah O. and Weiner, Zachary J.",
    title = "{Nonperturbative structure in coupled axion sectors and implications for direct detection}",
    eprint = "2208.05501",
    archivePrefix = "arXiv",
    primaryClass = "hep-ph",
    doi = "10.1103/PhysRevD.106.083503",
    journal = "Phys. Rev. D",
    volume = "106",
    number = "8",
    pages = "083503",
    year = "2022"
}

@article{Arkani-Hamed:2006emk,
    author = "Arkani-Hamed, Nima and Motl, Lubos and Nicolis, Alberto and Vafa, Cumrun",
    title = "{The String landscape, black holes and gravity as the weakest force}",
    eprint = "hep-th/0601001",
    archivePrefix = "arXiv",
    reportNumber = "HUTP-05-A0057",
    doi = "10.1088/1126-6708/2007/06/060",
    journal = "JHEP",
    volume = "06",
    pages = "060",
    year = "2007"
}

@article{Rudelius:2014wla,
    author = "Rudelius, Tom",
    title = "{On the Possibility of Large Axion Moduli Spaces}",
    eprint = "1409.5793",
    archivePrefix = "arXiv",
    primaryClass = "hep-th",
    doi = "10.1088/1475-7516/2015/04/049",
    journal = "JCAP",
    volume = "04",
    pages = "049",
    year = "2015"
}

@article{Brown:2015lia,
    author = "Brown, Jon and Cottrell, William and Shiu, Gary and Soler, Pablo",
    title = "{On Axionic Field Ranges, Loopholes and the Weak Gravity Conjecture}",
    eprint = "1504.00659",
    archivePrefix = "arXiv",
    primaryClass = "hep-th",
    reportNumber = "MAD-TH-15-05",
    doi = "10.1007/JHEP04(2016)017",
    journal = "JHEP",
    volume = "04",
    pages = "017",
    year = "2016"
}

@article{Heidenreich:2015wga,
    author = "Heidenreich, Ben and Reece, Matthew and Rudelius, Tom",
    title = "{Weak Gravity Strongly Constrains Large-Field Axion Inflation}",
    eprint = "1506.03447",
    archivePrefix = "arXiv",
    primaryClass = "hep-th",
    doi = "10.1007/JHEP12(2015)108",
    journal = "JHEP",
    volume = "12",
    pages = "108",
    year = "2015"
}

@preprint{Gavilan-Martin:2026,
    archiveprefix = {arXiv},
    author = {Gavilan-Martin, Dani and Simon, Olivier and Krishna Dhashin and F. Jackson Kimball, Derek and Budker, Dmitry and Wickenbrock, Arne},
    eprint = {26XX.XXXXX},
    primaryclass = {hep-ph},
    title = {{Black Hole Scalar Sirens in the Milky Way}},
    year = {2026}
    }

@article{Huang:2020etx,
    author = "Huang, Junwu and Madden, Amalia and Racco, Davide and Reig, Mario",
    title = "{Maximal axion misalignment from a minimal model}",
    eprint = "2006.07379",
    archivePrefix = "arXiv",
    primaryClass = "hep-ph",
    reportNumber = "IFIC/20-22",
    doi = "10.1007/JHEP10(2020)143",
    journal = "JHEP",
    volume = "10",
    pages = "143",
    year = "2020"
}

@article{Witten:1979vv,
    author = "Witten, Edward",
    title = "{Current Algebra Theorems for the U(1) Goldstone Boson}",
    reportNumber = "HUTP-79/A014",
    doi = "10.1016/0550-3213(79)90031-2",
    journal = "Nucl. Phys. B",
    volume = "156",
    pages = "269--283",
    year = "1979"
}

@article{Co:2018mho,
    author = "Co, Raymond T. and Gonzalez, Eric and Harigaya, Keisuke",
    title = "{Axion Misalignment Driven to the Hilltop}",
    eprint = "1812.11192",
    archivePrefix = "arXiv",
    primaryClass = "hep-ph",
    reportNumber = "LCTP-18-33",
    doi = "10.1007/JHEP05(2019)163",
    journal = "JHEP",
    volume = "05",
    pages = "163",
    year = "2019"
}

@article{Bachlechner:2015qja,
    author = "Bachlechner, Thomas C. and Long, Cody and McAllister, Liam",
    title = "{Planckian Axions and the Weak Gravity Conjecture}",
    eprint = "1503.07853",
    archivePrefix = "arXiv",
    primaryClass = "hep-th",
    doi = "10.1007/JHEP01(2016)091",
    journal = "JHEP",
    volume = "01",
    pages = "091",
    year = "2016"
}

@article{Hebecker:2015rya,
    author = "Hebecker, Arthur and Mangat, Patrick and Rompineve, Fabrizio and Witkowski, Lukas T.",
    title = "{Winding out of the Swamp: Evading the Weak Gravity Conjecture with F-term Winding Inflation?}",
    eprint = "1503.07912",
    archivePrefix = "arXiv",
    primaryClass = "hep-th",
    doi = "10.1016/j.physletb.2015.07.026",
    journal = "Phys. Lett. B",
    volume = "748",
    pages = "455--462",
    year = "2015"
}

\end{document}